\documentclass{aa}  

\usepackage{graphicx}
\usepackage{orcidlink}
\usepackage{amsmath, amscd, amssymb, mathrsfs,amsfonts}

\usepackage{txfonts}
\begin{document} 

\title{The Hot Neptune Initiative (HONEI)}
\subtitle{II. TOI-5795\,b: A hot super-Neptune orbiting a metal-poor star}
   \titlerunning{The super-Neptune TOI-5795\,b}
\author{F. Manni\inst{1,2}\fnmsep\thanks{Corresponding author: francesca.manni@inaf.it}
\and
L. Naponiello\inst{2}
\and
L. Mancini\inst{1,2,3}
\and
S. Vissapragada \inst{14}
\and
K. Biazzo \inst{4}
\and
A. S. Bonomo\inst{2}
\and
D. Polychroni\inst{2}
\and
D. Turrini\inst{2} 
\and
D. Locci\inst{13} 
\and
A. Maggio\inst{13} 
\and
V. D'Orazi\inst{1}
\and            
M. Damasso\inst{2}
\and 
C. Brice\~{n}o\inst{10}
\and
D. R. Ciardi\inst{8}
\and
C. A. Clark\inst{8}
\and  
K. A. Collins\inst{5}
\and
D. W. Latham\inst{5}
\and 
N. Law\inst{11}
\and
M. L\'{o}pez-Morales\inst{16}
\and
M. B. Lund\inst{8}
\and
L. Malavolta \inst{6,7}   
\and 
A. W. Mann\inst{11}
\and
G. Mantovan\inst{6,15} 
\and
D. Nardiello\inst{6,7}
\and
M. Pinamonti\inst{2}
\and 
D. J. Radford \inst{12}
\and
R. P. Schwarz \inst{5}
\and
A. Shporer\inst{16} 
\and
A. Sozzetti\inst{2}
\and 
C. N. Watkins \inst{5}
\and        
S.~W. Yee\inst{5} 
\and 
C. Ziegler\inst{9}
\and
T. Zingales\inst{6,7} 
}
\institute{
Department of Physics, University of Rome ``Tor Vergata'', Via della Ricerca Scientifica 1, 00133 Rome, Italy.
\and
INAF -- Turin Astrophysical Observatory, via Osservatorio 20, 10025 Pino Torinese, Italy
\and
Max Planck Institute for Astronomy, K\"{o}nigstuhl 17, 69117 Heidelberg, Germany 
\and
INAF -- Osservatorio Astronomico di Roma, via Frascati 33, 00040 Monte Porzio Catone (RM), Italy
\and
Center for Astrophysics \textbar \ Harvard \& Smithsonian, 60 Garden Street, Cambridge, MA 02138, USA
\and
INAF -- Padova Astronomical Observatory, Vicolo dell’Osservatorio 5, Padova 35122, Italy
\and
Department of Physics and Astronomy ``Galileo Galilei'', University of Padova, Vicolo dell’Osservatorio 3, 35122 Padova, Italy
\and
NASA Exoplanet Science Institute-Caltech/IPAC, Pasadena, CA 91125, USA
\and
Department of Physics, Engineering and Astronomy, S. F. Austin State University, 1936 North St, Nacogdoches, TX 75962, USA
\and
Cerro Tololo Inter-American Observatory, Casilla 603, La Serena, Chile
\and
Department of Physics and Astronomy, The University of North Carolina at Chapel Hill, Chapel Hill, NC 27599-3255, USA
\and
Brierfield Observatory, Bowral, NSW 2576, Australia
\and
INAF – Palermo Astronomical Observatory, Piazza del Parlamento 1, 90134 Palermo, Italy
\and
Carnegie Science Observatories, 813 Santa Barbara Street, Pasadena, CA 91101, USA.
\and
Centro di Ateneo di Studi e Attivit\`a Spaziali ``G. Colombo'' -- Universit\`a degli Studi di Padova, Via Venezia 15, IT-35131, Padova
\and 
Space Telescope Science Institute, 3700 San Martin Drive, Baltimore, MD 21218, USA
\and 
POC, GI Office, \& MAST 
}
\date{Received 24, 06 2025; accepted 26, 07 2025}
  \abstract
{
The formation of Neptune planets with orbital periods less than 10\,days remains uncertain. They might have developed similarly to their longer-period counterparts, emerged from rare collisions between smaller planets, or could be the remnant cores of stripped giant planets. Characterizing a large number of them is important to advance our understanding of how they form and evolve.}
{We aimed at confirming the planetary nature and characterizing the physical and orbital properties of a close-in Neptune-type transiting exoplanet candidate revealed by TESS around the star TOI-5795 (V = 10.7 mag), 162 pc away from the Sun.}
{We monitored TOI-5795 with the HARPS spectrograph for two months to quantify any periodic variations in its radial velocity (RV), necessary to estimate the mass of the smaller companion. We jointly analyzed these RV measurements and TESS photometry. Contaminating sources were excluded to be the origin of the detected signal thanks to high-angular-resolution speckle and adaptive optical imaging.}
{We found that the parent star is a metal-poor (${\rm [Fe/H]}=-0.27\pm0.07$), G3\,V star ($T_{\rm eff}=5718\pm50$\,K), with a radius of $R_{\star}=1.082\pm0.026\,R_{\sun}$, a mass of $M_{\star}=0.901^{+0.055}_{-0.037}\,M_{\sun}$ and an age of  $10.2^{+2.5}_{-3.3}$\,Gyr. We confirmed the planetary nature of the candidate, which can now be named TOI-5795\,b. We estimated that the planet has an orbital period of $P_{\rm orb}=6.1406325 \pm 0.0000054$ days and an orbital eccentricity compatible with zero.
Having a mass of $23.66^{+4.09}_{-4.60}\,M_{\oplus}$, a radius of $5.62\pm 0.11\,R_{\oplus}$ and an equilibrium temperature of $1136\pm18$\,K, it can be considered as a hot super-Neptune at the edge of the so-called Neptune desert. The transmission spectroscopy metric of TOI-5795\,b is $\approx 100$, which makes it an interesting target for probing the chemical composition of its atmosphere. We simulated planet-formation processes but found almost no successful matches to the observed planet's mass and orbit, suggesting that post-formation dynamical events may have shaped its current state. We also performed an atmospheric-evolution study of TOI-5795\,b finding that this planet likely experienced significant atmospheric stripping due to prolonged high-energy irradiation from its parent star.}
{}


\maketitle


\section{Introduction}
\label{sec:intro}
The {\it Neptune desert} \citep{2016A&A...589A..75M} is a region in the mass-period ($M_{\rm p}-P_{\rm orb}$) space of exoplanets where few Neptune-sized planets ($3\,R_\oplus \lesssim R_{\rm p} \lesssim 7\,R_\oplus$) with short orbital periods and high irradiance have been found, although observational bias should favor the detection of such planets. The dearth of both hot-Neptune and hot-Saturn planets ($10\,M_{\oplus}\lesssim M_{\rm p} \lesssim 100\,M_{\oplus}$) might be depleted as a result of the way planets form or migrate. For example, high-eccentricity migration can bring planets very close to their star, leading to tidal disruption \citep{2016ApJ...820L...8M,2018MNRAS.479.5012O}. Additionally, the Neptune desert could be partially explained by the fact that these planets are likely to lose their atmospheres because of intense ultraviolet and X-ray irradiation from their host stars (e.g., \citealt{2015Natur.522..459E,2018MNRAS.476.5639I,2023A&A...671A.132S}). 
Close-in Neptunes can be more susceptible than hot Jupiters to complete evaporation of their envelopes \citep{2012MNRAS.425.2931O}. 
Photoevaporation can be particularly strong for the puffiest planets because of their proximity to their parent stars and can be further enhanced by Roche-lobe overflow (e.g., \citealt{2022ApJ...929...52K,2023ApJ...945L..36T}). 
Recently, \citet{2024A&A...689A.250C} conducted a study focused on the Neptunian desert. Based on the Kepler DR25 catalog \citep{2018ApJS..235...38T}, these authors re-defined the boundaries of the desert ($P_{\rm orb}\lesssim3$\,days) and identified a new, highly populated area they term the {\it Neptunian ridge}, populated by planets with 3.2\,days\,$<P_{\rm orb}<$\,5.7\,days, which acts as a transition zone to the {\it savanna}, a more moderately populated region at greater orbital distances (5.7 to 100 days; also see \citealt{2018A&A...620A.147B}). They also stressed the role of high-eccentricity tidal migration (HEM) and photoevaporation as key processes in shaping this planetary landscape. 

Trends have been noted between planet bulk density and stellar metallicity for lightly irradiated ($F_{\star}<2\times10^8$\,erg\,s$^{-1}$\,cm$^{-2}$) sub-Neptunes \citep{2022MNRAS.511.1043W,2023MNRAS.520.3649H,2023AJ....166....7K}, suggesting that planets orbiting metal-rich stars have metal-rich atmospheres with reduced photo-evaporation. However, such a correlation is not evident for lightly irradiated gas giant planets, i.e. $0.1\,M_{\rm Jup}<M_{\rm p}<4\,M_{\rm Jup}$ \citep{2024A&A...691A..67M}.
Although interesting trends are emerging with the sample of Neptunes studied so far, an even larger sample is needed to understand these planets. In this context, we are running the HOt-NEptune Initiative (HONEI; \citealt{2025arXiv250510123N}), an observational program, based on high-resolution spectrographs, with the aim of finding ideal targets for atmospheric follow-up. This is the evolution of a previous GAPS\footnote{Global Architecture of Planetary Systems (e.g., \citealt{Esposito2014}).} sub-program \citep{2022A&A...667A...8N,2023Natur.622..255N,2025A&A...693A...7N} and is specifically designed to confirm and characterize hot- and warm Neptune-sized TESS Objects of Interest (TOIs) in different environments with the aim of investigating their origin. 

Here, we report the results of multi-instrument observations of the star TOI-5795, which allowed us to measure the mass of the planet candidate TOI\,5795.01, confirming its planetary nature and characterizing its main physical and orbital parameters.

This paper is structured as follows. Sect.~\ref{sec:observations} describes the photometric, spectroscopic, and high-angular-resolution-imaging observations on which this work is based. Section~\ref{sec:parentstar} details the physical and atmospheric parameters of the parent star. The analysis of the planetary signal and the subsequent characterization of the planetary system, by a joint analysis of different data sets, are described in Sect.~\ref{sec:analysis}. The possible formation and evolution scenarios of TOI-5795\,b are analyzed in Sect.~\ref{subsec:formation_history}, while the evolutionary history of its atmosphere is investigated in Sect.~\ref{sec:atmospheric_evaporation}. Our conclusions are summarized in Sect.~\ref{sec:summary}.

\section{Observations and data reduction}
\label{sec:observations}
\subsection{TESS photometry}
\label{subsec:TESS_photometry}
\begin{figure}
\centering
\includegraphics[width=0.43\textwidth]{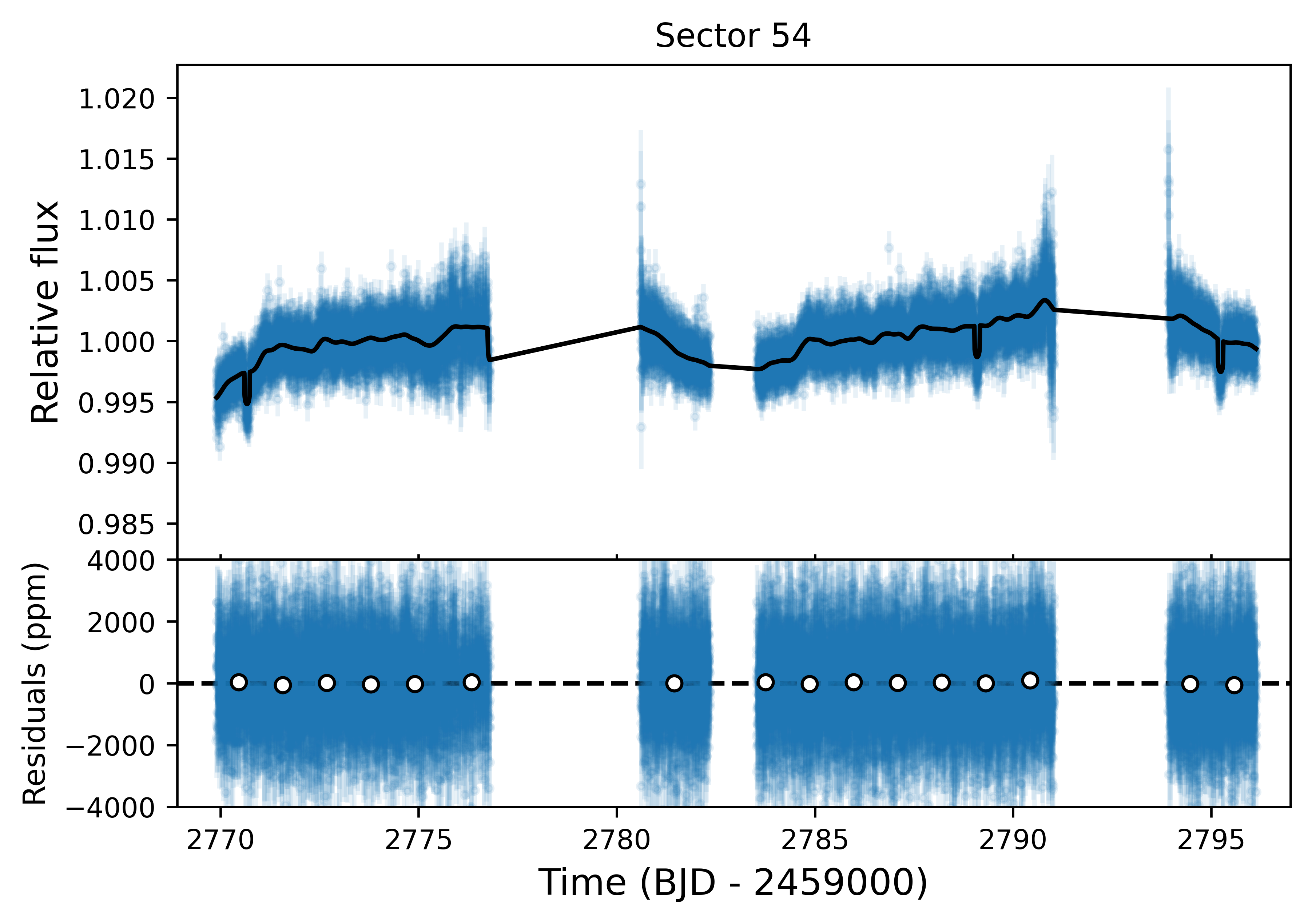}
\includegraphics[width=0.44\textwidth]{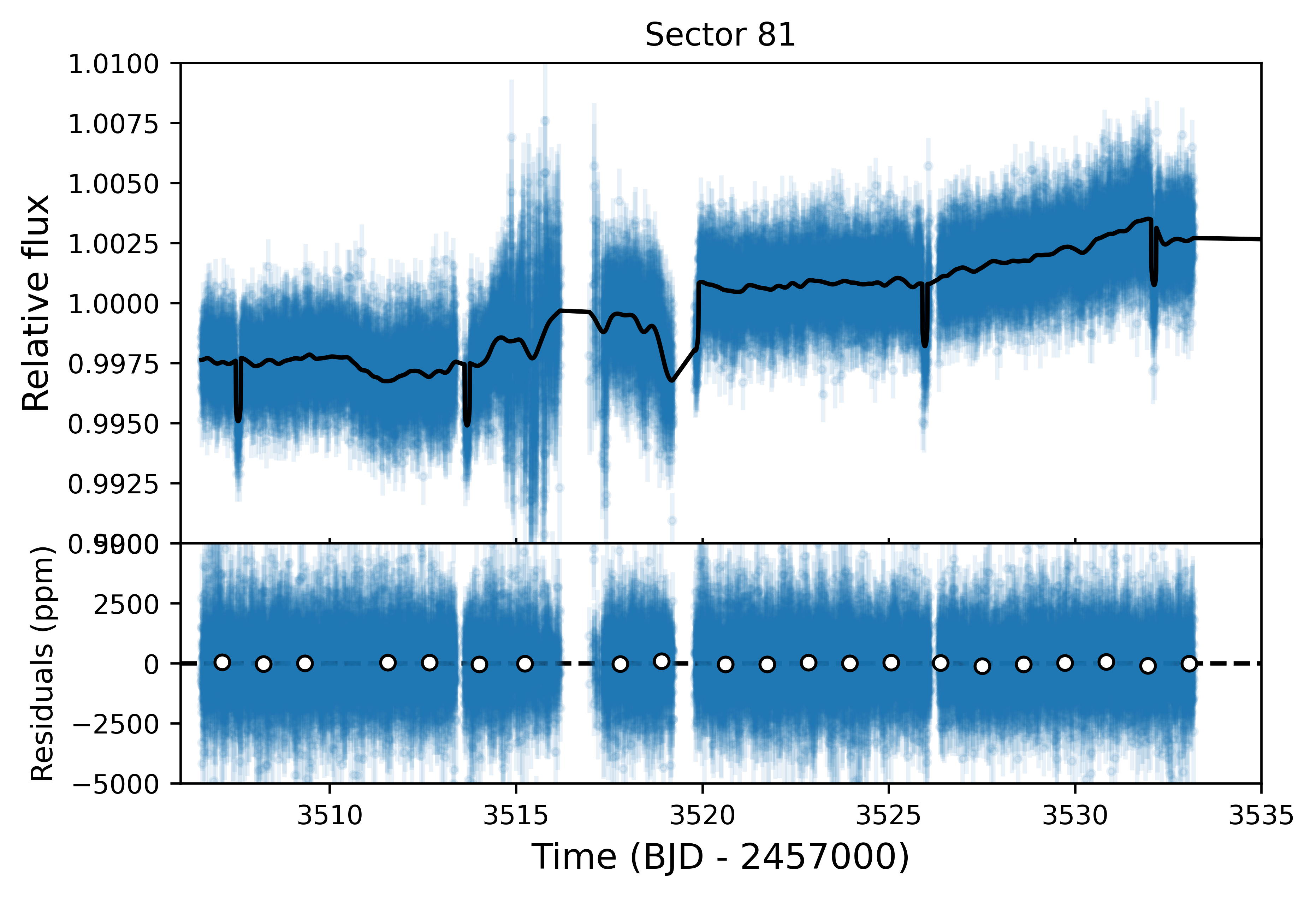}
\includegraphics[width=0.43\textwidth]{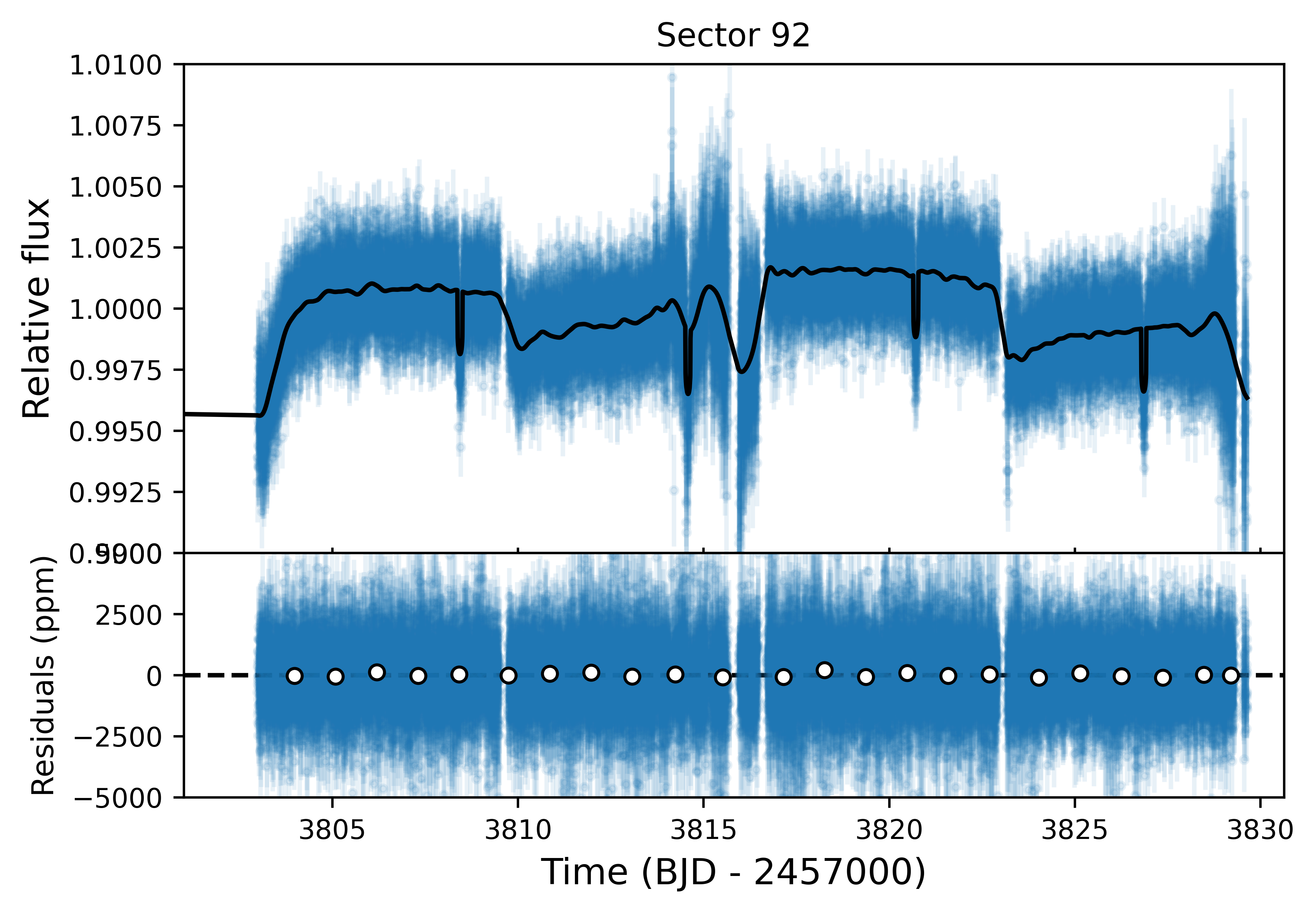}
\caption{The light curve of TOI-5795 obtained by TESS by monitoring the Sectors 54 (top panel), 81 (middle panel), and Sector 92 (bottom panel) with a 2-min cadence, as extracted by the SAP pipeline. No significant rotational modulation can be appreciated. The black line represents our best-fit transit model, while the white points are binned data. The residuals of the best-fit model are shown in parts per million in both the panels.}
\label{fig:confronto}
\end{figure}

The target of this study is the star TIC\,151724385 (aka 2MASS J20192849-0732511; $T mag = 10.19$\,mag; see Table~\ref{tab:star}), which is located $\approx 160$\,pc away and was observed by TESS, as a pre-selected target \citep{2019AJ....158..138S}, from July 9 to August 5, 2022, during Sector 54 of its first extended mission. It became a TOI on 2022-09-22 and was catalogued as TOI-5795. It was observed again by TESS from July 15 to August 10, 2024, during Sector 81 of its second extended mission, and then, from May 7 through June 3, 2025, during Sector 92. In all sectors, the target was observed with a 2-min cadence and a total of 30\,086 images were collected.
The three TESS light curves span about 763 days and contain 12 transit events\footnote{We note that a comment in the ExoFOP-TESS database \citep{Akeson2013} refers to a possible additional transit event at $\sim$2757 ($\mathrm{BJD}-2457000$); however, this timestamp is prior to the start of the available light curves, making such an event impossible. It likely refers instead to a dip at $\sim$2787. This feature is probably due to an off-target eclipsing binary blend, particularly since the first flagged event near $\sim$2772 is shallower, exhibits some depth–aperture correlation, and is likely a secondary eclipse.} of the planet candidate TOI\,5795.01 with an average depth of $2.86\pm0.16$ mmag (see Fig.~\ref{fig:confronto}).

We downloaded the TESS light curves of TOI-5795 as a \texttt{LightCurveFile} object, from the Mikulski Archive for Space Telescopes (MAST) using the Python package \texttt{lightkurve} \citep{lightkurve}. The \texttt{LightCurveFile} object contains both the Simple Aperture Photometry (SAP; \citealt{Twicken2010}) and the Presearch Data Conditioning SAP flux (PDC-SAP; \citealt{2012PASP..124.1000S,2012PASP..124..985S,2014PASP..126..100S}), which was extracted with the TESS SPOC pipeline \citep{2016SPIE.9913E..3EJ}. Potential contaminating sources within the TESS aperture, which could lead to a shallower transit depth than the actual value, are typically addressed using a dilution factor. Here, the PDC-SAP photometry was corrected for contamination from other objects using the Create Optimal Apertures module \citep{2010SPIE.7740E..1DB,2020ksci.rept....3B}.
However, in this case, there are no sources close to the borders of the SPOC pipeline aperture, as shown in Fig.~\ref{fig:tpfs}, where the apertures were drawn using \texttt{tpfplotter}
\footnote{\texttt{\url{https://github.com/jlillo/tpfplotter/}}} \citep{2020A&A...635A.128A} for Sectors~54, 81, and \textbf{92} together with the closest sources extracted from the Gaia satellite DR3 catalogue. 

%
\begin{table}
\centering %
\caption{Astrometric, spectroscopic, and photometric parameters for TOI-5795.} %
\label{tab:star} %
\resizebox{\hsize}{!}{
\begin{tabular}{l c c c}
\hline %
\hline  \\[-8pt] %
Parameter & Unit & Value & Source \\
\hline  \\[-6pt] %
\multicolumn{1}{l}{\large{{\bf Cross-identifications}}} \\ [2pt] %
TOI\,ID \dotfill & \dotfill& 5795 & TOI cat. \\
TIC\,ID \dotfill & \dotfill& 151724385 & TIC \\
TYC\,ID \dotfill & \dotfill& 5741-424-1 & TYC \\
GSC\,ID \dotfill & \dotfill& 05741-00424 & GSC \\
2MASS\,ID \dotfill & \dotfill& J20192849-0732511 & 2MASS \\
Gaia DR1\,ID \dotfill & \dotfill& 4216036345540368768 & Gaia~DR1 \\ 
Gaia DR3\,ID  \dotfill & \dotfill& 4216036349838071424 & Gaia~DR3 \\ [6pt] 
\multicolumn{1}{l}{\large{{\bf Astrometric properties}}} \\ [2pt] %
$\alpha$\,(J2000) \dotfill & h:m:s & $20:19:28.51$ & Gaia~DR3 \\
$\delta$\,(J2010)  \dotfill & d:m:s & $-07:32:51.57$ & Gaia~DR3 \\
$\pi$ \dotfill & mas & $6.1900 \pm 0.0213$ & Gaia~DR3 \\
$\mu_{\alpha}$ \dotfill & mas\,yr$^{-1}$  & $12.8097 \pm 0.0233$ & Gaia~DR3 \\
$\cos_{\delta}$ \dotfill & mas\,yr$^{-1}$  & $-29.7160 \pm 0.0152$ & Gaia~DR3 \\ [6pt] 
\multicolumn{1}{l}{\large{{\bf Photometric properties}}} \\ [2pt] %
$B_T$ \dotfill & mag & $11.466\pm0.135$ & Tycho-2 \\  
$V_T$ \dotfill & mag & $10.708\pm0.009$ & Tycho-2 \\ 
$g'$ \dotfill & mag & $11.107\pm0.026$ & APASS Sloan \\ 
$r'$ \dotfill & mag & $10.629\pm0.026$ & APASS Sloan \\ 
$i'$ \dotfill & mag & $10.512\pm0.073$ & APASS Sloan \\ 
$TESS$ \dotfill & mag & $10.187 \pm 0.006$ & TIC\,v8.2 \\ 
$G$ \dotfill & mag & $10.6376 \pm 0.000361$ & Gaia~DR3 \\
$BP$ \dotfill & mag & $10.9813 \pm 0.0007$ & Gaia~DR3 \\
$RP$ \dotfill & mag & $10.1264 \pm 0.0003$ & Gaia~DR3 \\
$J$ \dotfill & mag & $9.557\pm0.021$ & 2MASS \\
$H$ \dotfill & mag & $9.218\pm0.021$ & 2MASS \\
$K$ \dotfill  & mag & $9.168\pm0.021$ & 2MASS \\ 
$W1$\,3.4\,$\mu$m \dotfill & mag & $9.114\pm0.022$ & AllWISE \\
$W2$\,4.6\,$\mu$m \dotfill & mag & $9.146\pm0.020$ & AllWISE \\
$W3$\,12\,$\mu$m  \dotfill & mag & $9.109\pm0.033$ & AllWISE \\ [6pt]
\multicolumn{1}{l}{\large{{\bf Spectroscopic properties}}} \\ [2pt] %
Spectral type \dotfill & & G3\,V & This work$^{(a)}$ \\
$v\sin{i}$ \dotfill & km\,s$^{-1}$ & 1.9 $\pm$1.0& This work \\
$\rm{[Fe/H]}$ \dotfill & dex & $-0.27 \pm 0.07$ & This work \\%
$\rm{[C/H]}$ \dotfill & dex & $-0.27 \pm 0.05$ & This work \\%
$\rm{[S/H]}$ \dotfill & dex & $-0.21\pm 0.07$ & This work \\%
$\rm{[O/H]}$ \dotfill & dex & $-0.20 \pm 0.10$ & This work \\%
$\rm{[Mg/H]}$ \dotfill & dex & $-0.06 \pm 0.08$ & This work \\%
$\rm{[Si/H]}$ \dotfill & dex & $-0.14 \pm 0.08$ & This work \\ [6pt] %
\multicolumn{1}{l}{\large{{\bf Derived parameters}}} \\ [2pt] %
$L_{\star}$ \dotfill & $L_{\sun}$ & $1.130\pm0.043$ & This work \\ [2pt] %
$M_{\star}$ \dotfill & $M_{\sun}$ & $0.901^{+0.055}_{-0.037}$ & This work \\  [2pt] %
$R_{\star}$ \dotfill & $R_{\sun}$ & $1.082\pm 0.026$ & This work \\
$\log g_{\star}$ \dotfill & cgs & $4.32\pm0.13$ & This work \\
$\rho_{\star}$\dotfill & g\,cm$^{-3}$ & $1.008^{+0.091}_{-0.073}$ & This work \\  [2pt] %
$\log R^{\prime}_{\rm HK}$\dotfill & dex &  $-5.07\pm0.02$ & This work$^{(b)}$ \\
$T_{\rm eff}$ \dotfill & K & $5715 \pm 55$ & This work\\ %
Age\dotfill & Gyr & $10.2^{+2.5}_{-3.3}$ & This work \\ [2pt] %
$A_V$ \dotfill & mag & $<0.094$ & This work \\ [2pt] %
Distance \dotfill & pc & $161.5\pm0.6$ & This work \\ [2pt] %
$\gamma_{\rm RV}$ & m\,s$^{-1}$ & $-66962.3$ & This work\\  [2pt] %
\hline %
\end{tabular}
}

\tablebib{TESS Primary Mission TOI catalog \citep{2021ApJS..254...39G}; TIC \citep{2018AJ....156..102S,2019AJ....158..138S}; 
Tycho \citep{2000A&A...355L..27H};
APASS Johnson \citep{Henden2016}; 
2MASS \citep{2006AJ....131.1163S}; Gaia DR3 \citep{2023A&A...674A...1G}; AllWISE \citep{2014yCat.2328....0C}.}
\tablefoot{
\tablefoottext{a}{The spectral type has been derived from the tables of \cite{pecaut_mamajek_2013_spt} (version 2022) and using our $T_{\rm eff}$ value.}
\tablefoottext{b}{The $\log R^{\prime}_{\rm HK}$ has been derived using the \texttt{ACTIN2} code ({\tt https://github.com/gomesdasilva/ACTIN2}) v2.0 beta11 \citep{actin} from the combined HARPS spectra.}
}
\end{table}

%
\subsection{Ground-based photometric follow-up observations}
\label{subsec:ground}
Ground-based follow-up observations in search for transits of TOI-5795\,b were carried out by members of the TESS Follow-up Observing Program Working Group and are archived at ExoFOP.

A complete transit of TOI-5795\,b was observed on 25 June 2023 through a PanSTARRS $z_s$ filter (wavelength center $8700$\,\AA) using the $f/8$ Ritchey-Chretien 1-m telescope, from the Las Cumbres Observatory Global Telescope (LCOGT) \citep{2013PASP..125.1031B} at the Cerro Tololo Inter-American Observatory in Chile (CTIO). This telescope is equipped with the SINISTRO camera with $4096\times4096$ pixels and a pixel size of $15 \,\mu$m. The resulting plate scale is $0.389$\,arcsec\,pixel$^{-1}$, which determines a $26\arcmin\times26\arcmin$ field of view. The images were calibrated using the standard LCOGT {\tt BANZAI} pipeline \citep{2018SPIE10707E..0KM} and differential photometric data were extracted using {\tt AstroImageJ} \citep{2017AJ....153...77C}. A circular $5\farcs8$ photometric aperture was adopted to exclude all of the flux from the nearest known neighbour in the Gaia DR3 catalogue (Gaia\,DR3\,4216036345540514432), which is $17\farcs8$ northeast of TOI-5795.

Another complete transit was observed on 15 September 2024 through a Johnson/Cousins $B$ band using the 0.36\,m $f/7.2$ Corrected Dall-Kirkham Astrograph telescope from the Brierfield Observatory near Bowral, New S. Wales, Australia. This telescope is equipped with a Moravian G4-16000 KAF-16803 with $4096\times4096$ pixels and a pixel size of $9 \,\mu$m. The plate scale is 1.47\,arcsec\,pixel$^{-1}$, resulting in a $50\arcmin\times50\arcmin$ field of view. 
The differential photometric data were also extracted using {\tt AstroImageJ} by adopting a circular $5\farcs9$ photometric aperture. The final LCO and Brierfield light curves, from the global fit, are shown in Fig.~\ref{fig:lcs}, together with the phase-folded PDC-SAP light curve. 

\subsection{High-angular-resolution imaging}
High-angular-resolution imaging is needed to search for nearby sources that can contaminate the TESS photometry, resulting in an underestimated planetary radius, or be the source of astrophysical false positives, such as background eclipsing binaries \citep{2015ApJ...805...16C}. As part of our standard process for validating transiting exoplanets, TOI-5795 was observed with optical speckle and near-infrared adaptive-optical (AO) imaging.

\subsubsection{Optical Speckle Imaging}
We searched for stellar companions to TOI-5795 with speckle imaging on the 4.1-m Southern Astrophysical Research (SOAR) telescope \citep{2018AJ....155..235T} on November 4, 2022, observing in Cousins-$I$ band, a bandpass similar to TESS. This observation was sensitive to 5-$\sigma$ detections of 3.9-magnitude fainter stars at an angular distance of 1\arcsec from the target. More details of the observations within the SOAR TESS survey are available in \citet{2020AJ....159...19Z}. The 5-$\sigma$ detection sensitivity and speckle autocorrelation functions of the observations of TOI-5795 are shown in the top panel of Fig.~\ref{fig:har}. No nearby stars were detected within 3\arcsec of TOI-5795 in the SOAR observations.
%

\subsubsection{Near-Infrared AO Imaging}
Observations of TOI-5795 were made on  August 29, 2023, with the PHARO instrument \citep{2001PASP..113..105H} on the Palomar Hale 5.1-m telescope behind the P3K natural guide star AO system \citep{2013ApJ...776..130D}. The pixel scale for PHARO is $0.025\arcsec$. Palomar data were collected in a standard 5-point quincunx dither pattern in the Br-$\gamma$ filter ($\lambda_0=2166\, \mu$m; $\Delta \lambda=0.020\, \mu$m). The reduced science frames were combined into a single mosaic image with final resolutions of $\sim 0.095\arcsec$.
	
\begin{figure}
\centering
\includegraphics[width=0.43\textwidth]{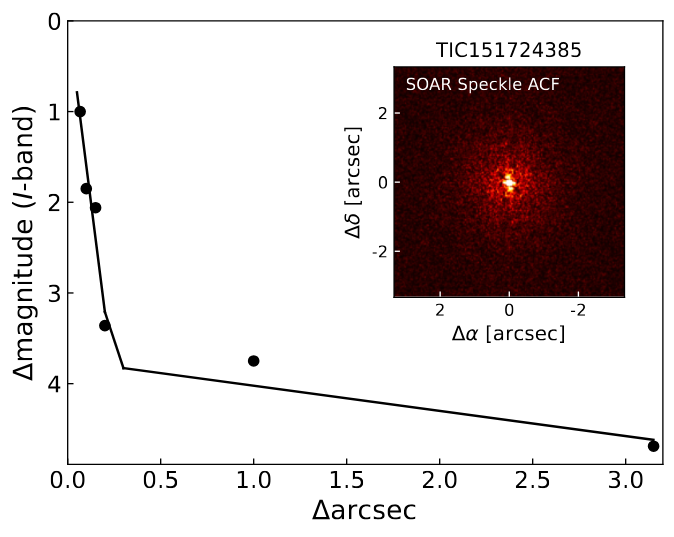}
\includegraphics[width=0.48\textwidth]{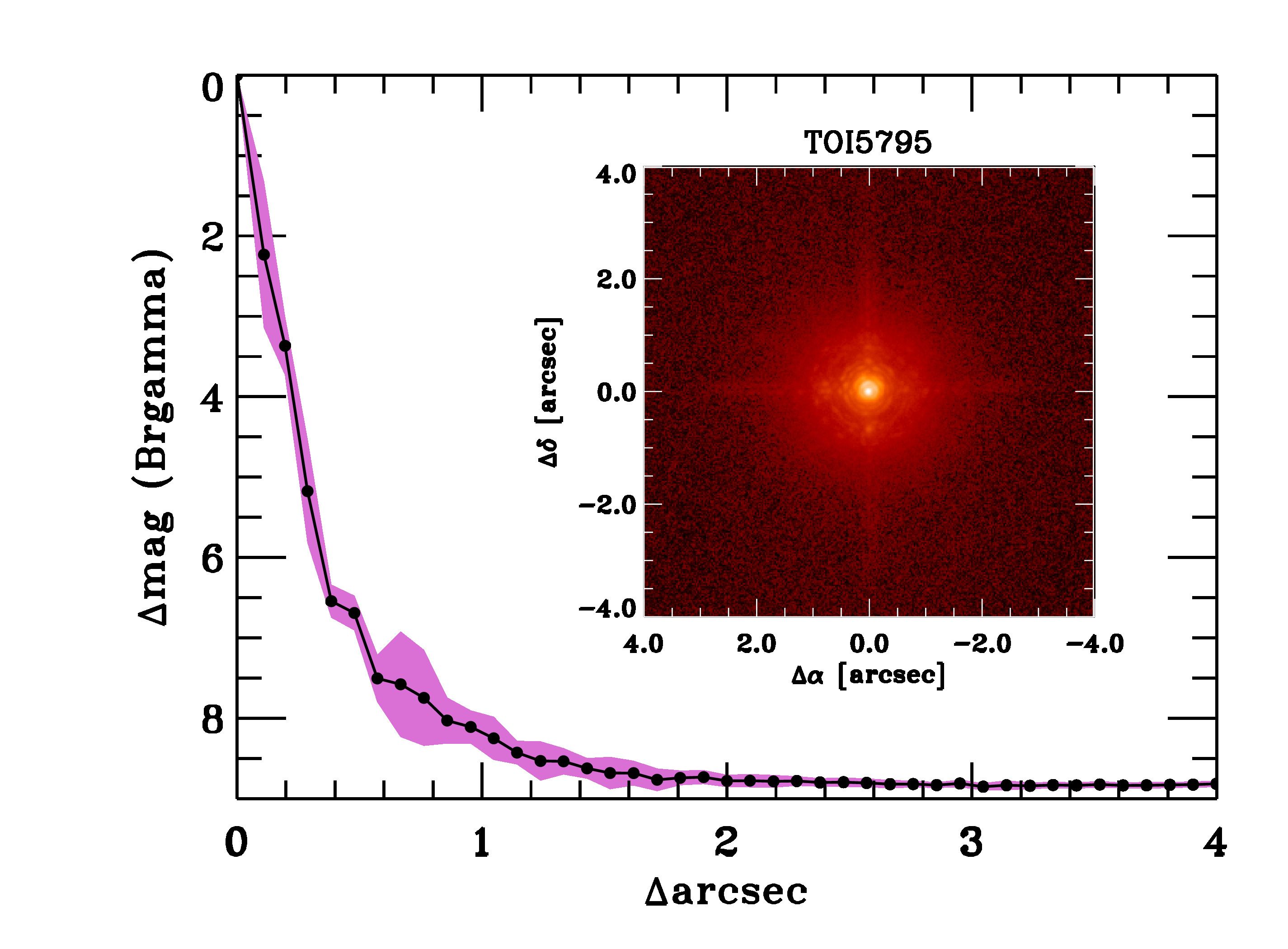}
\caption{High-resolution sensitivity curves. Final sensitivity of SOAR (top panel) and Palomar Hale (bottom panel), plotted as a function of angular separation from the host star. Insets in the respective panels display images of the central region of the data.}
\label{fig:har}
\end{figure}

The sensitivity of the final combined AO images was determined by injecting simulated sources azimuthally around the primary target every $20^\circ$ at separations of integer multiples of the central source FWHM \citep{2017AJ....153...71F}. The brightness of each injected source was scaled until standard aperture photometry detected it with 5-$\sigma$ significance. The final 5-$\sigma$ limit at each separation was determined from the average of all limits determined at that separation, and the uncertainty on the limit was established by the rms dispersion of the azimuthal slices at a given radial distance. No stellar companions were detected, with the images reaching a contrast of $\sim 3.9$ (SOAR) and $\sim 7.5$ (Palomar) magnitudes fainter than TOI-5795 within $0.5\arcsec$ in each respective band (see Fig.~\ref{fig:har}).

\subsection{Spectroscopic Observations}
\label{subsec:spectra}
Spectroscopic follow-up observations and radial-velocity (RV) measurements of TOI-5795 were obtained with TRES (Tillinghast Reflector \'{E}chelle
Spectrograph; \citealt{2007RMxAC..28..129S}) and HARPS (High Accuracy Radial velocity Planet Searcher; \citealt{2002Msngr.110....9P,2003Msngr.114...20M}). 
\subsubsection{TRES data}
\label{subsubsec:tres}
TRES is a fiber echelle spectrograph ($R=44\,000$) mounted on the 1.5-meter Tillinghast telescope at the Smithsonian Astrophysical Observatory's Fred L. Whipple Observatory on Mt. Hopkins in Arizona (USA). Two reconnaissance spectra of TOI-5795 were obtained in October 2022, revealing a G dwarf. No significant RV variations in the two initial spectra were revealed, discarding binary scenarios. Other spectra of this star were collected for detecting possible long RV trends. In total, 13 TRES observations were recorded over a two-year span of time. 
The derived RVs have an average precision of $27.7$\,m\,s$^{-1}$ (see Table~\ref{tab:RV_data}), consistent with no detectable RV variation due to the TESS planet candidate or other external long-period objects orbiting around TOI-5795. Details regarding the detection sensitivity of this dataset are provided in Appendix \ref{sec:sensitivity}, together with those of HARPS.

\subsubsection{HARPS data}
\label{subsubsec:harps}
HARPS is a high-resolution ($R=115\,000$), visible-light, fibre-fed echelle spectrograph mounted on the ESO 3.6-meter telescope at La Silla (Chile). 
Due to its long-term stability and simultaneous wavelength calibration, HARPS is one of the few instruments currently in operation that can achieve, in optimal conditions, RV measurements with an accuracy of $\sim 1$\,m\,s$^{-1}$ \citep{2014Natur.513..358P}. We monitored TOI-5795 in summer 2024, from 29 June to 13 August, under our ESO programme 113.26UJ (PI: Naponiello), collecting a total of 19 RV measurements with an average SNR of $27$ (see Table~\ref{tab:RV_data}). We have used the fibre AB spectroscopy mode (object and sky) and set an exposure time of 20 min. The spectra were calibrated and the RVs derived using the SpEctrum Radial Velocity analyzer (SERVAL) pipeline \citep{2018A&A...609A..12Z}, and the RV measurements resulted in an average uncertainty of $2.7$\,m\,s$^{-1}$.

We also investigated various stellar activity indicators in this dataset, including the bisector inverse slope (BIS), the full width at half maximum (FWHM), the line contrast of the cross-correlation function (CCF), and the chromospheric activity index $\log R'_{\mathrm{HK}}$, as well as the cores of the sodium (Na\,\textsc{i}), helium (He\,\textsc{i}), and H$\alpha$ spectral lines (extracted with \texttt{ACTIN2}; \citealt{actin}). None of these indicators showed significant periodic signals, with all generalized Lomb-Scargle periodograms returning false alarm probabilities (FAP) above 15\%.

\section{Host-star characterization}
\label{sec:parentstar}
To estimate the photospheric parameters of the host star, we produced a coadded spectrum from the HARPS spectra (see Sect.\,\ref{subsubsec:harps}). The signal-to-noise ratio of the coadded spectrum was $\sim 160$ at $\lambda \sim 6000\,\AA$. From the coadded spectrum, we derived the spectroscopic atmospheric parameters, i.e. effective temperature ($T_{\rm eff}$), surface gravity ($\log{g}$), microturbulence velocity ($\xi$), and iron abundance ([Fe/H]) using equivalent widths (EWs) of iron lines (see \citealt{Biazzoetal2022} for details). Line equivalents were measured using the ARESv2 code \citep{sousa_2015_ares_v2}. We then adopted the \citet{castelli_kurukz_2003} grid of model atmospheres with new opacities (ODFNEW) and the spectral package pymoogi \citep{adamow_2017_pymoogi}, which is a wrapper of the MOOG code (version 2019; \citealt{sneden1973}). We also computed the elemental abundances of magnesium, silicon, carbon, and sulfur using the procedure based on line EWs and the same codes and grid of models as above. Finally, following the spectral synthesis method described by \citet{Biazzoetal2022} and assuming the macroturbulence velocity $v_{\rm macro}=3.3$\,km\,s$^{-1}$ by \citet{doyle_2014}, we find a projected rotational velocity $v\sin{i}=1.9\pm1.0$ km\,s$^{-1}$, which is at the limit of the HARPS spectral resolution. The same procedure was applied to the TRES spectrum with the highest signal-to-noise ratio to derive the oxygen abundance from the infrared triplet lines (at $\sim 7772-7775$\,\AA) and applied the non-LTE (NLTE) corrections by \citet{amarsi_2015_OI_nlte}. All stellar parameters and elemental abundances are listed in Table\,\ref{tab:star}. It is worth noting that the star is metal poor, i.e. ${\rm [Fe/H]} = -0.27 \pm 0.07$.

\begin{figure}
\centering
\includegraphics[width=1\linewidth]{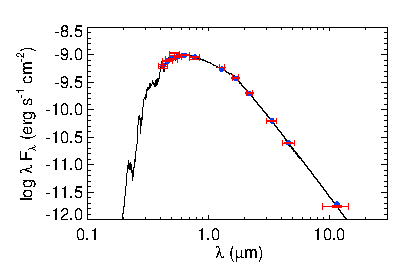}
\caption{Stellar spectral energy distribution (SED). The broad band measurements from the Tycho, APASS Johnson and Sloan, 2MASS and WISE magnitudes are shown in red, and the corresponding theoretical values with blue circles. The non-averaged best-fit model is displayed with a black solid line.} 
\label{fig:stellarSED}
\end{figure}

We determined the stellar mass, radius, and age using the {\tt EXOFASTv2} differential evolution Markov chain Monte Carlo tool \citep{2017ascl.soft10003E, Eastman2019}, by simultaneously modeling the stellar spectral energy distribution (SED)  and the MESA Isochrones and Stellar Tracks (MIST, \citealt{Paxton2015}); see \citet{2025A&A...693A...7N} for more details. We sampled the SED with the Tycho-2 $B_{\rm T}$ and $V_{\rm T}$, APASS Johnson $B$, $V$, $g^{\prime}$, $r^{\prime}$, and $i^{\prime}$, 2MASS $J$, $H$, and $K_{\rm s}$, and WISE $W1$, $W2$, and $W3$ magnitudes (see Table~\ref{tab:star} and Fig.~\ref{fig:stellarSED}). 
We imposed Gaussian priors on the Gaia DR3 parallax, as well as on the $T_{\rm eff}$ and [Fe/H] as derived from our spectral analysis.  
The fitted and derived parameters of the host star are given in Table~\ref{tab:star}. The estimated old age of $\sim 10$~Gyr is consistent with both the low $\log{R^{\prime}_{HK}}=-5.07\pm0.02$ and the absence of significant variations due to stellar activity in the light curve and in the activity spectra indices. 

\section{Physical and orbital characterization of the TOI-5795 planetary system}
\label{sec:analysis}
\subsection{Periodogram of the RV time series}
\begin{figure}
\centering
\includegraphics[width=0.9\linewidth]{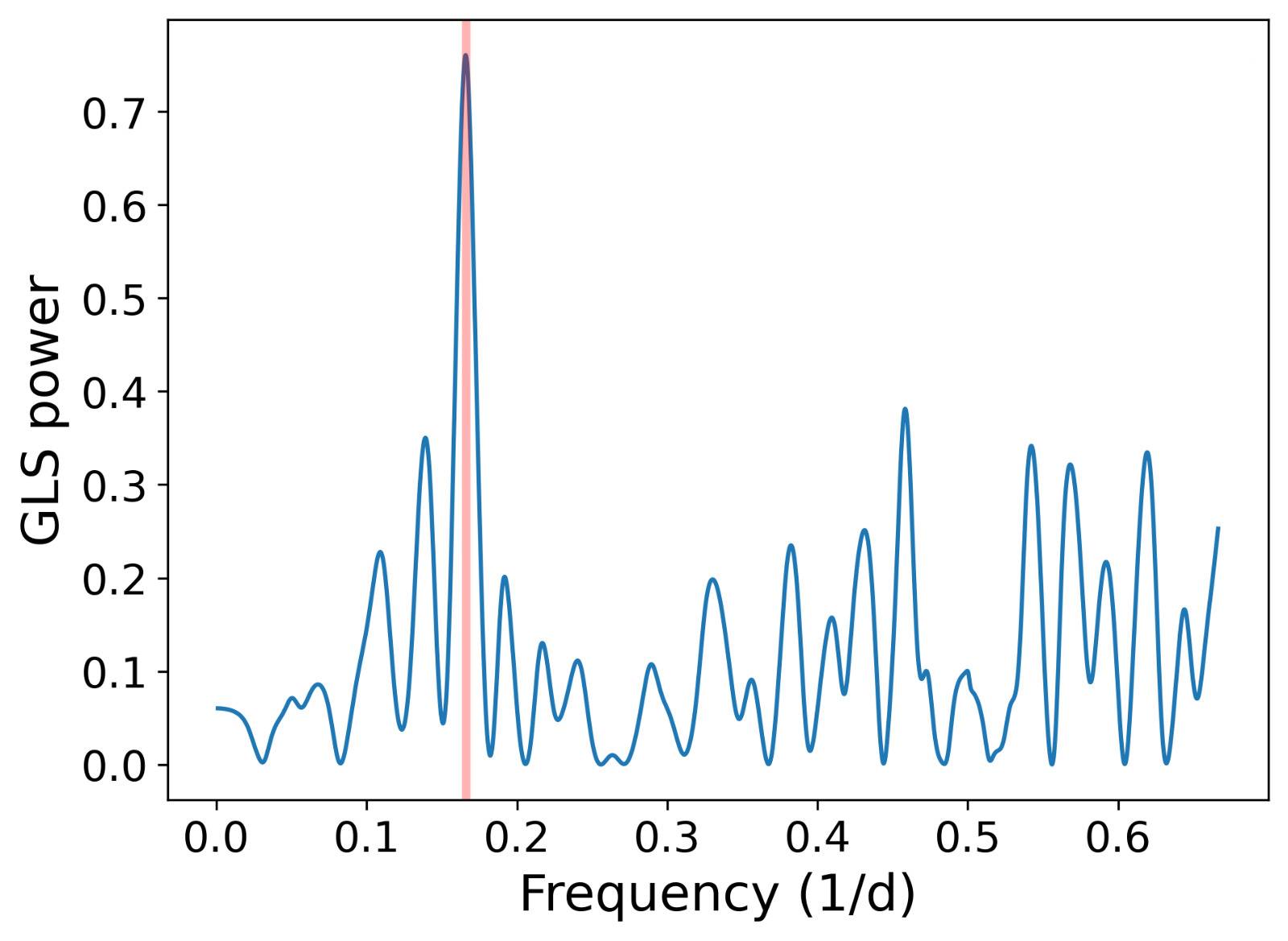}
\caption{GLS periodogram of TOI-5795 HARPS RVs, with the main peak correctly identified at $\sim6$ days.} 
\label{fig:GLS}
\end{figure}

In the RVs obtained with HARPS we searched for the same periodic signal present in the TESS light curve using the Generalized Lomb-Scargle periodogram (GLS; \citealt{2009A&A...496..577Z}), using the Python package \texttt{astropy} v.5.2.2 \citep{2018AJ....156..123A}. The orbital period of the planet candidate TOI\,5795.01 was clearly identified as the highest peak in the periodogram (see Fig.\,\ref{fig:GLS}). It often happens that no further signals come out of the first RV periodogram because the principal signal is too strong, but once this is removed, secondary signals come out in the residual periodogram. However, while the residuals of the 1-planet model show some scatter, the limited number of RV data points makes it difficult to confidently identify any additional signals that might be present, and indeed no significant signals were revealed in the GLS periodogram of the residuals of the 1-planet model. Furthermore, no other transit signals were detected in the light curve.

\subsection{Global modeling of the data}
We performed a joint modeling of TESS photometry, ground-based follow-up photometry, and HARPS-RV data following our consolidated procedure, as described in \citet{2022A&A...667A...8N,2023Natur.622..255N,2025A&A...693A...7N}. Briefly, we used the Python wrapper \texttt{juliet}\footnote{\texttt{\url{https://juliet.readthedocs.io/}}} \citep{2019MNRAS.490.2262E}, which makes use of \texttt{batman}\footnote{\url{https://github.com/lkreidberg/batman}.} \citep{2015PASP..127.1161K} and \texttt{RadVel}\footnote{\url{https://radvel.readthedocs.io}.} \citep{2018PASP..130d4504F}, for the modelling of transit and RV data, respectively. The systematics (red noise) in photometric time series were modeled by means of Gaussian Processes (GPs) via the \texttt{celerite} package \citep{2017AJ....154..220F}.
A Bayesian approach, based on the dynamic nested sampling package \texttt{dynesty} \citep{2020MNRAS.493.3132S}, is then adopted to explore the posterior distribution of the model parameters and estimate the Bayesian evidence, ensuring consistency with the data.

The RV/photometry joint analysis was run for a simple 1-planet model. While for the RV dataset we account only for uncorrelated (white) noise. 
Large uniform (uninformative) priors were used for the distribution of all parameters. In particular, the priors for $P_{\rm orb}$ and $T_0$ were centered around the values reported in the data validation report produced by the TESS Science Processing Operations Center (SPOC) \citep{2016SPIE.9913E..3EJ} pipeline. The best-fitting values and uncertainties for the transit, RV, and Keplerian parameters were, then, derived from the posterior probability distributions.
Following \citet{2013PASP..125...83E}, the eccentricity, $e$, and the argument of periastron, $\omega$, were parametrized as $\sqrt{e}\sin{\omega}$ and $\sqrt{e}\cos{\omega}$.
The impact parameter, $b$ and the ratio of the radii of the star and the planet, $k=R_{\rm p}/R_{\star}$, were parametrized as $r_1$ and $r_2$ (see \citealt{2018RNAAS...2..209E}).
For the limb-darkening (LD) coefficients, we adopted the same parameterization as \citet{2013MNRAS.435.2152K}, i.e. $q_1 \equiv (u_1+u_2)^2$ and $q_2 \equiv 0.5\,(u_1+u_2)^{-1}$, where $u_1$ and $u_2$ are the LD coefficients of the quadratic law.

\begin{figure}
\centering
\includegraphics[width=1\linewidth]{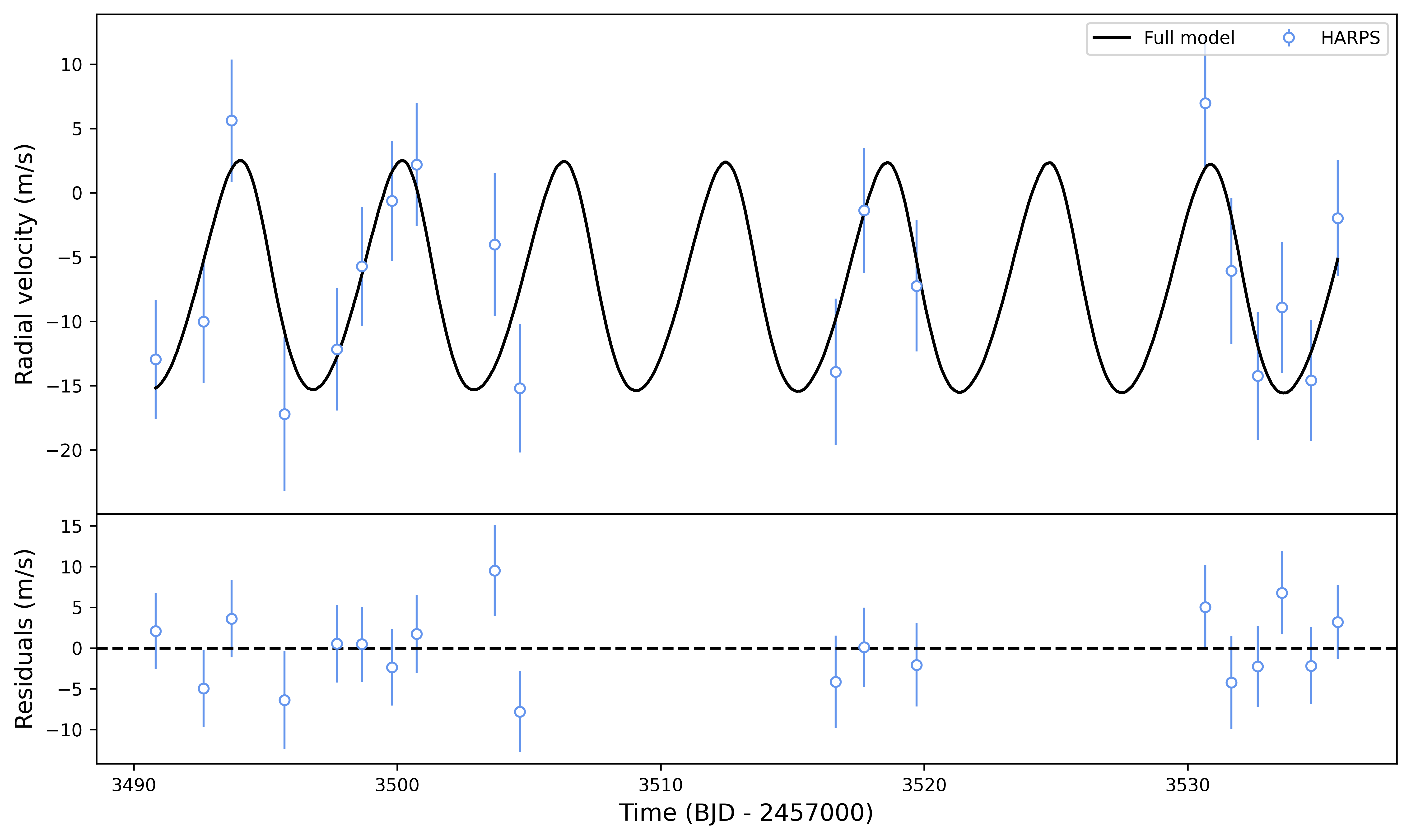}
\caption{{\it Top panel}: HARPS RV measurements of TOI-5795 in blue and the preferred model fit in black. {\it Bottom panel}: RV residuals over the model fit.} 
\label{fig:RV}
\end{figure}

\begin{figure}
\centering
\includegraphics[width=0.9\linewidth]{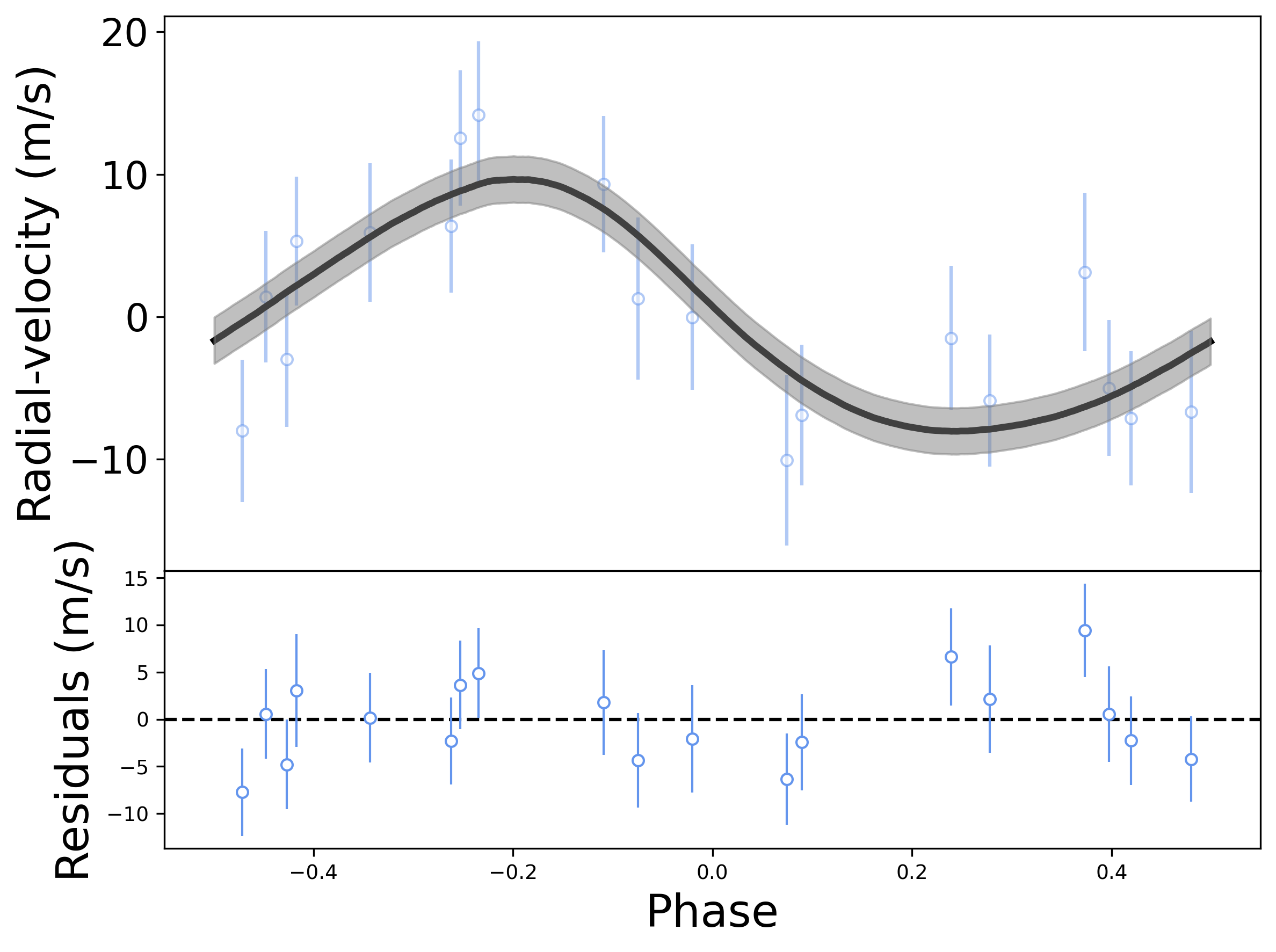}
\caption{{\it Top panel}: Phase-folded HARPS RV measurements of TOI-5795 in blue and the preferred model fit in black. {\it Bottom panel}: RV residuals over the model fit. The error bars include both the data uncertainty and the jitter derived from the analysis.} 
\label{fig:RV_ph}
\end{figure}

\begin{figure}
\centering
\includegraphics[width=1\linewidth]{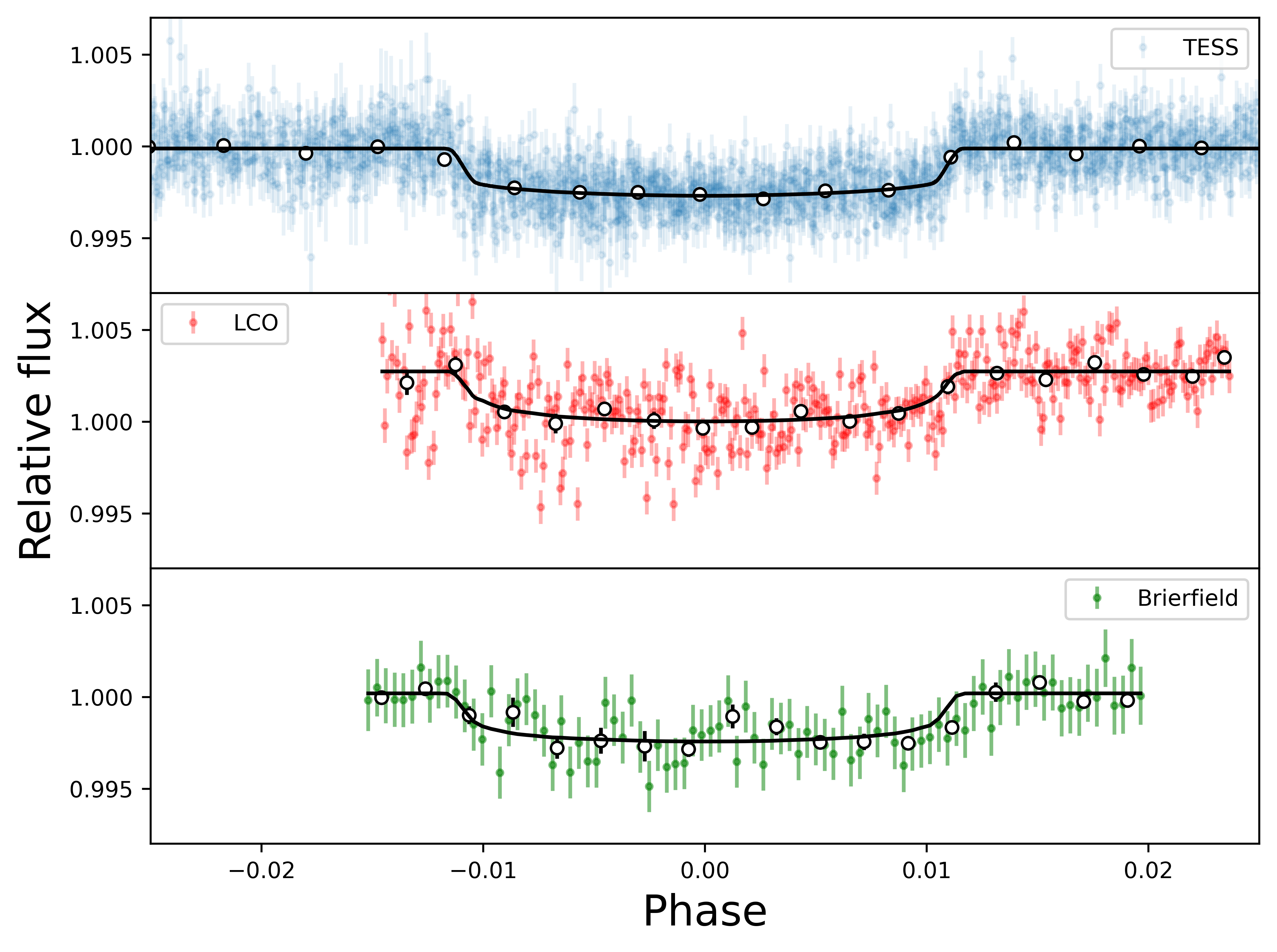}
\caption{Global fit result for TESS and the ground-based observations. The superimposed points correspond to phase-bins of 22 minutes, 18 minutes, and 17 minutes for TESS, LCO and Brierfield, respectively. The error bars include both the data uncertainty and the jitter derived from the analysis.
} 
\label{fig:lcs}
\end{figure}

\begin{figure}
\centering
\includegraphics[width=0.9\linewidth]{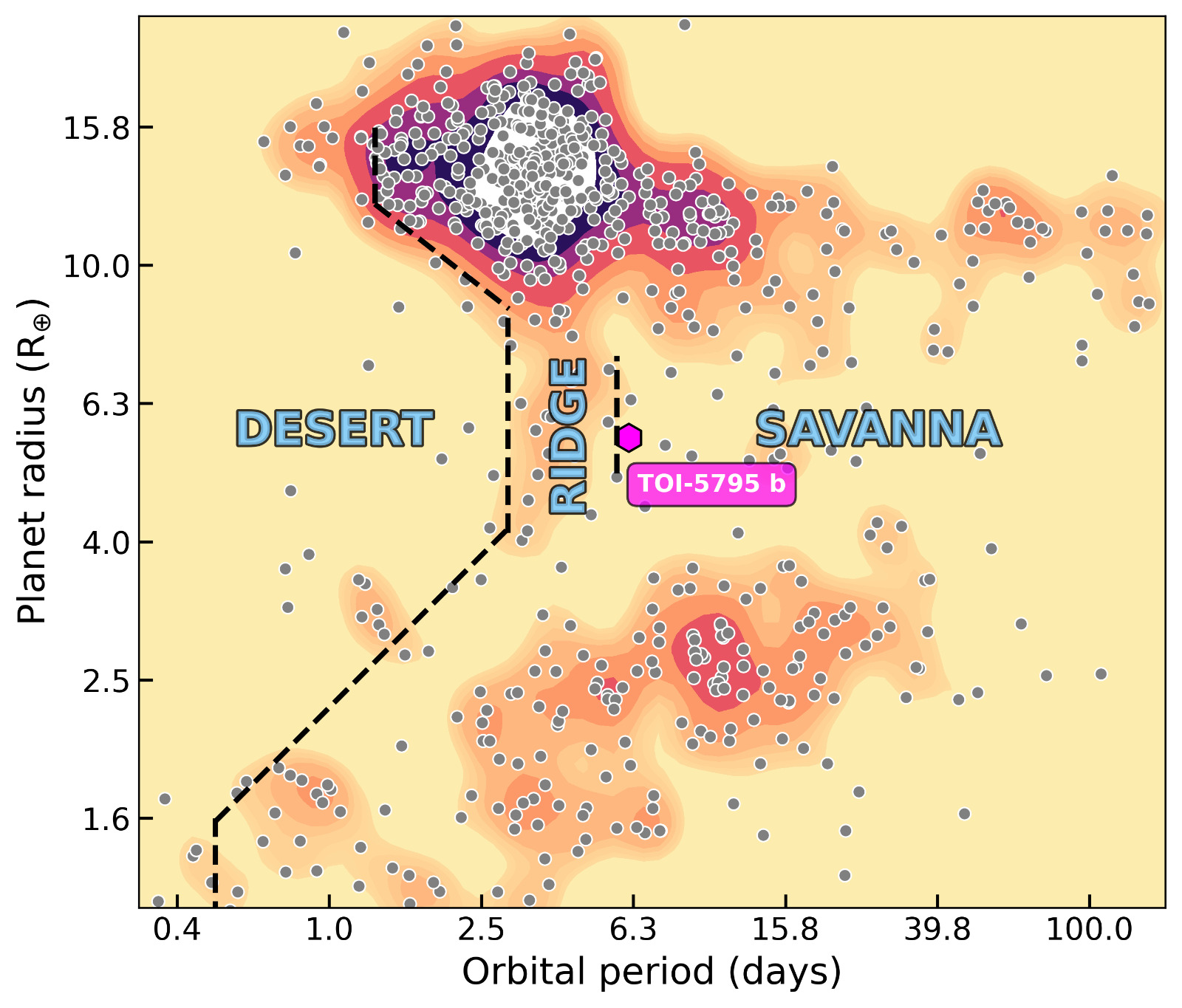}
\caption{Radius-period diagram of close-in exoplanets with mass and radius known with an accuracy of at least $5\,\sigma$. The data were collected from the NASA Exoplanet Archive on 13/02/2025. The error bars have been suppressed for clarity. The position of TOI-5795\,b is highlighted together with the population-based boundaries of the Neptunian desert, ridge, and savanna, as derived by \citet{2024A&A...689A.250C}. This plot was generated with \texttt{nep-des} ({\url{github.com/castro-gzlz/nep-des}}).
} 
\label{fig:desert}
\end{figure}

HARPS RVs are shown in Fig.~\ref{fig:RV} and Fig.~\ref{fig:RV_ph} (phase-folded) together with the preferred global model (top panels) and its residuals (bottom panels). The TESS, LCO and Brierfield light curves of TOI-5795, folded with the planet's orbital period, are plotted in Fig.~\ref{fig:lcs} together with the best-fit transit models resulting from the global fit.
The posterior estimates of the main parameters of the joint fit are reported in Table~\ref{tab:planet}. Therefore, here we confirm the planetary nature of the TESS candidate TOI\,5795.01, which we will call TOI-5795\,b from now on, with a mass of $M_{\rm p}=23.66^{+4.09}_{-4.60}\, M_{\oplus}$ (e.g. $>5\sigma$ significance), a radius of $R_{\rm p}=5.62\pm0.11 \, R_{\oplus}$, corresponding to a bulk density of $0.73\pm0.13$\,g\,cm$^{-3}$ (compatible with that of Saturn, $0.69\, g\,cm^{-3}$) and an equilibrium temperature of $T_{\rm eq}=1136\pm18$\,K.
Therefore, it can be considered as a hot super Neptune and its orbital period ($6.14$\,days) places it at the edge of the Neptune ridge (according to the picture by \citealt{2024A&A...689A.250C}), i.e. at the beginning of the so-called {\it savanna} (see Fig.~\ref{fig:desert} and the discussion in Sect.~\ref{sec:summary}). The orbital eccentricity of TOI-5795\,b was estimated to be $e=0.15 \pm 0.06$ and is, thus, compatible with zero at $95.5\%$ confidence, though the Bayesian evidence is slightly in favor of the circular scenario ($\Delta\ln{\mathcal{Z}}^{e=0}_{e\geq0}<3$).
Finally, we note that the measured RV jitter is higher than anticipated ($3.75^{+1.23}_{-1.08}$; see Table\,\ref{tab:planet}), particularly given the relatively quiet nature of TOI-5795. This may be attributable to unresolved signals not captured within our limited temporal baseline. Continued RV monitoring is therefore recommended to better characterize any additional variability.

\begin{table} %
\centering %
\caption{Orbital and physical parameters for  TOI-5795\,b.} %
\label{tab:planet} %
\begin{tabular}{l c c}
\hline\hline \\[-8pt]
Parameter & Unit & Value \\ [2pt]
\hline \\[-8pt]%
\multicolumn{1}{l}{~~~~~~~~~~~\textbf{Light curve}} \\ 
\multicolumn{1}{l}{~~~~~~~~~~~\textbf{parameters}} \\ [2pt]
$P_{\rm orb}$\dotfill & day & $6.1406325 \pm 0.0000054$ \\
$T_{\rm 0}$\dotfill & BJD$_{\rm TDB}$ & $2\,459\,770.67397_{-0.00061}^{+0.00060}$ \\[2pt]
$T_{\rm 14}$\dotfill & hr & $3.407^{+0.028}_{-0.026}$ \\ [2pt]
$R_{\rm p}/R_{\star}$\dotfill & & $0.0477\pm0.0006$ \\ [2pt]
$b$\dotfill & & $0.29^{+0.12}_{-0.14}$ \\  [2pt]
$i$\dotfill & deg & $88.54^{+0.68}_{-0.55}$ \\  [2pt]
$a/R_{\star}$\dotfill & & $12.65^{+0.34}_{-0.35}$ \\ [6pt]
\multicolumn{1}{l}{~~~~~~~\textbf{Limb-darkening}} \\ 
\multicolumn{1}{l}{~~~~~~~~~~~\textbf{coefficients$^{(a)}$}} \\ [2pt]
$q_1$,\,\textsc{tess}\dotfill & & $0.20^{+0.15}_{-0.08}$ \\ [2pt]
$q_2$,\,\textsc{tess}\dotfill & & $0.41^{+0.33}_{-0.24}$ \\ [2pt]
$q_1$,\,$B$\dotfill & & $0.42^{+0.36}_{-0.29}$ \\ [2pt]
$q_2$,\,$B$\dotfill & & $0.52^{+0.31}_{-0.29}$ \\ [2pt] %
$q_1$,\,\textsc{$\rm LCO$}\dotfill & & $0.50^{+0.33}_{-0.32}$ \\ [2pt] %
$q_2$,\,\textsc{$\rm LCO$}\dotfill & & $0.54^{+0.32}_{-0.33}$ \\ [6pt] %
\multicolumn{1}{l}{~~~~~~~~\textbf{RV parameters}} \\ [2pt] %
$K$\dotfill & m\,s$^{-1}$ & $8.97^{+1.51}_{-1.74}$ \\ [2pt] %
$\sqrt{e}\sin\omega$\dotfill & & $0.197^{+0.101}_{-0.105}$ \\ [2pt] %
$\sqrt{e}\cos\omega$\dotfill & & $0.317^{+0.107}_{-0.142}$ \\ [2pt] %
$e^{(b)}$\dotfill & & $<0.25$ \\ [2pt] %
$\omega$\dotfill & deg & $31.67^{+25.07}_{-17.83}$ \\   [2pt] %
RV jitter HARPS & m\,s$^{-1}$ & $3.75^{+1.23}_{-1.08}$ \\  [2pt] %
\multicolumn{1}{l}{~~~~~~~~~~~~\textbf{Derived}} \\ [2pt] %
$M_{\rm p}$\dotfill & $M_{\oplus}$ & $23.66^{+4.09}_{-4.61}$ \\ [2pt]
$R_{\rm p}$\dotfill & $R_{\oplus}$ & $5.62\pm0.11$ \\
$\rho_{\rm p}$\dotfill & g\,cm$^{-3}$ & $0.73 \pm 0.13$ \\ 
$\log{g_{p}}$\dotfill & cgs & $7.33\pm1.29$ \\
$a$\dotfill & au & $0.064\pm0.002$ \\ 
$T_{\rm eq}^{(c)}$\dotfill & K & $1136.78^{+18.89}_{-18.02}$ \\  [2pt] %
TSM$^{(d)}$\dotfill & & $103^{+25}_{-15}$ \\ [2pt]
\multicolumn{1}{l}{~~~~~~~~~~~~\textbf{GP coefficients}} \\ [2pt] %
$\sigma_{\rm GP}$\dotfill &  & $0.00160 \pm 0.00014$ \\ 
$\rho_{\rm GP}$\dotfill &  & $0.98 \pm 0.09$ \\ 
\hline
\end{tabular}
\tablefoot{The median values of the best-fit parameters for TOI-5795\,b, along with their upper and lower $68\%$ credibility intervals as uncertainties. These values, both fitted and derived, were obtained from the posterior distributions of the corresponding models.
\tablefoottext{a}{$q_1 \equiv (u_1+u_2)^2$ and $q_2 \equiv 0.5\,(u_1+u_2)^{-1}$, where $u_1$ and $u_2$ are the LD coefficients of the quadratic law \citep{2013MNRAS.435.2152K}.}
\tablefoottext{b}{The $95\%$ ($2\,\sigma$) confidence upper limit on the eccentricity determined when $\sqrt{e} \cos{\omega}$ and $\sqrt{e} \sin{\omega}$ are allowed to vary in the fit.}
\tablefoottext{c}{This represents the equilibrium temperature assuming a Bond albedo of zero and an uniform redistribution of heat to the night side.}
\tablefoottext{d}{Transmission spectroscopy metric (TSM; \citealt{2018PASP..130k4401K}).}
}
\end{table}

\section{The formation and evolution of TOI-5795\,b}
\label{subsec:formation_history}
We used the Monte Carlo version of the \texttt{GroMiT} (planetary GROwth and MIgration Tracks) code \citep{polychroni_2023_10593198} to investigate the possible formation history of TOI-5795\,b, following \citet{2024A&A...691A..67M} and \citet{2025A&A...693A...7N}. The simulations are based on the pebble accretion model adopting the treatment for the growth and migration of solids- and gas-accreting planets from \citet{johansen2019planetary} and \citet{tanaka2020final}, respectively. We use the scaling law for the pebble isolation mass from \citet{bitsch2018pebble} to explore the formation of 10$^{5}$ planetary seeds embedded in a plausible native protoplanetary disk with mass equal to 5\% that of the star and characteristic radius of 60 au (i.e. following the Solar Nebula-like template from \cite{turrini2023gaps}. The protoplanetary disk is described by a steady state viscous disk \citep{johansen2019planetary, armitage2020astrophysics} characterized by mass loss due to photoevaporation of $10^{-9}$ M$_{\odot}$ yr$^{-1}$ \citep{tanaka2020final} and thermal profile due to viscous heating and stellar irradiation following the prescription of \citet{ida2016radial}. The pebble-to-gas ratio in the different regions of the disk is computed from the stellar metallicity, given in Table \ref{tab:star}, using the condensation profile from \citet{turrini2023gaps}. We use the stellar evolutionary models of \citet{baraffe2015new} to set the pre-main sequence luminosity to $1.23\,L_{\odot}$ as predicted for a star with a mass like that of TOI-5795 at 1.5\,Myr. 

\begin{figure}
\centering
\includegraphics[width=1\linewidth]{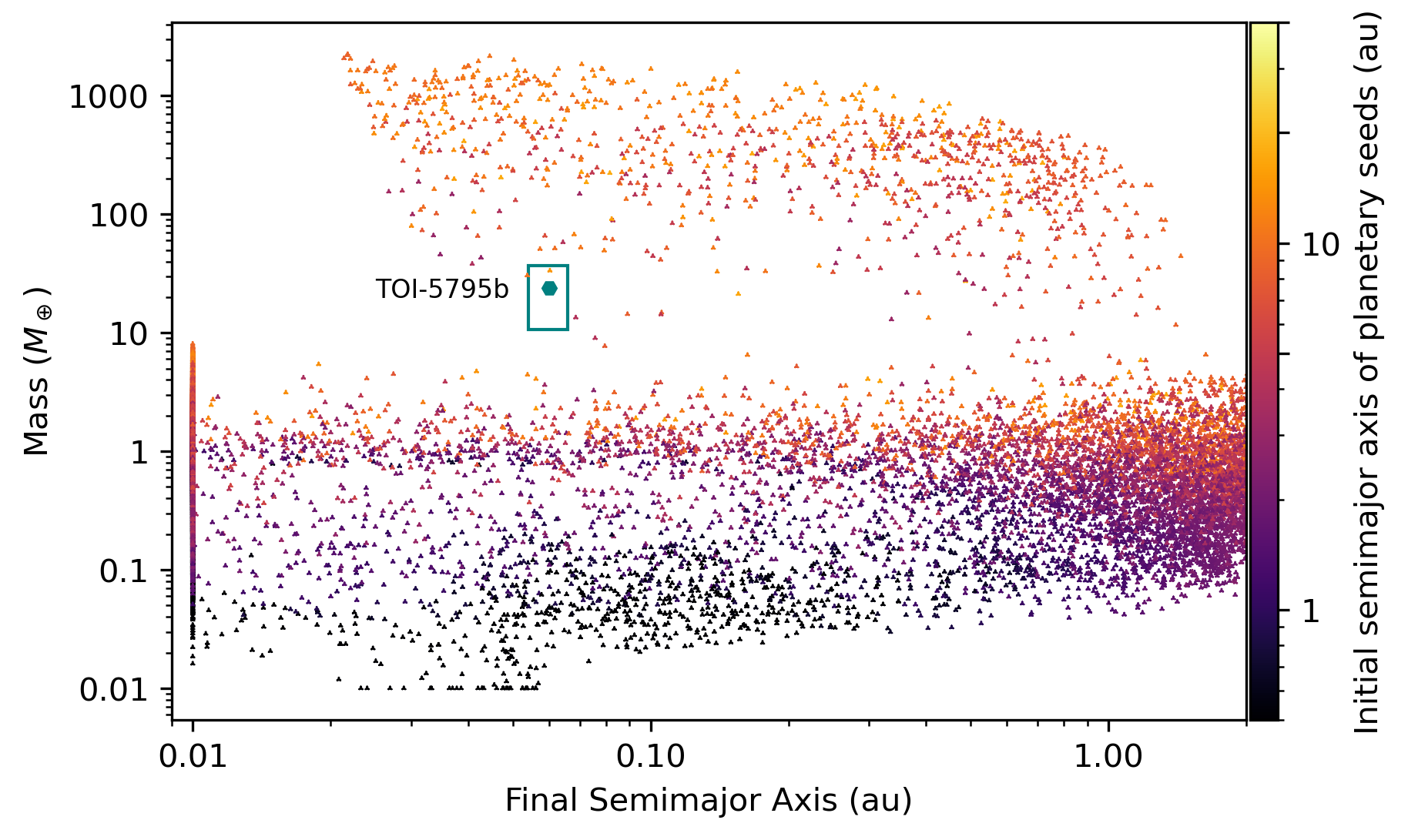}
\caption{Synthetic population of 4$\times$10$^5$ exoplanets generated by a plausible native protoplanetary disk of TOI-5795\,b. The synthetic planets are plotted in the final semi-major axis versus planet mass plane. The color-scale traces the initial formation region of the planetary seeds. Planet TOI-5795\,b and its 3$\sigma$ uncertainty range are shown in cyan. } 
\label{fig:popsyn}
\end{figure}

The simulations follow the planet formation process for 5\,Myr. The planetary seeds are randomly placed in the disk varying ($i$) their initial semi-major axis (0.1-60\,au), ($ii$) their input time ($0.1\,{\rm yr} - 3\,{\rm Myr}$) and ($iii$) the disk viscosity coefficient within the range reported in \cite{rosotti2023}. We considered both mm- and cm-size pebble-dominated disks. The success rate in producing synthetic planets within $3\,\sigma$ of observed semi-major axis and mass of the real planet proved extremely limited. This prompted us in running further sets of simulations. We first explored the impact of more efficient pebble accretion using pebble sizes from 1\,cm to 1\,dm for a total of $3\times 10^5$ additional simulations. In Fig. \ref{fig:popsyn} we plot the cumulative results of all these simulations, over-plotting the current semi-major axis and mass of the planet. Once again reproducing TOI-5795\,b proved difficult in the framework of pebble accretion, as we found only one object within $3\,\sigma$ of the planet out of a total of 4$\times$10$^5$ synthetic planets.

\begin{figure}
\centering
\includegraphics[width=1\linewidth]{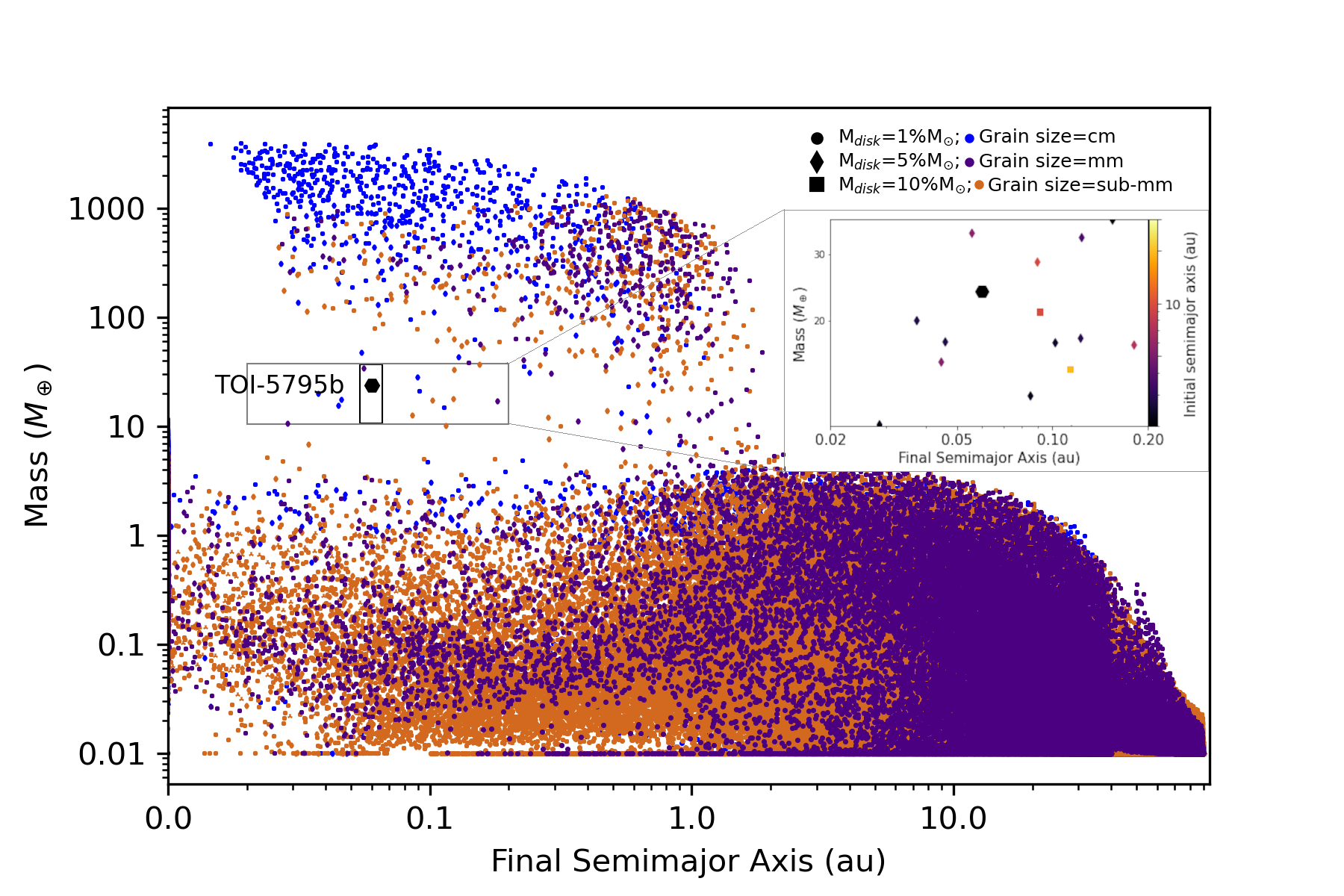}
\caption{Synthetic populations of 7.9$\times$10$^5$ exoplanets generated sampling different initial circumstellar disk configurations (disk mass: 1, 5, 10\% of the stellar mass; R$_0$: 40 and 90\,au; pebble size: cm, mm and sub-mm). Planet TOI-5795\,b and its $3\sigma$ uncertainty range are shown in black. The grey box indicates the range inwards of 0.2\,au and the $3\sigma$ mass uncertainty range of the planet. The inset plot is a zoom in of this window where we show as a color bar the initial semi-major axis of the seeds as the color scale. } 
\label{fig:zoomin}
\end{figure}

We then explored whether this lack of matching outcomes could be due to chosen disk conditions, as pebble accretion strongly depends on the properties of the environment in which it takes place. We run further simulation campaigns changing the initial mass and characteristic radius of the circumstellar disk based on the results of observational surveys from \citet{Testi2022} and \citet{andrews2020}. Specifically, we considered disk masses of 1\%, 5\% and 10\% of the stellar mass and characteristic radii R$_0$=40\,au (compact disk) and R$_0$=90\,au (extended disk). While the resulting planetary populations contain more objects in the mass range of TOI-5795\,b, once again only one synthetic planet falls within $3\,\sigma$ of the real one (see Figure \ref{fig:zoomin}).

Finally, we explored the possibility that the migration prescription adopted in \texttt{GroMiT} does not provide an accurate description of the dynamical history of TOI-5795\,b due to the uncertainty on migration efficiency \citep[e.g.][]{pirani2019consequences}. Specifically, we focused on all synthetic planets with final semi-major axis inwards of $0.2\,au$. We found 14 synthetic planets that fit TOI-5795\,b under this relaxed fitting criterion (see zoom-in window in Figure \ref{fig:zoomin}), translating into a comparatively higher success rate that, however, remains quite low in absolute terms ($0.0018\%$). These best-fitting solutions arise from disks populated by pebbles either cm-sized or sub-mm sized and with initial mass of either 5\% or 10\% of the stellar mass of the system. The larger pebble size favors starting conditions beyond the water snowline while the smaller one favor initial conditions closer to the host star, suggesting differences in the bulk composition of the core (ice-rich vs rock-dominated).

Our results appear to indicate that pebble accretion was not the dominant mechanism in forming this planet, or that it worked in combination with planetesimal accretion. This finding is consistent with the conclusions of previous studies on exoplanets with similar masses \citep[see][]{mantovan2024,2025A&A...693A...7N,zingales2025} and with the challenges met by population synthesis codes in capturing the full diversity of exoplanets. In their recent review, \citet{Burn2025} compared several planet formation population synthesis codes, including some based on the pebble formation theory, and found that no model is currently capable of fully reproducing the observed mass-distance diagram of known planets (both among exoplanets and in the Solar System). 

Given the mature age of the star, we also cannot exclude that this is not the planet as originally formed. Specifically, the present TOI-5795\,b could be the product of a stochastic event, like a merger between two or more smaller primordial planets in a system originally characterized by higher multiplicity \citep{turrini2020}, as its modal eccentricity could be compatible with collisional damping after dynamical instability \citep{chambers2001,laskar2017}. Another explanation is that this planet may have formed at a larger orbital distance and been scattered inwards later, with its orbit becoming circularized after initially being eccentric. Given the still large uncertainty on the eccentricity of the planet, exploring such a scenario is beyond the scope of this paper. Regardless, such an event would confuse, or even erase the original formation path of this planet.

\section{Atmospheric evaporation of TOI-5795\,b}
\label{sec:atmospheric_evaporation}
We investigated the evolutionary history of the atmosphere of TOI-5795\,b since, as is well known, a close-in planet can be stripped of part of its atmosphere due to UV radiation coming from its parent star. This kind of analysis is useful for evaluating how much atmosphere such a super-Neptune planet could have retained after an evolution of several millions of years.

\begin{figure}
\centering
\includegraphics[width=0.49\textwidth]{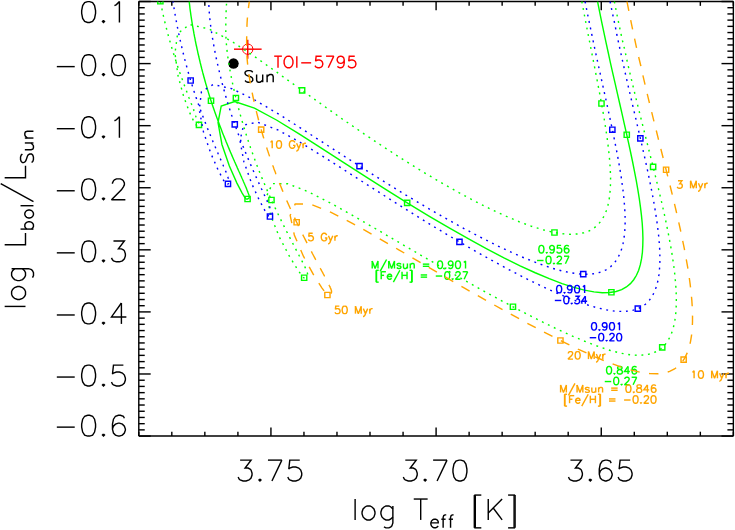}
\caption{Evolutionary track of TOI-5795 in the effective temperature-bolometric luminosity plane. The red dot marks the current location of the star on the track.} 
\label{fig:track_MESA}
\end{figure}

First, we have selected a MESA evolutionary track \citep{2016ApJ...823..102C}, which recovers the stellar position at its nominal age in the $L_{\rm bol}$ vs. $T_{\rm eff}$ diagram, where $L_{\rm bol}$ is the bolometric luminosity of the host star. To this aim, we considered a grid of evolutionary tracks for stars with masses and metallicities equal to the nominal values for TOI-5795, or equal to
$\pm 1\sigma$ values in both parameters (9 tracks in total).
Some of them are shown in Fig.~\ref{fig:track_MESA}. We found that the three tracks with nominal ${\rm [Fe/H]}=-0.27$ (in green) appear to be too `cool' with respect to the parent star. The same occurred for the tracks with $M_{\star} = 0.9 \, M_{\sun}$ (nominal value, in blue). Eventually, we preferred the track with $M_{\star} = 0.846 \, M_{\sun}$ ($-1\sigma$) and ${\rm [Fe/H]} = -0.20$ ($+1\sigma$) (in orange), that provides the best match with the position of star in the luminosity--temperature diagram at its nominal age. This track is used in our modeling to determine the bolometric irradiation of the planet and therefore its equilibrium temperature and core-envelope structure during evolution, taking into account the gravitational shrinking \citep{2007ApJ...659.1661F}.

To model the evolution of high-energy irradiation, we needed to determine an anchor point of the X-ray luminosity at the current age. Since there is no X-ray detection available for this old, low-activity star, we considered as proxies the measured chromospheric $R^{\prime}_{\rm HK}=-5.07 \pm 0.02$ index and stellar age.
Using the relationships by \citet{2008ApJ...687.1264M}, the chromospheric index yields an X-ray luminosity $L_{\rm X} \sim 7 \times 10^{26}$\,erg/s, that is compatible with an age of $\sim 8$\,Gyr. Assuming a rotational velocity of $2$\,km\,s$^{-1}$, we computed a lower limit on the rotational period of $\approx 30$\,d, which indicates an upper limit $L_{\rm X} < 7 \times 10^{27}$\,erg\,s$^{-1}$, using the activity-rotation relationship of \citet{2003A&A...397..147P}. We also determined an uncertainty range in $L_{\rm X}$ from the relationship of the X-ray to bolometric luminosity ratio vs. the Rossby number proposed by \citet{2018MNRAS.479.2351W}. Taking into account the uncertainties on both the stellar mass and the rotational period, we determined a conservative lower limit of $2 \times 10^{26}$\,erg\,s$^{-1}$ for the X-ray luminosity.

\begin{figure}
\centering
\includegraphics[width=1\linewidth]{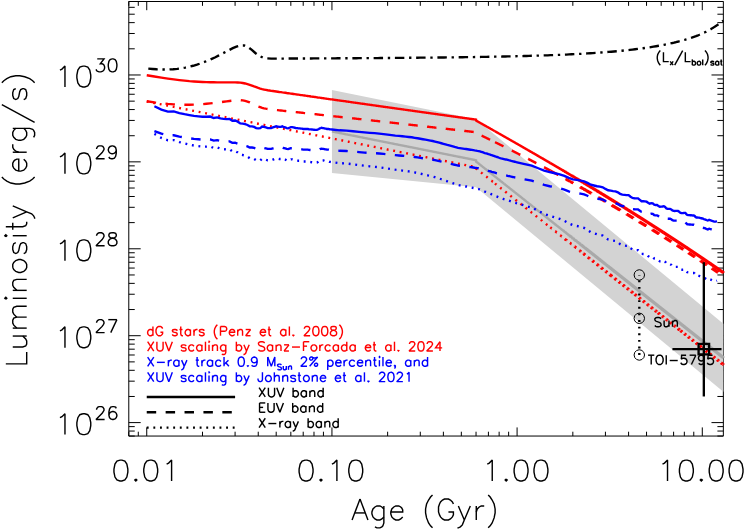}
\caption{Time evolution of X-ray ($5-100$\,\AA), EUV ($100-920$\,\AA), and total XUV luminosity of TOI-5795, according to \citet{2008A&A...477..309P} and the X-ray/EUV scaling by \citet{2025A&A...693A.285S} (red lines) 
and according to \citet{2021A&A...649A..96J} (blue lines). Uncertainties on the age and X-ray luminosity of TOI-5795 are also indicated. The gray area is the original locus for dG stars in \citet{2008A&A...477..309P}.} 
\label{fig:LKUVevol}
\end{figure}

In Fig.~\ref{fig:LKUVevol}  we show the time evolution of the X-ray ($5-100$\,\AA), EUV ($100-920$\,\AA), and XUV (X+EUV) luminosities, according to \citet{2008A&A...479..579P} X-ray luminosity evolution coupled with the \citet{2025A&A...693A.285S} X-ray to the EUV scaling law or, alternatively, adopting the \citet{2021A&A...649A..96J} prescription. 
In the former case, the current X-ray luminosity appears to be near the median for solar-mass stars at ages of $\sim 10$\,Gyr, such as TOI-5795.
As an alternative description, we selected the evolutionary track corresponding to the lowest $2\%$ percentile of the activity distribution appropriate for stars with $0.9\,M_{\sun}$. Nonetheless, the \citet{2021A&A...649A..96J} evolutionary path yields an X-ray luminosity that is about a factor 4 higher than the nominal value determined above, but within the large uncertainty range. 
On the other hand, given the very old age of TOI-5795, its current activity-rotation state is also compatible with the evolution of a star that started with a much higher rotation rate and activity level. For this reason, we also explored the results of photoevaporation assuming the \citet{2021A&A...649A..96J} track corresponding to the highest $98\%$ percentile of the activity distribution. With this assumption, the high-energy irradiation of the planet is possibly overestimated at late ages ($>1$\,Gyr). 

\begin{figure*}[h!]
\centering
\begin{tabular}{@{}c@{}c@{}c@{}}
    \includegraphics[width=6.5cm]{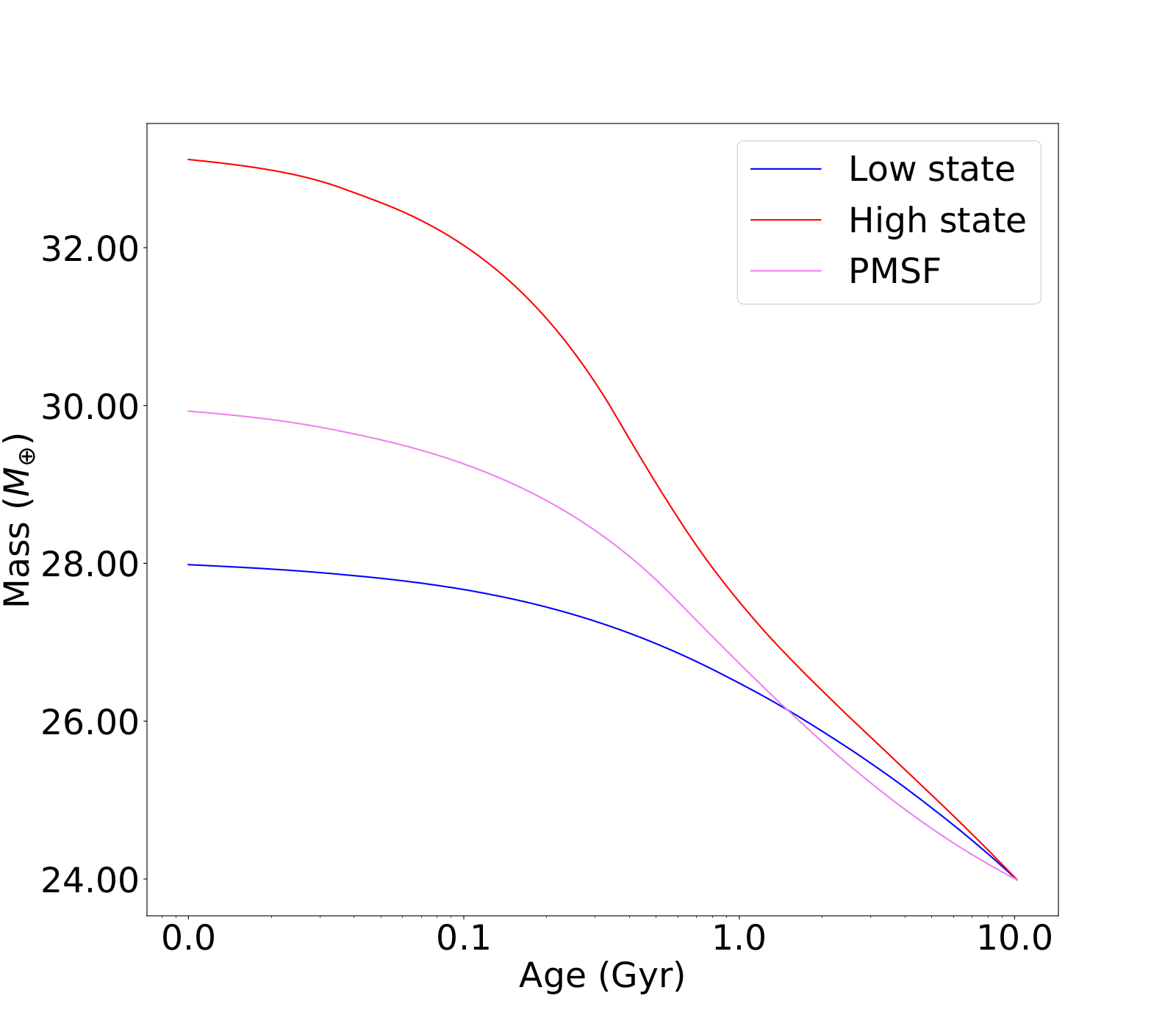} & 
    \includegraphics[width=6.5cm]{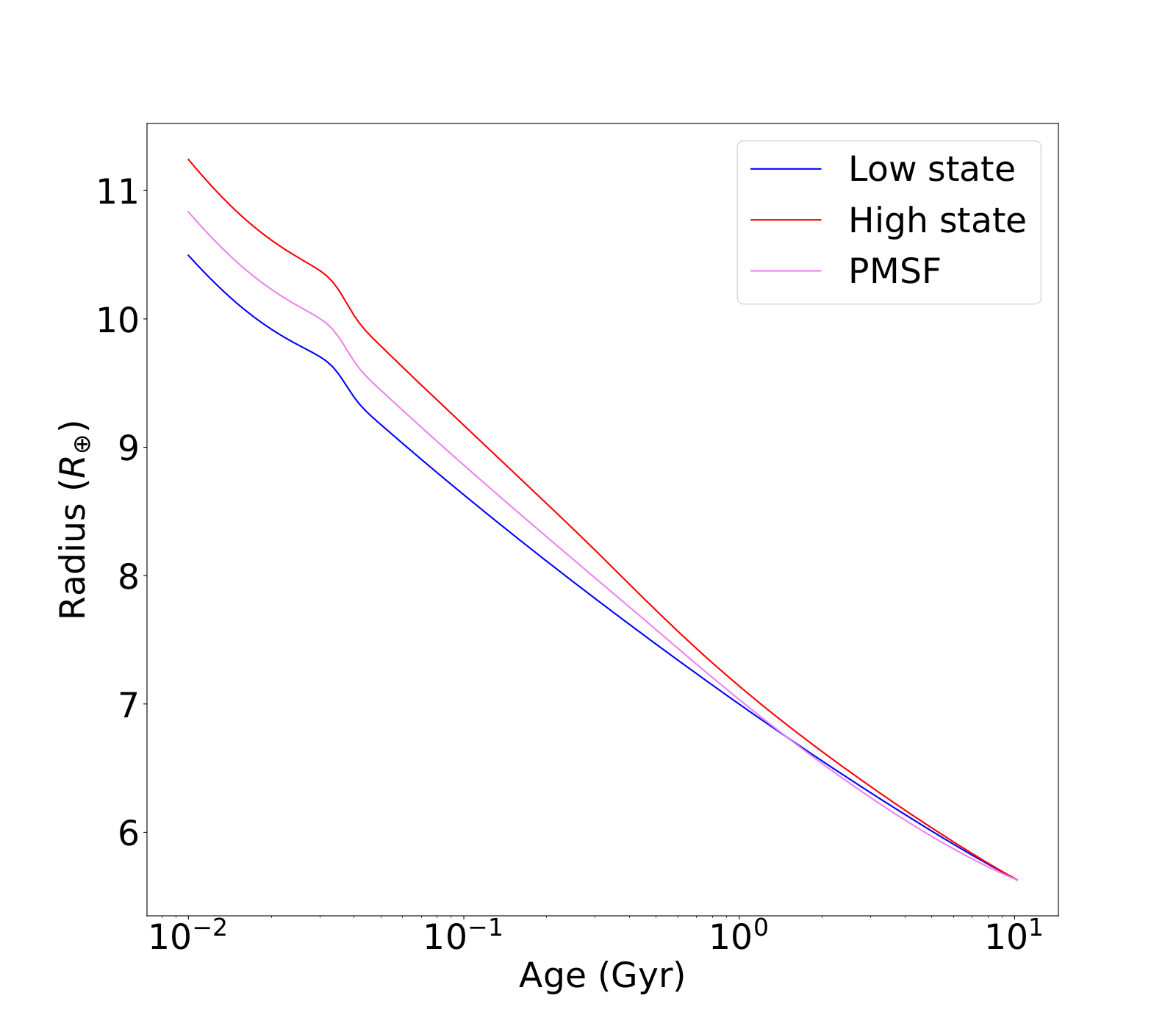} & 
    \includegraphics[width=6.5cm]{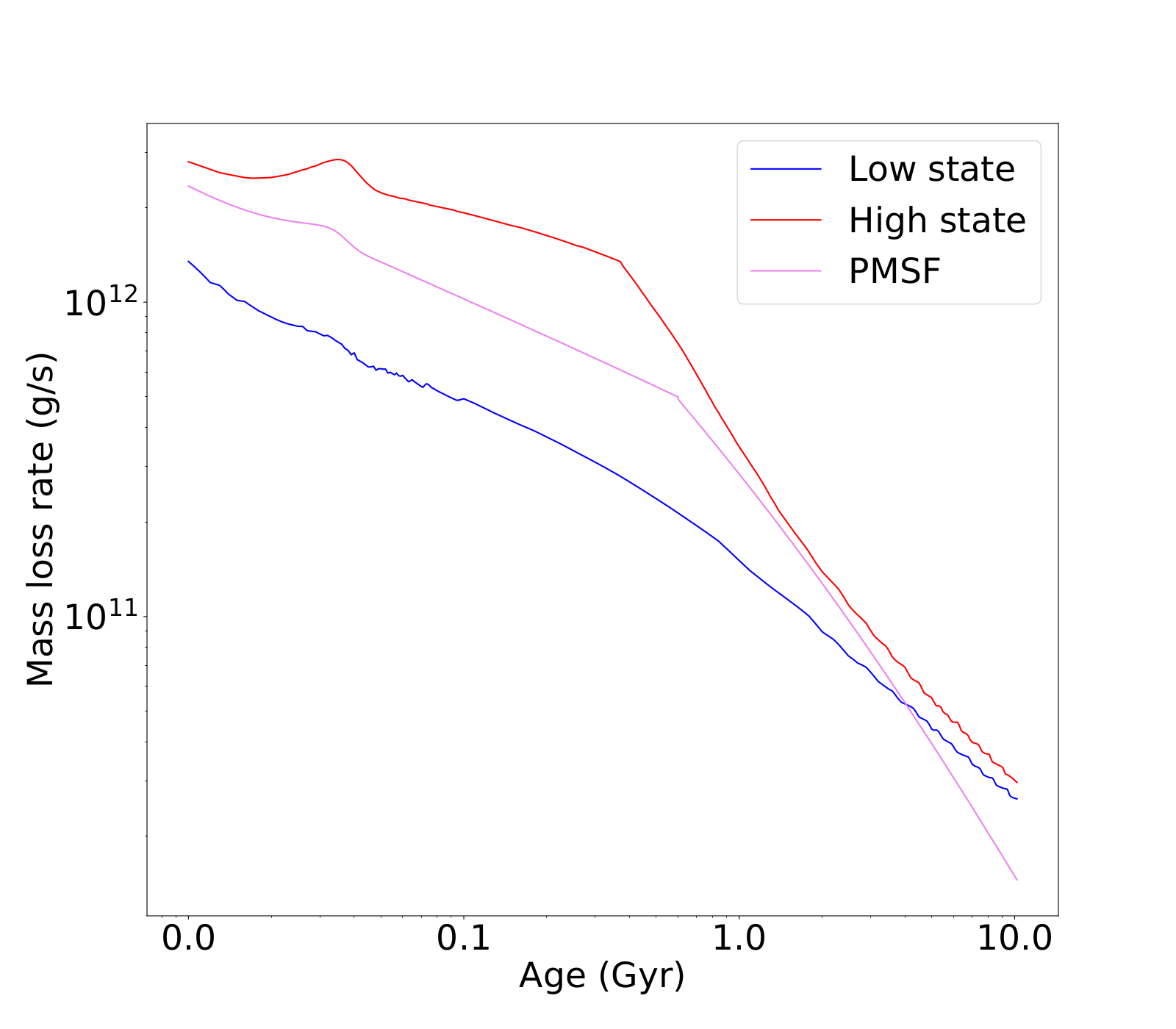} \\
\end{tabular}
\caption{Temporal evolution of mass, radius, and mass-loss rate of TOI-5795\,b. The left panel shows the evolution of planetary mass, the middle panel the evolution of the radius, and the right panel the evolution of the mass-loss rate. The colors refers to the different high energy evolutionary tracks: blue line refers to the track called low state (track corresponding to the lowest 2\% percentile of the activity distribution), red line refers to the track called high state (track corresponding to the highest 2\% percentile), and the pink line refers to the the track called PMSF (track based on the X-ray luminosity evolution of \citealt{2008A&A...479..579P}, combined with the X-ray to EUV conversion from \citealt{2025A&A...693A.285S}).}
\label{fig:evap}
\end{figure*}

For the analysis of the past evaporative history, we employed the numerical code presented in \citet{Locci19} and subsequently used it for the study of single systems (e.g., \citet{Maggio22}). In this study, we considered three distinct evolutionary tracks for stellar XUV luminosity, referred to as the low state (track corresponding to the lowest 2\% percentile of the activity distribution), the high state (track corresponding to the highest 2\% percentile), and the PMSF track (based on the X-ray luminosity evolution of \citealt{2008A&A...479..579P}, combined with the X-ray to EUV conversion from \citealt{2025A&A...693A.285S}). In Fig. \ref{fig:evap} we show the evolution of the mass, radius, and mass loss rate for the three scenarios.

Assuming an Earth-like core composition, we first calculated the planet’s internal structure, finding a core mass of approximately $16.5\,M_\oplus$ and a radius of around $2\,R_\oplus$, which implies an atmospheric fraction of approximately 30\%. We then proceeded to compute the backward evolution of key planetary parameters such as mass and radius.

The mass-loss rates were estimated using the ATES analytical approximation \citep{Caldiroli21}. At the present time, we found mass-loss rates of approximately $1.4 \times 10^{10}$\,g\,s$^{-1}$  for the PMSF track, $2.6 \times 10^{10}$\,g\,s$^{-1}$ for the low state, and $3 \times 10^{10}$\,g\,s$^{-1}$ for the high state.

We followed the evolution back to an age of 10 million years, which we assumed as the time when the accretion disk is fully dissipated and the planet has reached its final orbit. The inferred initial masses were approximately $30\,M_\oplus$ for the PMSF track, $28\,M_\oplus$ for the low state, and $33\,M_\oplus$ for the high state. The current mass of the planet is about $24\,M_\oplus$, indicating a mass loss of $9\,M_\oplus$ in the high state scenario, $6\,M_\oplus$ in the PMSF track, and $4\,M_\oplus$ in the low state.

We also observed a significant contraction of the planetary radius in all three cases. The predicted initial radii were approximately $11.2\,R_\oplus$ for the high state, $10.8\,R_\oplus$ for the PMSF track, and $10.5\,R_\oplus$ for the low state. The current planetary radius is around $6\,R_\oplus$, suggesting a large contraction mostly driven by gravitational compression and possibly also by atmospheric loss.

\section{Summary and discussion}
\label{sec:summary}
\begin{figure}
\centering
\includegraphics[width=0.8\linewidth]{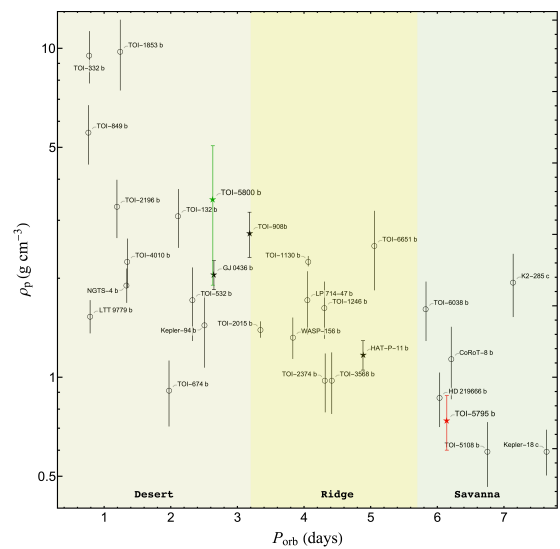}
\caption{$\rho_{\rm p}-P_{\rm orb}$ $\log$-linear diagram of known close-in transiting exoplanets with $3\,R_{\oplus}<R_{\rm p}<7\,R_{\oplus}$ and mean density measured with an accuracy of no more than $30\%$. Circles display the position of planets with eccentricity $e \leq 0.1$, while five-pointed stars those with $e > 0.1$. The values of the parameters were taken from TEPCat. Horizontal error bars have been suppressed for clarity. Vertical error bars represent one standard deviation. The red star indicates the position of the new planet TOI-5795\,b, see Sect.~\ref{sec:analysis}. The green star indicates the position of the new planet TOI-5800\,b from our HONEI I paper \citep{2025arXiv250510123N}; its mean density has been measured with an accuracy $<50\%$. The three colored regions of the diagram are those recognised by \citet{2024A&A...689A.250C}.
} 
\label{fig:diagram_1}
\end{figure}

Thanks to the continued discovery of planet candidates by the Transiting Exoplanet Survey Satellite (TESS) \citep{2015JATIS...1a4003R} and the effort of follow-up groups, a number of planets have been found deep within the Neptune desert, as well as at its boundaries, over the past few years. These surprising discoveries include both low-density gaseous planets as well as high-density Neptunes, which should be unusually composed of a remarkable fraction of rocks, see Fig.~\ref{fig:diagram_1}\footnote{Data taken from the Transiting Extrasolar Planet Catalog (TEPCat), available at \url{https://www.astro.keele.ac.uk/jkt/tepcat/} \citep{2011MNRAS.417.2166S} as of June 2025.}. The values of the mean density, $\rho_{\rm p}$, of the known Neptune-size planets range from those of HATS-38\,b ($0.40\pm0.07$\,g\,cm$^{-3}$; \citealt{2024AJ....168..185E}) and TIC\,365102760\,b ($0.44\pm0.15$\,g\,cm$^{-3}$; \citealt{2024AJ....168....1G}) to the extreme ones of TOI-332\,b ($9.60\pm1.10$\,g\,cm$^{-3}$; \citealt{2023MNRAS.526..548O}) and TOI-1853\,b ($9.74\pm0.76$\,g\,cm$^{-3}$; \citealt{2023Natur.622..255N}). This diversity challenges conventional theories of planetary formation and evolution and calls for other hypotheses, such as catastrophic formation scenarios (e.g., multiple planetary collisions; \citealt{2023Natur.622..255N}). 

We have confirmed and characterized TOI-5795\,b, a new super-Neptune planet \textbf{($M_{\rm p}=23.66^{+4.09}_{-4.60}\, M_{\oplus}$; $R_{\rm p}=5.62\pm0.11 \, R_{\oplus}$)} orbiting a metal-poor G3\,V star on a 6.14-day orbit compatible with a circular one. Its position in the mass-period plane is at the border between the ridge and the savanna; see Fig.~\ref{fig:desert}.

\begin{figure}
\centering
\includegraphics[width=1\linewidth]{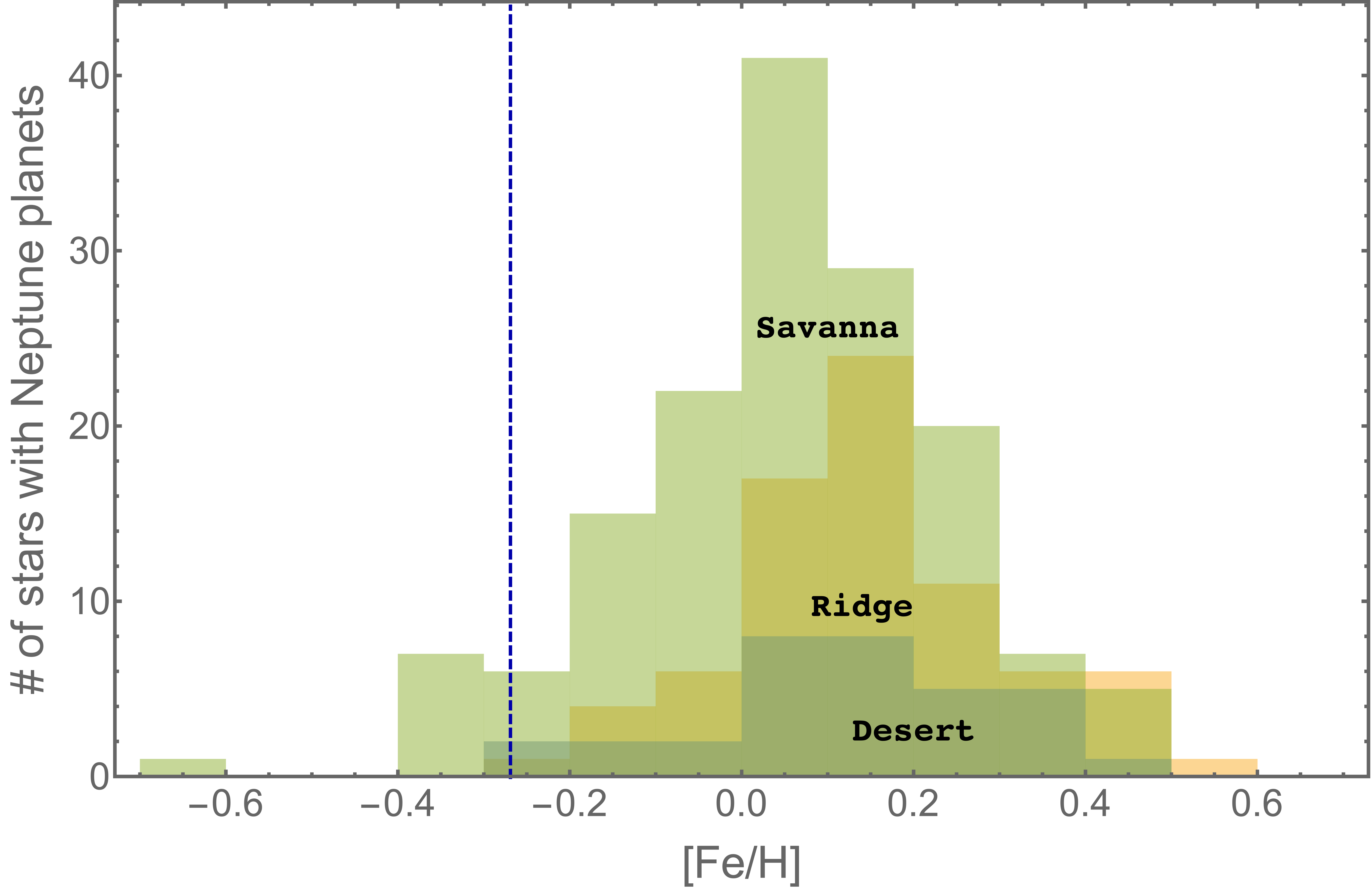}
\caption{Frequency distribution of parent-star metallicity for transiting exoplanets with 10\,$M_{\oplus}$<$M_{\rm p}$<100\,$M_{\oplus}$ \citep{2025AJ....169..117V}. The histograms are shown with a 0.1 bin and for the three different populations pointed out by \citet{2024A&A...689A.250C}. ``Desert'' refers to Neptune planets with orbital period $P_{\rm orb}$<\,3.2\,days; ``Ridge'' with 3.2\,days\,<\,$P_{\rm orb}$<\,5.7\,days; ``Savanna'' with 5.7\,days\,<\,$P_{\rm orb}$<100\,days. Data taken on 2025/06/06 from TEPCat. The dashed vertical line refers to the metallicity that we measured for the parent star of TOI-5795\,b.
} 
\label{fig:metallicity}
\end{figure}

Very recently, \citet{2025AJ....169..117V} presented a demographic study, based on data from Gaia DR3 \citep{2023A&A...674A...1G}, by considering the metallicities of stars hosting Neptune-sized planets. They found that planets in the Neptune desert and in the ridge orbit stars significantly more metal-rich than hosts of smaller planets. The {\it top-down} theory, which proposes that Neptune desert planets are the exposed interiors of former gas giants, predicts that their host stars should be metal-rich, similar to hot Jupiters. The host-star population of Neptune-savanna planets, instead, resulted to be significantly less metal-rich. The metal-poor nature of TOI-5795 (${\rm [Fe/H]}=-0.27\pm0.07$) aligns with the characteristics expected for Neptune savanna hosts, contrasting sharply with the metal-rich hosts typically found for Neptune desert planets \citep{2024A&A...691A.233C}, see Fig.~\ref{fig:metallicity}. Another property of savanna planets is that they are typically low density compared to the higher-density ridge and desert planets \citep{2024A&A...691A.233C}, and 5795\,b fits in with this paradigm as well (see Fig.~\ref{fig:diagram_1}). 

The characteristics of the TOI-5795 planetary system support the idea that planets in the savanna are unlikely to have formed through the top-down mechanism and likely have a distinct origin and evolution history compared to the hottest Neptunes in the desert. In this context, we carried out further investigation by simulating numerous planetary formation scenarios within a protoplanetary disk, varying factors such as initial position, seed input time, and disk viscosity; see Sect.~\ref{subsec:formation_history}. Despite extensive simulations, few simulated planets accurately matched the observed characteristics of TOI-5795\,b, suggesting challenges in explaining its formation through current pebble accretion theory. This discrepancy might be due to uncertainties in migration prescriptions within the code or the inherent difficulty in forming planets of TOI-5795\,b's mass via pebble accretion, also acknowledging the possibility of stochastic events like three-body interactions or mergers.

By determining the time evolution of the parent star's high-energy irradiation, we also investigated the impact of UV radiation from the parent star in stripping a part of a planet's atmosphere over billions of years. Since TOI-5795 is very old ($10.2_{-3.3}^{+2.5}$\,Gyr), its initial rotation period is poorly constrained. This is why we chose to consider three different evolutionary tracks for XUV radiation, which produce significantly different results, as discussed in Sect.~\ref{sec:atmospheric_evaporation}.
The evolution was tracked back to an age of 10 million years, which is assumed to be when the accretion disk dissipated and the planet reached its final orbit. We estimated that TOI-5795\,b may have lost a significant portion ($14\%-27\%$) of its initial mass due to atmospheric evaporation.
In particular, in the high-state scenario, the planet experienced the largest mass loss, estimated at $9\,M_\oplus$ from an initial mass of approximately $33\,M_\oplus$. This scenario might overestimate high-energy irradiation at late ages. Instead, under the PMSF track,  the planet lost $6\,M_\oplus$ from an initial mass of approximately $30\,M_\oplus$. In the low state scenario, finally, the planet lost $4\,M_\oplus$ from an initial mass of approximately $28\,M_\oplus$. However, TOI-5795\,b, due to its large mass, manages to retain a large part of its atmosphere, ending its evolution with about $32\%$ of its mass in the atmosphere. Along with mass loss, a significant contraction in the planetary radius was observed across all three scenarios. The predicted initial radii were approximately $11.2\,R_\oplus$ (high state), $10.8\,R_\oplus$ (PMSF track), and $10.5\,R_\oplus$ (low state). The current planetary radius is around $6\,R_\oplus$, indicating a strong contraction driven primarily by gravitational compression.

Using the parameters of the star (Table~\ref{tab:star}) and the planet (Table~\ref{tab:planet}), we estimated the transmission spectroscopy metric (TSM; \citealt{2018PASP..130k4401K}) of TOI-5795\,b, which, considering its equilibrium temperature ($T_{\rm eq}=1136\pm18$\,K.), is quite high, i.e. equal to $103_{-15}^{+25}$. This makes TOI-5795\,b a good candidate for possible atmospheric characterization with space telescopes. In this context, we note that TOI\,5795.01 is included in a list of TESS candidates, which are potentially suitable for Ariel observations.\footnote{\url{https://exofop.ipac.caltech.edu/tess/view_toi_ariel.php}.} The present confirmation of TOI\,5795.01 as a planet is, therefore, helpful for the ongoing preparatory science work carried out by the Ariel Science Consortium \citep{2022EPSC...16.1114T}.

\begin{acknowledgements}
This work is based on observations collected at the European Organization for Astronomical Research in the Southern Hemisphere under ESO programme 113.26UJ.001.
This work includes data collected with the TESS mission, obtained from the MAST data archive at the Space Telescope Science Institute (STScI). Funding for the TESS mission is provided by the NASA Explorer Program. STScI is operated by the Association of Universities for Research in Astronomy, Inc., under the NASA contract NAS 5–26555. The authors acknowledge the use of public TESS data from pipelines at the TESS Science Office and at the TESS Science Processing Operations Center. Resources supporting this work were provided by the NASA High-End Computing (HEC) Program through the NASA Advanced Supercomputing (NAS) Division at Ames Research Center for the production of the SPOC data products. 
This research has used the Exoplanet Follow-up Observation Program (ExoFOP; DOI: 10.26134/ExoFOP5) website, which is operated by the California Institute of Technology, under contract with the National Aeronautics and Space Administration under the Exoplanet Exploration Program.
This research has used the NASA Exoplanet Archive, which is operated by the California Institute of Technology, under contract with the National Aeronautics and Space Administration under the Exoplanet Exploration Program.
This work makes use of observations from the LCOGT network. Part of the LCOGT telescope time was granted by NOIRLab through the Mid-Scale Innovation Program (MSIP). MSIP is funded by NSF.
L.M. acknowledges the financial contribution from the PRIN MUR 2022 project 2022J4H55R. KAC and CNW acknowledge support from the TESS mission via subaward s3449 from MIT.
L.N. acknowledges financial contribution from the INAF Large Grant 2023 ``EXODEMO''.
This work has made use of data from the European Space Agency (ESA) mission {\it Gaia} (\url{https://www.cosmos.esa.int/gaia}), processed by the {\it Gaia} Data Processing and Analysis Consortium (DPAC,
\url{https://www.cosmos.esa.int/web/gaia/dpac/consortium}). Funding for the DPAC has been provided by national institutions, in particular the institutions participating in the {\it Gaia} Multilateral Agreement.
MP acknowledges financial support from the European Union – NextGenerationEU (PRIN MUR 2022 20229R43BH) and the ``Programma di Ricerca Fondamentale INAF 2023''.
The authors gratefully acknowledge the coordination of the HARPS time sharing, which was essential to this work.
\end{acknowledgements}

\bibliographystyle{aa} 
\bibliography{bib.bib}

\begin{thebibliography}{114}
\expandafter\ifx\csname natexlab\endcsname\relax\def\natexlab#1{#1}\fi

\bibitem[{{Adamow}(2017)}]{adamow_2017_pymoogi}
{Adamow}, M.~M. 2017, in American Astronomical Society Meeting Abstracts, Vol. 230, American Astronomical Society Meeting Abstracts \#230, 216.07

\bibitem[{{Akeson} {et~al.}(2013){Akeson}, {Chen}, {Ciardi}, {Crane}, {Good}, {Harbut}, {Jackson}, {Kane}, {Laity}, {Leifer}, {Lynn}, {McElroy}, {Papin}, {Plavchan}, {Ram{\'\i}rez}, {Rey}, {von Braun}, {Wittman}, {Abajian}, {Ali}, {Beichman}, {Beekley}, {Berriman}, {Berukoff}, {Bryden}, {Chan}, {Groom}, {Lau}, {Payne}, {Regelson}, {Saucedo}, {Schmitz}, {Stauffer}, {Wyatt}, \& {Zhang}}]{Akeson2013}
{Akeson}, R.~L., {Chen}, X., {Ciardi}, D., {et~al.} 2013, \pasp, 125, 989

\bibitem[{{Aller} {et~al.}(2020){Aller}, {Lillo-Box}, {Jones}, {Miranda}, \& {Barcel{\'o} Forteza}}]{2020A&A...635A.128A}
{Aller}, A., {Lillo-Box}, J., {Jones}, D., {Miranda}, L.~F., \& {Barcel{\'o} Forteza}, S. 2020, \aap, 635, A128

\bibitem[{{Amarsi} {et~al.}(2015){Amarsi}, {Asplund}, {Collet}, \& {Leenaarts}}]{amarsi_2015_OI_nlte}
{Amarsi}, A.~M., {Asplund}, M., {Collet}, R., \& {Leenaarts}, J. 2015, \mnras, 454, L11

\bibitem[{{Andrews}(2020)}]{andrews2020}
{Andrews}, S.~M. 2020, \araa, 58, 483

\bibitem[{Armitage(2020)}]{armitage2020astrophysics}
Armitage, P.~J. 2020

\bibitem[{{Astropy Collaboration} {et~al.}(2018){Astropy Collaboration}, {Price-Whelan}, {Sip{\H{o}}cz}, {G{\"u}nther}, {Lim}, {Crawford}, {Conseil}, {Shupe}, {Craig}, {Dencheva}, {Ginsburg}, {VanderPlas}, {Bradley}, {P{\'e}rez-Su{\'a}rez}, {de Val-Borro}, {Aldcroft}, {Cruz}, {Robitaille}, {Tollerud}, {Ardelean}, {Babej}, {Bach}, {Bachetti}, {Bakanov}, {Bamford}, {Barentsen}, {Barmby}, {Baumbach}, {Berry}, {Biscani}, {Boquien}, {Bostroem}, {Bouma}, {Brammer}, {Bray}, {Breytenbach}, {Buddelmeijer}, {Burke}, {Calderone}, {Cano Rodr{\'\i}guez}, {Cara}, {Cardoso}, {Cheedella}, {Copin}, {Corrales}, {Crichton}, {D'Avella}, {Deil}, {Depagne}, {Dietrich}, {Donath}, {Droettboom}, {Earl}, {Erben}, {Fabbro}, {Ferreira}, {Finethy}, {Fox}, {Garrison}, {Gibbons}, {Goldstein}, {Gommers}, {Greco}, {Greenfield}, {Groener}, {Grollier}, {Hagen}, {Hirst}, {Homeier}, {Horton}, {Hosseinzadeh}, {Hu}, {Hunkeler}, {Ivezi{\'c}}, {Jain}, {Jenness}, {Kanarek}, {Kendrew}, {Kern}, {Kerzendorf}, {Khvalko}, {King}, {Kirkby}, {Kulkarni},
  {Kumar}, {Lee}, {Lenz}, {Littlefair}, {Ma}, {Macleod}, {Mastropietro}, {McCully}, {Montagnac}, {Morris}, {Mueller}, {Mumford}, {Muna}, {Murphy}, {Nelson}, {Nguyen}, {Ninan}, {N{\"o}the}, {Ogaz}, {Oh}, {Parejko}, {Parley}, {Pascual}, {Patil}, {Patil}, {Plunkett}, {Prochaska}, {Rastogi}, {Reddy Janga}, {Sabater}, {Sakurikar}, {Seifert}, {Sherbert}, {Sherwood-Taylor}, {Shih}, {Sick}, {Silbiger}, {Singanamalla}, {Singer}, {Sladen}, {Sooley}, {Sornarajah}, {Streicher}, {Teuben}, {Thomas}, {Tremblay}, {Turner}, {Terr{\'o}n}, {van Kerkwijk}, {de la Vega}, {Watkins}, {Weaver}, {Whitmore}, {Woillez}, {Zabalza}, \& {Astropy Contributors}}]{2018AJ....156..123A}
{Astropy Collaboration}, {Price-Whelan}, A.~M., {Sip{\H{o}}cz}, B.~M., {et~al.} 2018, \aj, 156, 123

\bibitem[{Baraffe {et~al.}(2015)Baraffe, Homeier, Allard, \& Chabrier}]{baraffe2015new}
Baraffe, I., Homeier, D., Allard, F., \& Chabrier, G. 2015, Astronomy \& Astrophysics, 577, A42

\bibitem[{{Biazzo} {et~al.}(2022){Biazzo}, {D'Orazi}, {Desidera}, {Turrini}, {Benatti}, {Gratton}, {Magrini}, {Sozzetti}, {Baratella}, {Bonomo}, {Borsa}, {Claudi}, {Covino}, {Damasso}, {Di Mauro}, {Lanza}, {Maggio}, {Malavolta}, {Maldonado}, {Marzari}, {Micela}, {Poretti}, {Vitello}, {Affer}, {Bignamini}, {Carleo}, {Cosentino}, {Fiorenzano}, {Giacobbe}, {Harutyunyan}, {Leto}, {Mancini}, {Molinari}, {Molinaro}, {Nardiello}, {Nascimbeni}, {Pagano}, {Pedani}, {Piotto}, {Rainer}, \& {Scandariato}}]{Biazzoetal2022}
{Biazzo}, K., {D'Orazi}, V., {Desidera}, S., {et~al.} 2022, \aap, 664, A161

\bibitem[{Bitsch {et~al.}(2018)Bitsch, Morbidelli, Johansen, Lega, Lambrechts, \& Crida}]{bitsch2018pebble}
Bitsch, B., Morbidelli, A., Johansen, A., {et~al.} 2018, Astronomy \& Astrophysics, 612, A30

\bibitem[{{Bourrier} {et~al.}(2018){Bourrier}, {Lecavelier des Etangs}, {Ehrenreich}, {Sanz-Forcada}, {Allart}, {Ballester}, {Buchhave}, {Cohen}, {Deming}, {Evans}, {Garc{\'\i}a Mu{\~n}oz}, {Henry}, {Kataria}, {Lavvas}, {Lewis}, {L{\'o}pez-Morales}, {Marley}, {Sing}, \& {Wakeford}}]{2018A&A...620A.147B}
{Bourrier}, V., {Lecavelier des Etangs}, A., {Ehrenreich}, D., {et~al.} 2018, \aap, 620, A147

\bibitem[{{Brown} {et~al.}(2013){Brown}, {Baliber}, {Bianco}, {Bowman}, {Burleson}, {Conway}, {Crellin}, {Depagne}, {De Vera}, {Dilday}, {Dragomir}, {Dubberley}, {Eastman}, {Elphick}, {Falarski}, {Foale}, {Ford}, {Fulton}, {Garza}, {Gomez}, {Graham}, {Greene}, {Haldeman}, {Hawkins}, {Haworth}, {Haynes}, {Hidas}, {Hjelstrom}, {Howell}, {Hygelund}, {Lister}, {Lobdill}, {Martinez}, {Mullins}, {Norbury}, {Parrent}, {Paulson}, {Petry}, {Pickles}, {Posner}, {Rosing}, {Ross}, {Sand}, {Saunders}, {Shobbrook}, {Shporer}, {Street}, {Thomas}, {Tsapras}, {Tufts}, {Valenti}, {Vander Horst}, {Walker}, {White}, \& {Willis}}]{2013PASP..125.1031B}
{Brown}, T.~M., {Baliber}, N., {Bianco}, F.~B., {et~al.} 2013, \pasp, 125, 1031

\bibitem[{{Bryson} {et~al.}(2020){Bryson}, {Jenkins}, {Klaus}, {Cote}, {Quintana}, {Campbell}, {Zamudio}, {Chandrasekaran}, {Caldwell}, {Van Cleve}, \& {Haas}}]{2020ksci.rept....3B}
{Bryson}, S.~T., {Jenkins}, J.~M., {Klaus}, T.~C., {et~al.} 2020, {Kepler Data Processing Handbook: Target and Aperture Definitions: Selecting Pixels for Kepler Downlink}, Kepler Science Document KSCI-19081-003, id. 3. Edited by Jon M. Jenkins.

\bibitem[{{Bryson} {et~al.}(2010){Bryson}, {Jenkins}, {Klaus}, {Cote}, {Quintana}, {Hall}, {Ibrahim}, {Chandrasekaran}, {Caldwell}, {Van Cleve}, \& {Haas}}]{2010SPIE.7740E..1DB}
{Bryson}, S.~T., {Jenkins}, J.~M., {Klaus}, T.~C., {et~al.} 2010, in Society of Photo-Optical Instrumentation Engineers (SPIE) Conference Series, Vol. 7740, Software and Cyberinfrastructure for Astronomy, ed. N.~M. {Radziwill} \& A.~{Bridger}, 77401D

\bibitem[{Burn \& Mordasini(2025)}]{Burn2025}
Burn, R. \& Mordasini, C. 2025, Planetary Population Synthesis, ed. H.~J. Deeg \& J.~A. Belmonte (Cham: Springer Nature Switzerland), 1--60

\bibitem[{{Caldiroli} {et~al.}(2021){Caldiroli}, {Haardt}, {Gallo}, {Spinelli}, {Malsky}, \& {Rauscher}}]{Caldiroli21}
{Caldiroli}, A., {Haardt}, F., {Gallo}, E., {et~al.} 2021, \aap, 655, A30

\bibitem[{{Castelli} \& {Kurucz}(2003)}]{castelli_kurukz_2003}
{Castelli}, F. \& {Kurucz}, R.~L. 2003, in Modelling of Stellar Atmospheres, ed. N.~{Piskunov}, W.~W. {Weiss}, \& D.~F. {Gray}, Vol. 210, A20

\bibitem[{{Castro-Gonz{\'a}lez} {et~al.}(2024{\natexlab{a}}){Castro-Gonz{\'a}lez}, {Bourrier}, {Lillo-Box}, {Delisle}, {Armstrong}, {Barrado}, \& {Correia}}]{2024A&A...689A.250C}
{Castro-Gonz{\'a}lez}, A., {Bourrier}, V., {Lillo-Box}, J., {et~al.} 2024{\natexlab{a}}, \aap, 689, A250

\bibitem[{{Castro-Gonz{\'a}lez} {et~al.}(2024{\natexlab{b}}){Castro-Gonz{\'a}lez}, {Lillo-Box}, {Armstrong}, {Acu{\~n}a}, {Aguichine}, {Bourrier}, {Gandhi}, {Sousa}, {Delgado-Mena}, {Moya}, {Adibekyan}, {Correia}, {Barrado}, {Damasso}, {Winn}, {Santos}, {Barkaoui}, {Barros}, {Benkhaldoun}, {Bouchy}, {Brice{\~n}o}, {Caldwell}, {Collins}, {Essack}, {Ghachoui}, {Gillon}, {Hounsell}, {Jehin}, {Jenkins}, {Keniger}, {Law}, {Mann}, {Nielsen}, {Pozuelos}, {Schanche}, {Seager}, {Tan}, {Timmermans}, {Villase{\~n}or}, {Watkins}, \& {Ziegler}}]{2024A&A...691A.233C}
{Castro-Gonz{\'a}lez}, A., {Lillo-Box}, J., {Armstrong}, D.~J., {et~al.} 2024{\natexlab{b}}, \aap, 691, A233

\bibitem[{{Chambers}(2001)}]{chambers2001}
{Chambers}, J.~E. 2001, \icarus, 152, 205

\bibitem[{{Choi} {et~al.}(2016){Choi}, {Dotter}, {Conroy}, {Cantiello}, {Paxton}, \& {Johnson}}]{2016ApJ...823..102C}
{Choi}, J., {Dotter}, A., {Conroy}, C., {et~al.} 2016, \apj, 823, 102

\bibitem[{{Ciardi} {et~al.}(2015){Ciardi}, {Beichman}, {Horch}, \& {Howell}}]{2015ApJ...805...16C}
{Ciardi}, D.~R., {Beichman}, C.~A., {Horch}, E.~P., \& {Howell}, S.~B. 2015, \apj, 805, 16

\bibitem[{{Collins} {et~al.}(2017){Collins}, {Kielkopf}, {Stassun}, \& {Hessman}}]{2017AJ....153...77C}
{Collins}, K.~A., {Kielkopf}, J.~F., {Stassun}, K.~G., \& {Hessman}, F.~V. 2017, \aj, 153, 77

\bibitem[{{Cutri} {et~al.}(2021){Cutri}, {Wright}, {Conrow}, {Fowler}, {Eisenhardt}, {Grillmair}, {Kirkpatrick}, {Masci}, {McCallon}, {Wheelock}, {Fajardo-Acosta}, {Yan}, {Benford}, {Harbut}, {Jarrett}, {Lake}, {Leisawitz}, {Ressler}, {Stanford}, {Tsai}, {Liu}, {Helou}, {Mainzer}, {Gettngs}, {Gonzalez}, {Hoffman}, {Marsh}, {Padgett}, {Skrutskie}, {Beck}, {Papin}, \& {Wittman}}]{2014yCat.2328....0C}
{Cutri}, R.~M., {Wright}, E.~L., {Conrow}, T., {et~al.} 2021, {VizieR Online Data Catalog: AllWISE Data Release (Cutri+ 2013)}, VizieR On-line Data Catalog: II/328. Originally published in: IPAC/Caltech (2013)

\bibitem[{{Dekany} {et~al.}(2013){Dekany}, {Roberts}, {Burruss}, {Bouchez}, {Truong}, {Baranec}, {Guiwits}, {Hale}, {Angione}, {Trinh}, {Zolkower}, {Shelton}, {Palmer}, {Henning}, {Croner}, {Troy}, {McKenna}, {Tesch}, {Hildebrandt}, \& {Milburn}}]{2013ApJ...776..130D}
{Dekany}, R., {Roberts}, J., {Burruss}, R., {et~al.} 2013, \apj, 776, 130

\bibitem[{Doyle {et~al.}(2014)Doyle, Davies, Smalley, Chaplin, \& Elsworth}]{doyle_2014}
Doyle, A.~P., Davies, G.~R., Smalley, B., Chaplin, W.~J., \& Elsworth, Y. 2014, Monthly Notices of the Royal Astronomical Society, 444, 3592

\bibitem[{{Eastman}(2017)}]{2017ascl.soft10003E}
{Eastman}, J. 2017, {EXOFASTv2: Generalized publication-quality exoplanet modeling code}, Astrophysics Source Code Library, record ascl:1710.003

\bibitem[{{Eastman} {et~al.}(2013){Eastman}, {Gaudi}, \& {Agol}}]{2013PASP..125...83E}
{Eastman}, J., {Gaudi}, B.~S., \& {Agol}, E. 2013, \pasp, 125, 83

\bibitem[{{Eastman} {et~al.}(2019){Eastman}, {Rodriguez}, {Agol}, {Stassun}, {Beatty}, {Vanderburg}, {Gaudi}, {Collins}, \& {Luger}}]{Eastman2019}
{Eastman}, J.~D., {Rodriguez}, J.~E., {Agol}, E., {et~al.} 2019, arXiv e-prints, arXiv:1907.09480

\bibitem[{{Ehrenreich} {et~al.}(2015){Ehrenreich}, {Bourrier}, {Wheatley}, {Lecavelier des Etangs}, {H{\'e}brard}, {Udry}, {Bonfils}, {Delfosse}, {D{\'e}sert}, {Sing}, \& {Vidal-Madjar}}]{2015Natur.522..459E}
{Ehrenreich}, D., {Bourrier}, V., {Wheatley}, P.~J., {et~al.} 2015, \nat, 522, 459

\bibitem[{{Espinoza}(2018)}]{2018RNAAS...2..209E}
{Espinoza}, N. 2018, Research Notes of the American Astronomical Society, 2, 209

\bibitem[{{Espinoza} {et~al.}(2019){Espinoza}, {Kossakowski}, \& {Brahm}}]{2019MNRAS.490.2262E}
{Espinoza}, N., {Kossakowski}, D., \& {Brahm}, R. 2019, \mnras, 490, 2262

\bibitem[{{Espinoza-Retamal} {et~al.}(2024){Espinoza-Retamal}, {Stef{\'a}nsson}, {Petrovich}, {Brahm}, {Jord{\'a}n}, {Sedaghati}, {Lucero}, {Tala Pinto}, {Mu{\~n}oz}, {Boyle}, {Leiva}, \& {Suc}}]{2024AJ....168..185E}
{Espinoza-Retamal}, J.~I., {Stef{\'a}nsson}, G., {Petrovich}, C., {et~al.} 2024, \aj, 168, 185

\bibitem[{{Esposito} {et~al.}(2014){Esposito}, {Covino}, {Mancini}, {Harutyunyan}, {Southworth}, {Biazzo}, {Gandolfi}, {Lanza}, {Barbieri}, {Bonomo}, {Borsa}, {Claudi}, {Cosentino}, {Desidera}, {Gratton}, {Pagano}, {Sozzetti}, {Boccato}, {Maggio}, {Micela}, {Molinari}, {Nascimbeni}, {Piotto}, {Poretti}, \& {Smareglia}}]{Esposito2014}
{Esposito}, M., {Covino}, E., {Mancini}, L., {et~al.} 2014, \aap, 564, L13

\bibitem[{{Foreman-Mackey} {et~al.}(2017){Foreman-Mackey}, {Agol}, {Ambikasaran}, \& {Angus}}]{2017AJ....154..220F}
{Foreman-Mackey}, D., {Agol}, E., {Ambikasaran}, S., \& {Angus}, R. 2017, \aj, 154, 220

\bibitem[{{Fortney} {et~al.}(2007){Fortney}, {Marley}, \& {Barnes}}]{2007ApJ...659.1661F}
{Fortney}, J.~J., {Marley}, M.~S., \& {Barnes}, J.~W. 2007, \apj, 659, 1661

\bibitem[{{Fulton} {et~al.}(2018){Fulton}, {Petigura}, {Blunt}, \& {Sinukoff}}]{2018PASP..130d4504F}
{Fulton}, B.~J., {Petigura}, E.~A., {Blunt}, S., \& {Sinukoff}, E. 2018, \pasp, 130, 044504

\bibitem[{{Furlan} {et~al.}(2017){Furlan}, {Ciardi}, {Everett}, {Saylors}, {Teske}, {Horch}, {Howell}, {van Belle}, {Hirsch}, {Gautier}, {Adams}, {Barrado}, {Cartier}, {Dressing}, {Dupree}, {Gilliland}, {Lillo-Box}, {Lucas}, \& {Wang}}]{2017AJ....153...71F}
{Furlan}, E., {Ciardi}, D.~R., {Everett}, M.~E., {et~al.} 2017, \aj, 153, 71

\bibitem[{{Gaia Collaboration} {et~al.}(2023){Gaia Collaboration}, {Vallenari}, {Brown}, {Prusti}, {de Bruijne}, {Arenou}, {Babusiaux}, {Biermann}, {Creevey}, {Ducourant}, {Evans}, {Eyer}, {Guerra}, {Hutton}, {Jordi}, {Klioner}, {Lammers}, {Lindegren}, {Luri}, {Mignard}, {Panem}, {Pourbaix}, {Randich}, {Sartoretti}, {Soubiran}, {Tanga}, {Walton}, {Bailer-Jones}, {Bastian}, {Drimmel}, {Jansen}, {Katz}, {Lattanzi}, {van Leeuwen}, {Bakker}, {Cacciari}, {Casta{\~n}eda}, {De Angeli}, {Fabricius}, {Fouesneau}, {Fr{\'e}mat}, {Galluccio}, {Guerrier}, {Heiter}, {Masana}, {Messineo}, {Mowlavi}, {Nicolas}, {Nienartowicz}, {Pailler}, {Panuzzo}, {Riclet}, {Roux}, {Seabroke}, {Sordo}, {Th{\'e}venin}, {Gracia-Abril}, {Portell}, {Teyssier}, {Altmann}, {Andrae}, {Audard}, {Bellas-Velidis}, {Benson}, {Berthier}, {Blomme}, {Burgess}, {Busonero}, {Busso}, {C{\'a}novas}, {Carry}, {Cellino}, {Cheek}, {Clementini}, {Damerdji}, {Davidson}, {de Teodoro}, {Nu{\~n}ez Campos}, {Delchambre}, {Dell'Oro}, {Esquej},
  {Fern{\'a}ndez-Hern{\'a}ndez}, {Fraile}, {Garabato}, {Garc{\'\i}a-Lario}, {Gosset}, {Haigron}, {Halbwachs}, {Hambly}, {Harrison}, {Hern{\'a}ndez}, {Hestroffer}, {Hodgkin}, {Holl}, {Jan{\ss}en}, {Jevardat de Fombelle}, {Jordan}, {Krone-Martins}, {Lanzafame}, {L{\"o}ffler}, {Marchal}, {Marrese}, {Moitinho}, {Muinonen}, {Osborne}, {Pancino}, {Pauwels}, {Recio-Blanco}, {Reyl{\'e}}, {Riello}, {Rimoldini}, {Roegiers}, {Rybizki}, {Sarro}, {Siopis}, {Smith}, {Sozzetti}, {Utrilla}, {van Leeuwen}, {Abbas}, {{\'A}brah{\'a}m}, {Abreu Aramburu}, {Aerts}, {Aguado}, {Ajaj}, {Aldea-Montero}, {Altavilla}, {{\'A}lvarez}, {Alves}, {Anders}, {Anderson}, {Anglada Varela}, {Antoja}, {Baines}, {Baker}, {Balaguer-N{\'u}{\~n}ez}, {Balbinot}, {Balog}, {Barache}, {Barbato}, {Barros}, {Barstow}, {Bartolom{\'e}}, {Bassilana}, {Bauchet}, {Becciani}, {Bellazzini}, {Berihuete}, {Bernet}, {Bertone}, {Bianchi}, {Binnenfeld}, {Blanco-Cuaresma}, {Blazere}, {Boch}, {Bombrun}, {Bossini}, {Bouquillon}, {Bragaglia}, {Bramante}, {Breedt},
  {Bressan}, {Brouillet}, {Brugaletta}, {Bucciarelli}, {Burlacu}, {Butkevich}, {Buzzi}, {Caffau}, {Cancelliere}, {Cantat-Gaudin}, {Carballo}, {Carlucci}, {Carnerero}, {Carrasco}, {Casamiquela}, {Castellani}, {Castro-Ginard}, {Chaoul}, {Charlot}, {Chemin}, {Chiaramida}, {Chiavassa}, {Chornay}, {Comoretto}, {Contursi}, {Cooper}, {Cornez}, {Cowell}, {Crifo}, {Cropper}, {Crosta}, {Crowley}, {Dafonte}, {Dapergolas}, {David}, {David}, {de Laverny}, {De Luise}, \& {De March}}]{2023A&A...674A...1G}
{Gaia Collaboration}, {Vallenari}, A., {Brown}, A.~G.~A., {et~al.} 2023, \aap, 674, A1

\bibitem[{{Gomes da Silva} {et~al.}(2018){Gomes da Silva}, {Figueira}, {Santos}, \& {Faria}}]{actin}
{Gomes da Silva}, J., {Figueira}, P., {Santos}, N., \& {Faria}, J. 2018, The Journal of Open Source Software, 3, 667

\bibitem[{{Grunblatt} {et~al.}(2024){Grunblatt}, {Saunders}, {Huber}, {Thorngren}, {Vissapragada}, {Yoshida}, {Schlaufman}, {Giacalone}, {Macdougall}, {Chontos}, {Turtelboom}, {Beard}, {Murphy}, {Rice}, {Isaacson}, {Angus}, \& {Howard}}]{2024AJ....168....1G}
{Grunblatt}, S.~K., {Saunders}, N., {Huber}, D., {et~al.} 2024, \aj, 168, 1

\bibitem[{{Guerrero} {et~al.}(2021){Guerrero}, {Seager}, {Huang}, {Vanderburg}, {Garcia Soto}, {Mireles}, {Hesse}, {Fong}, {Glidden}, {Shporer}, {Latham}, {Collins}, {Quinn}, {Burt}, {Dragomir}, {Crossfield}, {Vanderspek}, {Fausnaugh}, {Burke}, {Ricker}, {Daylan}, {Essack}, {G{\"u}nther}, {Osborn}, {Pepper}, {Rowden}, {Sha}, {Villanueva}, {Yahalomi}, {Yu}, {Ballard}, {Batalha}, {Berardo}, {Chontos}, {Dittmann}, {Esquerdo}, {Mikal-Evans}, {Jayaraman}, {Krishnamurthy}, {Louie}, {Mehrle}, {Niraula}, {Rackham}, {Rodriguez}, {Rowden}, {Sousa-Silva}, {Watanabe}, {Wong}, {Zhan}, {Zivanovic}, {Christiansen}, {Ciardi}, {Swain}, {Lund}, {Mullally}, {Fleming}, {Rodriguez}, {Boyd}, {Quintana}, {Barclay}, {Col{\'o}n}, {Rinehart}, {Schlieder}, {Clampin}, {Jenkins}, {Twicken}, {Caldwell}, {Coughlin}, {Henze}, {Lissauer}, {Morris}, {Rose}, {Smith}, {Tenenbaum}, {Ting}, {Wohler}, {Bakos}, {Bean}, {Berta-Thompson}, {Bieryla}, {Bouma}, {Buchhave}, {Butler}, {Charbonneau}, {Doty}, {Ge}, {Holman}, {Howard}, {Kaltenegger}, {Kane},
  {Kjeldsen}, {Kreidberg}, {Lin}, {Minsky}, {Narita}, {Paegert}, {P{\'a}l}, {Palle}, {Sasselov}, {Spencer}, {Sozzetti}, {Stassun}, {Torres}, {Udry}, \& {Winn}}]{2021ApJS..254...39G}
{Guerrero}, N.~M., {Seager}, S., {Huang}, C.~X., {et~al.} 2021, \apjs, 254, 39

\bibitem[{{Hawthorn} {et~al.}(2023){Hawthorn}, {Bayliss}, {Wilson}, {Bonfanti}, {Adibekyan}, {Alibert}, {Sousa}, {Collins}, {Bryant}, {Osborn}, {Armstrong}, {Abe}, {Acton}, {Addison}, {Agabi}, {Alonso}, {Alves}, {Anglada-Escud{\'e}}, {B{\'a}rczy}, {Barclay}, {Barrado}, {Barros}, {Baumjohann}, {Bendjoya}, {Benz}, {Bieryla}, {Bonfils}, {Bouchy}, {Brandeker}, {Broeg}, {Brown}, {Burleigh}, {Buttu}, {Cabrera}, {Caldwell}, {Casewell}, {Charbonneau}, {Charnoz}, {Cloutier}, {Collier Cameron}, {Collins}, {Conti}, {Crouzet}, {Czismadia}, {Davies}, {Deleuil}, {Delgado-Mena}, {Delrez}, {Demangeon}, {Demory}, {Dransfield}, {Dumusque}, {Egger}, {Ehrenreich}, {Eigm{\"u}ller}, {Erickson}, {Essack}, {Fortier}, {Fossati}, {Fridlund}, {G{\"u}nther}, {G{\"u}del}, {Gandolfi}, {Gillard}, {Gillon}, {Gnilka}, {Goad}, {Goeke}, {Guillot}, {Hadjigeorghiou}, {Hellier}, {Henderson}, {Heng}, {Hooton}, {Horne}, {Howell}, {Hoyer}, {Irwin}, {Jenkins}, {Jenkins}, {Jensen}, {Kane}, {Kendall}, {Kielkopf}, {Kiss}, {Lacedelli}, {Laskar},
  {Latham}, {Etangs}, {Leleu}, {Lendl}, {Lillo-Box}, {Lovis}, {M{\'e}karnia}, {Massey}, {Masters}, {Maxted}, {Nascimbeni}, {Nielsen}, {O'Brien}, {Olofsson}, {Osborn}, {Pagano}, {Pall{\'e}}, {Persson}, {Piotto}, {Plavchan}, {Pollacco}, {Queloz}, {Ragazzoni}, {Rauer}, {Ribas}, {Ricker}, {S{\'e}gransan}, {Salmon}, {Santerne}, {Santos}, {Scandariato}, {Schmider}, {Schwarz}, {Seager}, {Shporer}, {Simon}, {Smith}, {Srdoc}, {Steller}, {Suarez}, {Szab{\'o}}, {Teske}, {Thomas}, {Tilbrook}, {Triaud}, {Udry}, {Van Grootel}, {Walton}, {Wang}, {Wheatley}, {Winn}, {Wittenmyer}, \& {Zhang}}]{2023MNRAS.520.3649H}
{Hawthorn}, F., {Bayliss}, D., {Wilson}, T.~G., {et~al.} 2023, \mnras, 520, 3649

\bibitem[{{Hayward} {et~al.}(2001){Hayward}, {Brandl}, {Pirger}, {Blacken}, {Gull}, {Schoenwald}, \& {Houck}}]{2001PASP..113..105H}
{Hayward}, T.~L., {Brandl}, B., {Pirger}, B., {et~al.} 2001, \pasp, 113, 105

\bibitem[{{Henden} {et~al.}(2016){Henden}, {Templeton}, {Terrell}, {Smith}, {Levine}, \& {Welch}}]{Henden2016}
{Henden}, A.~A., {Templeton}, M., {Terrell}, D., {et~al.} 2016, {VizieR Online Data Catalog: AAVSO Photometric All Sky Survey (APASS) DR9 (Henden+, 2016)}, VizieR On-line Data Catalog: II/336. Originally published in: 2015AAS...22533616H

\bibitem[{{H{\o}g} {et~al.}(2000){H{\o}g}, {Fabricius}, {Makarov}, {Urban}, {Corbin}, {Wycoff}, {Bastian}, {Schwekendiek}, \& {Wicenec}}]{2000A&A...355L..27H}
{H{\o}g}, E., {Fabricius}, C., {Makarov}, V.~V., {et~al.} 2000, \aap, 355, L27

\bibitem[{Ida {et~al.}(2016)Ida, Guillot, \& Morbidelli}]{ida2016radial}
Ida, S., Guillot, T., \& Morbidelli, A. 2016, Astronomy \& Astrophysics, 591, A72

\bibitem[{{Ionov} {et~al.}(2018){Ionov}, {Pavlyuchenkov}, \& {Shematovich}}]{2018MNRAS.476.5639I}
{Ionov}, D.~E., {Pavlyuchenkov}, Y.~N., \& {Shematovich}, V.~I. 2018, \mnras, 476, 5639

\bibitem[{{Jenkins} {et~al.}(2016){Jenkins}, {Twicken}, {McCauliff}, {Campbell}, {Sanderfer}, {Lung}, {Mansouri-Samani}, {Girouard}, {Tenenbaum}, {Klaus}, {Smith}, {Caldwell}, {Chacon}, {Henze}, {Heiges}, {Latham}, {Morgan}, {Swade}, {Rinehart}, \& {Vanderspek}}]{2016SPIE.9913E..3EJ}
{Jenkins}, J.~M., {Twicken}, J.~D., {McCauliff}, S., {et~al.} 2016, in Society of Photo-Optical Instrumentation Engineers (SPIE) Conference Series, Vol. 9913, Software and Cyberinfrastructure for Astronomy IV, ed. G.~{Chiozzi} \& J.~C. {Guzman}, 99133E

\bibitem[{Johansen {et~al.}(2019)Johansen, Ida, \& Brasser}]{johansen2019planetary}
Johansen, A., Ida, S., \& Brasser, R. 2019, Astronomy \& Astrophysics, 622, A202

\bibitem[{{Johnstone} {et~al.}(2021){Johnstone}, {Bartel}, \& {G{\"u}del}}]{2021A&A...649A..96J}
{Johnstone}, C.~P., {Bartel}, M., \& {G{\"u}del}, M. 2021, \aap, 649, A96

\bibitem[{{Kempton} {et~al.}(2018){Kempton}, {Bean}, {Louie}, {Deming}, {Koll}, {Mansfield}, {Christiansen}, {L{\'o}pez-Morales}, {Swain}, {Zellem}, {Ballard}, {Barclay}, {Barstow}, {Batalha}, {Beatty}, {Berta-Thompson}, {Birkby}, {Buchhave}, {Charbonneau}, {Cowan}, {Crossfield}, {de Val-Borro}, {Doyon}, {Dragomir}, {Gaidos}, {Heng}, {Hu}, {Kane}, {Kreidberg}, {Mallonn}, {Morley}, {Narita}, {Nascimbeni}, {Pall{\'e}}, {Quintana}, {Rauscher}, {Seager}, {Shkolnik}, {Sing}, {Sozzetti}, {Stassun}, {Valenti}, \& {von Essen}}]{2018PASP..130k4401K}
{Kempton}, E. M.~R., {Bean}, J.~L., {Louie}, D.~R., {et~al.} 2018, \pasp, 130, 114401

\bibitem[{{Kipping}(2013)}]{2013MNRAS.435.2152K}
{Kipping}, D.~M. 2013, \mnras, 435, 2152

\bibitem[{{Koskinen} {et~al.}(2022){Koskinen}, {Lavvas}, {Huang}, {Bergsten}, {Fernandes}, \& {Young}}]{2022ApJ...929...52K}
{Koskinen}, T.~T., {Lavvas}, P., {Huang}, C., {et~al.} 2022, \apj, 929, 52

\bibitem[{{Kreidberg}(2015)}]{2015PASP..127.1161K}
{Kreidberg}, L. 2015, \pasp, 127, 1161

\bibitem[{{Kunimoto} {et~al.}(2023){Kunimoto}, {Vanderburg}, {Huang}, {Davis}, {Affer}, {Cameron}, {Charbonneau}, {Cosentino}, {Damasso}, {Dumusque}, {Fiorenzano}, {Ghedina}, {Haywood}, {Lienhard}, {L{\'o}pez-Morales}, {Mayor}, {Pepe}, {Pinamonti}, {Poretti}, {Maldonado}, {Rice}, {Sozzetti}, {Wilson}, {Udry}, {Baptista}, {Barkaoui}, {Becker}, {Benni}, {Bieryla}, {Bosch-Cabot}, {Ciardi}, {Collins}, {Collins}, {Evans}, {Dupuy}, {Goliguzova}, {Guerra}, {Kraus}, {Lissauer}, {Huber}, {Murgas}, {Palle}, {Quinn}, {Safonov}, {Schwarz}, {Shporer}, {Stassun}, {Jenkins}, {Latham}, {Ricker}, {Seager}, {Vanderspek}, {Winn}, {Essack}, {Lewis}, \& {Rose}}]{2023AJ....166....7K}
{Kunimoto}, M., {Vanderburg}, A., {Huang}, C.~X., {et~al.} 2023, \aj, 166, 7

\bibitem[{{Laskar} \& {Petit}(2017)}]{laskar2017}
{Laskar}, J. \& {Petit}, A.~C. 2017, \aap, 605, A72

\bibitem[{{Lightkurve Collaboration} {et~al.}(2018){Lightkurve Collaboration}, {Cardoso}, {Hedges}, {Gully-Santiago}, {Saunders}, {Cody}, {Barclay}, {Hall}, {Sagear}, {Turtelboom}, {Zhang}, {Tzanidakis}, {Mighell}, {Coughlin}, {Bell}, {Berta-Thompson}, {Williams}, {Dotson}, \& {Barentsen}}]{lightkurve}
{Lightkurve Collaboration}, {Cardoso}, J. V. d.~M., {Hedges}, C., {et~al.} 2018, {Lightkurve: Kepler and TESS time series analysis in Python}, Astrophysics Source Code Library, record ascl:1812.013

\bibitem[{{Locci} {et~al.}(2019){Locci}, {Cecchi-Pestellini}, \& {Micela}}]{Locci19}
{Locci}, D., {Cecchi-Pestellini}, C., \& {Micela}, G. 2019, \aap, 624, A101

\bibitem[{{Maggio} {et~al.}(2022){Maggio}, {Locci}, {Pillitteri}, {Benatti}, {Claudi}, {Desidera}, {Micela}, {Damasso}, {Sozzetti}, \& {Suarez Mascare{\~n}o}}]{Maggio22}
{Maggio}, A., {Locci}, D., {Pillitteri}, I., {et~al.} 2022, \apj, 925, 172

\bibitem[{{Mamajek} \& {Hillenbrand}(2008)}]{2008ApJ...687.1264M}
{Mamajek}, E.~E. \& {Hillenbrand}, L.~A. 2008, \apj, 687, 1264

\bibitem[{{Mantovan} {et~al.}(2024{\natexlab{a}}){Mantovan}, {Malavolta}, {Locci}, {Polychroni}, {Turrini}, {Maggio}, {Desidera}, {Spinelli}, {Benatti}, {Piotto}, {Lanza}, {Marzari}, {Sozzetti}, {Damasso}, {Nardiello}, {Cabona}, {D'Arpa}, {Guilluy}, {Mancini}, {Micela}, {Nascimbeni}, \& {Zingales}}]{mantovan2024}
{Mantovan}, G., {Malavolta}, L., {Locci}, D., {et~al.} 2024{\natexlab{a}}, \aap, 684, L17

\bibitem[{{Mantovan} {et~al.}(2024{\natexlab{b}}){Mantovan}, {Wilson}, {Borsato}, {Zingales}, {Biazzo}, {Nardiello}, {Malavolta}, {Desidera}, {Marzari}, {Collier Cameron}, {Nascimbeni}, {Majidi}, {Montalto}, {Piotto}, {Stassun}, {Winn}, {Jenkins}, {Mignon}, {Bieryla}, {Latham}, {Barkaoui}, {Collins}, {Evans}, {Fausnaugh}, {Granata}, {Kostov}, {Mann}, {Pozuelos}, {Radford}, {Relles}, {Rowden}, {Seager}, {Tan}, {Timmermans}, \& {Watkins}}]{2024A&A...691A..67M}
{Mantovan}, G., {Wilson}, T.~G., {Borsato}, L., {et~al.} 2024{\natexlab{b}}, \aap, 691, A67

\bibitem[{{Matsakos} \& {K{\"o}nigl}(2016)}]{2016ApJ...820L...8M}
{Matsakos}, T. \& {K{\"o}nigl}, A. 2016, \apjl, 820, L8

\bibitem[{{Mayor} {et~al.}(2003){Mayor}, {Pepe}, {Queloz}, {Bouchy}, {Rupprecht}, {Lo Curto}, {Avila}, {Benz}, {Bertaux}, {Bonfils}, {Dall}, {Dekker}, {Delabre}, {Eckert}, {Fleury}, {Gilliotte}, {Gojak}, {Guzman}, {Kohler}, {Lizon}, {Longinotti}, {Lovis}, {Megevand}, {Pasquini}, {Reyes}, {Sivan}, {Sosnowska}, {Soto}, {Udry}, {van Kesteren}, {Weber}, \& {Weilenmann}}]{2003Msngr.114...20M}
{Mayor}, M., {Pepe}, F., {Queloz}, D., {et~al.} 2003, The Messenger, 114, 20

\bibitem[{{Mazeh} {et~al.}(2016){Mazeh}, {Holczer}, \& {Faigler}}]{2016A&A...589A..75M}
{Mazeh}, T., {Holczer}, T., \& {Faigler}, S. 2016, \aap, 589, A75

\bibitem[{{McCully} {et~al.}(2018){McCully}, {Volgenau}, {Harbeck}, {Lister}, {Saunders}, {Turner}, {Siiverd}, \& {Bowman}}]{2018SPIE10707E..0KM}
{McCully}, C., {Volgenau}, N.~H., {Harbeck}, D.-R., {et~al.} 2018, in Society of Photo-Optical Instrumentation Engineers (SPIE) Conference Series, Vol. 10707, Software and Cyberinfrastructure for Astronomy V, ed. J.~C. {Guzman} \& J.~{Ibsen}, 107070K

\bibitem[{{Naponiello} {et~al.}(2025{\natexlab{a}}){Naponiello}, {Bonomo}, {Mancini}, {Steinmeyer}, {Biazzo}, {Polychroni}, {Dorn}, {Turrini}, {Lanza}, {Sozzetti}, {Desidera}, {Damasso}, {Collins}, {Carleo}, {Collins}, {Colombo}, {D'Arpa}, {Dumusque}, {Gonz{\'a}lez}, {Guilluy}, {Lorenzi}, {Mantovan}, {Nardiello}, {Pinamonti}, {Schwarz}, {Singh}, {Watkins}, \& {Zingales}}]{2025A&A...693A...7N}
{Naponiello}, L., {Bonomo}, A.~S., {Mancini}, L., {et~al.} 2025{\natexlab{a}}, \aap, 693, A7

\bibitem[{{Naponiello} {et~al.}(2022){Naponiello}, {Mancini}, {Damasso}, {Bonomo}, {Sozzetti}, {Nardiello}, {Biazzo}, {Stognone}, {Lillo-Box}, {Lanza}, {Poretti}, {Lissauer}, {Zeng}, {Bieryla}, {H{\'e}brard}, {Basilicata}, {Benatti}, {Bignamini}, {Borsa}, {Claudi}, {Cosentino}, {Covino}, {de Gurtubai}, {Delfosse}, {Desidera}, {Dragomir}, {Eastman}, {Essack}, {Fiorenzano}, {Giacobbe}, {Harutyunyan}, {Heidari}, {Hellier}, {Jenkins}, {Knapic}, {K{\"o}nig}, {Latham}, {Magazz{\`u}}, {Maggio}, {Maldonado}, {Micela}, {Molinari}, {Molinaro}, {Morgan}, {Moutou}, {Nascimbeni}, {Pace}, {Pagano}, {Pedani}, {Piotto}, {Pinamonti}, {Quintana}, {Rainer}, {Ricker}, {Seager}, {Twicken}, {Vanderspek}, \& {Winn}}]{2022A&A...667A...8N}
{Naponiello}, L., {Mancini}, L., {Damasso}, M., {et~al.} 2022, \aap, 667, A8

\bibitem[{{Naponiello} {et~al.}(2023){Naponiello}, {Mancini}, {Sozzetti}, {Bonomo}, {Morbidelli}, {Dou}, {Zeng}, {Leinhardt}, {Biazzo}, {Cubillos}, {Pinamonti}, {Locci}, {Maggio}, {Damasso}, {Lanza}, {Lissauer}, {Collins}, {Carter}, {Jensen}, {Bignamini}, {Boschin}, {Bouma}, {Ciardi}, {Cosentino}, {Crossfield}, {Desidera}, {Dumusque}, {Fiorenzano}, {Fukui}, {Giacobbe}, {Gnilka}, {Ghedina}, {Guilluy}, {Harutyunyan}, {Howell}, {Jenkins}, {Lund}, {Kielkopf}, {Lester}, {Malavolta}, {Mann}, {Matson}, {Matthews}, {Nardiello}, {Narita}, {Pace}, {Pagano}, {Palle}, {Pedani}, {Seager}, {Schlieder}, {Schwarz}, {Shporer}, {Twicken}, {Winn}, {Ziegler}, \& {Zingales}}]{2023Natur.622..255N}
{Naponiello}, L., {Mancini}, L., {Sozzetti}, A., {et~al.} 2023, \nat, 622, 255

\bibitem[{{Naponiello} {et~al.}(2025{\natexlab{b}}){Naponiello}, {Vissapragada}, {Bonomo}, {Steinmeyer}, {Filomeno}, {D'Orazi}, {Dorn}, {Sozzetti}, {Mancini}, {Lanza}, {Biazzo}, {Watkins}, {H{\'e}brard}, {Lissauer}, {Howell}, {Ciardi}, {Mantovan}, {Baker}, {Bourrier}, {Buchhave}, {Clark}, {Collins}, {Cosentino}, {Damasso}, {Dumusque}, {Fiorenzano}, {Forveille}, {Heidari}, {Latham}, {Littlefield}, {L{\'o}pez-Morales}, {Lund}, {Malavolta}, {Manni}, {Nardiello}, {Pinamonti}, {Yee}, {Zambelli}, {Ziegler}, \& {Zingales}}]{2025arXiv250510123N}
{Naponiello}, L., {Vissapragada}, S., {Bonomo}, A.~S., {et~al.} 2025{\natexlab{b}}, arXiv e-prints, arXiv:2505.10123

\bibitem[{{Osborn} {et~al.}(2023){Osborn}, {Armstrong}, {Fern{\'a}ndez Fern{\'a}ndez}, {Knierim}, {Adibekyan}, {Collins}, {Delgado-Mena}, {Fridlund}, {Gomes da Silva}, {Hellier}, {Jackson}, {King}, {Lillo-Box}, {Matson}, {Matthews}, {Santos}, {Sousa}, {Stassun}, {Tan}, {Ricker}, {Vanderspek}, {Latham}, {Seager}, {Winn}, {Jenkins}, {Bayliss}, {Bouma}, {Ciardi}, {Collins}, {Col{\'o}n}, {Crossfield}, {Demangeon}, {D{\'\i}az}, {Dorn}, {Dumusque}, {Keniger}, {Figueira}, {Gan}, {Goeke}, {Hadjigeorghiou}, {Hawthorn}, {Helled}, {Howell}, {Nielsen}, {Osborn}, {Quinn}, {Sefako}, {Shporer}, {Str{\o}m}, {Twicken}, {Vanderburg}, \& {Wheatley}}]{2023MNRAS.526..548O}
{Osborn}, A., {Armstrong}, D.~J., {Fern{\'a}ndez Fern{\'a}ndez}, J., {et~al.} 2023, \mnras, 526, 548

\bibitem[{{Owen} \& {Jackson}(2012)}]{2012MNRAS.425.2931O}
{Owen}, J.~E. \& {Jackson}, A.~P. 2012, \mnras, 425, 2931

\bibitem[{{Owen} \& {Lai}(2018)}]{2018MNRAS.479.5012O}
{Owen}, J.~E. \& {Lai}, D. 2018, \mnras, 479, 5012

\bibitem[{{Paxton} {et~al.}(2015){Paxton}, {Marchant}, {Schwab}, {Bauer}, {Bildsten}, {Cantiello}, {Dessart}, {Farmer}, {Hu}, {Langer}, {Townsend}, {Townsley}, \& {Timmes}}]{Paxton2015}
{Paxton}, B., {Marchant}, P., {Schwab}, J., {et~al.} 2015, \apjs, 220, 15

\bibitem[{{Pecaut} \& {Mamajek}(2013)}]{pecaut_mamajek_2013_spt}
{Pecaut}, M.~J. \& {Mamajek}, E.~E. 2013, \apjs, 208, 9

\bibitem[{{Penz} \& {Micela}(2008)}]{2008A&A...479..579P}
{Penz}, T. \& {Micela}, G. 2008, \aap, 479, 579

\bibitem[{{Penz} {et~al.}(2008){Penz}, {Micela}, \& {Lammer}}]{2008A&A...477..309P}
{Penz}, T., {Micela}, G., \& {Lammer}, H. 2008, \aap, 477, 309

\bibitem[{{Pepe} {et~al.}(2014){Pepe}, {Ehrenreich}, \& {Meyer}}]{2014Natur.513..358P}
{Pepe}, F., {Ehrenreich}, D., \& {Meyer}, M.~R. 2014, \nat, 513, 358

\bibitem[{{Pepe} {et~al.}(2002){Pepe}, {Mayor}, {Rupprecht}, {Avila}, {Ballester}, {Beckers}, {Benz}, {Bertaux}, {Bouchy}, {Buzzoni}, {Cavadore}, {Deiries}, {Dekker}, {Delabre}, {D'Odorico}, {Eckert}, {Fischer}, {Fleury}, {George}, {Gilliotte}, {Gojak}, {Guzman}, {Koch}, {Kohler}, {Kotzlowski}, {Lacroix}, {Le Merrer}, {Lizon}, {Lo Curto}, {Longinotti}, {Megevand}, {Pasquini}, {Petitpas}, {Pichard}, {Queloz}, {Reyes}, {Richaud}, {Sivan}, {Sosnowska}, {Soto}, {Udry}, {Ureta}, {van Kesteren}, {Weber}, {Weilenmann}, {Wicenec}, {Wieland}, {Christensen-Dalsgaard}, {Dravins}, {Hatzes}, {K{\"u}rster}, {Paresce}, \& {Penny}}]{2002Msngr.110....9P}
{Pepe}, F., {Mayor}, M., {Rupprecht}, G., {et~al.} 2002, The Messenger, 110, 9

\bibitem[{Pirani {et~al.}(2019)Pirani, Johansen, Bitsch, Mustill, \& Turrini}]{pirani2019consequences}
Pirani, S., Johansen, A., Bitsch, B., Mustill, A.~J., \& Turrini, D. 2019, Astronomy \& Astrophysics, 623, A169

\bibitem[{{Pizzolato} {et~al.}(2003){Pizzolato}, {Maggio}, {Micela}, {Sciortino}, \& {Ventura}}]{2003A&A...397..147P}
{Pizzolato}, N., {Maggio}, A., {Micela}, G., {Sciortino}, S., \& {Ventura}, P. 2003, \aap, 397, 147

\bibitem[{Polychroni {et~al.}(2023)Polychroni, Turrini, \& Pirani}]{polychroni_2023_10593198}
Polychroni, D., Turrini, D., \& Pirani, S. 2023, GroMiT: Planet Growth and Migration Track code

\bibitem[{{Ricker} {et~al.}(2015){Ricker}, {Winn}, {Vanderspek}, {Latham}, {Bakos}, {Bean}, {Berta-Thompson}, {Brown}, {Buchhave}, {Butler}, {Butler}, {Chaplin}, {Charbonneau}, {Christensen-Dalsgaard}, {Clampin}, {Deming}, {Doty}, {De Lee}, {Dressing}, {Dunham}, {Endl}, {Fressin}, {Ge}, {Henning}, {Holman}, {Howard}, {Ida}, {Jenkins}, {Jernigan}, {Johnson}, {Kaltenegger}, {Kawai}, {Kjeldsen}, {Laughlin}, {Levine}, {Lin}, {Lissauer}, {MacQueen}, {Marcy}, {McCullough}, {Morton}, {Narita}, {Paegert}, {Palle}, {Pepe}, {Pepper}, {Quirrenbach}, {Rinehart}, {Sasselov}, {Sato}, {Seager}, {Sozzetti}, {Stassun}, {Sullivan}, {Szentgyorgyi}, {Torres}, {Udry}, \& {Villasenor}}]{2015JATIS...1a4003R}
{Ricker}, G.~R., {Winn}, J.~N., {Vanderspek}, R., {et~al.} 2015, Journal of Astronomical Telescopes, Instruments, and Systems, 1, 014003

\bibitem[{{Rosotti}(2023)}]{rosotti2023}
{Rosotti}, G.~P. 2023, \nar, 96, 101674

\bibitem[{{Sanz-Forcada} {et~al.}(2025){Sanz-Forcada}, {L{\'o}pez-Puertas}, {Lamp{\'o}n}, {Czesla}, {Nortmann}, {Caballero}, {Zapatero Osorio}, {Amado}, {Murgas}, {Orell-Miquel}, {Pall{\'e}}, {Quirrenbach}, {Reiners}, {Ribas}, {S{\'a}nchez-L{\'o}pez}, \& {Solano}}]{2025A&A...693A.285S}
{Sanz-Forcada}, J., {L{\'o}pez-Puertas}, M., {Lamp{\'o}n}, M., {et~al.} 2025, \aap, 693, A285

\bibitem[{{Skrutskie} {et~al.}(2006){Skrutskie}, {Cutri}, {Stiening}, {Weinberg}, {Schneider}, {Carpenter}, {Beichman}, {Capps}, {Chester}, {Elias}, {Huchra}, {Liebert}, {Lonsdale}, {Monet}, {Price}, {Seitzer}, {Jarrett}, {Kirkpatrick}, {Gizis}, {Howard}, {Evans}, {Fowler}, {Fullmer}, {Hurt}, {Light}, {Kopan}, {Marsh}, {McCallon}, {Tam}, {Van Dyk}, \& {Wheelock}}]{2006AJ....131.1163S}
{Skrutskie}, M.~F., {Cutri}, R.~M., {Stiening}, R., {et~al.} 2006, \aj, 131, 1163

\bibitem[{{Smith} {et~al.}(2012){Smith}, {Stumpe}, {Van Cleve}, {Jenkins}, {Barclay}, {Fanelli}, {Girouard}, {Kolodziejczak}, {McCauliff}, {Morris}, \& {Twicken}}]{2012PASP..124.1000S}
{Smith}, J.~C., {Stumpe}, M.~C., {Van Cleve}, J.~E., {et~al.} 2012, \pasp, 124, 1000

\bibitem[{{Sneden}(1973)}]{sneden1973}
{Sneden}, C. 1973, \apj, 184, 839

\bibitem[{{Sousa} {et~al.}(2015){Sousa}, {Santos}, {Adibekyan}, {Delgado-Mena}, \& {Israelian}}]{sousa_2015_ares_v2}
{Sousa}, S.~G., {Santos}, N.~C., {Adibekyan}, V., {Delgado-Mena}, E., \& {Israelian}, G. 2015, \aap, 577, A67

\bibitem[{{Southworth}(2011)}]{2011MNRAS.417.2166S}
{Southworth}, J. 2011, \mnras, 417, 2166

\bibitem[{{Speagle}(2020)}]{2020MNRAS.493.3132S}
{Speagle}, J.~S. 2020, \mnras, 493, 3132

\bibitem[{{Stassun} {et~al.}(2019){Stassun}, {Oelkers}, {Paegert}, {Torres}, {Pepper}, {De Lee}, {Collins}, {Latham}, {Muirhead}, {Chittidi}, {Rojas-Ayala}, {Fleming}, {Rose}, {Tenenbaum}, {Ting}, {Kane}, {Barclay}, {Bean}, {Brassuer}, {Charbonneau}, {Ge}, {Lissauer}, {Mann}, {McLean}, {Mullally}, {Narita}, {Plavchan}, {Ricker}, {Sasselov}, {Seager}, {Sharma}, {Shiao}, {Sozzetti}, {Stello}, {Vanderspek}, {Wallace}, \& {Winn}}]{2019AJ....158..138S}
{Stassun}, K.~G., {Oelkers}, R.~J., {Paegert}, M., {et~al.} 2019, \aj, 158, 138

\bibitem[{{Stassun} {et~al.}(2018){Stassun}, {Oelkers}, {Pepper}, {Paegert}, {De Lee}, {Torres}, {Latham}, {Charpinet}, {Dressing}, {Huber}, {Kane}, {L{\'e}pine}, {Mann}, {Muirhead}, {Rojas-Ayala}, {Silvotti}, {Fleming}, {Levine}, \& {Plavchan}}]{2018AJ....156..102S}
{Stassun}, K.~G., {Oelkers}, R.~J., {Pepper}, J., {et~al.} 2018, \aj, 156, 102

\bibitem[{{Stumpe} {et~al.}(2014){Stumpe}, {Smith}, {Catanzarite}, {Van Cleve}, {Jenkins}, {Twicken}, \& {Girouard}}]{2014PASP..126..100S}
{Stumpe}, M.~C., {Smith}, J.~C., {Catanzarite}, J.~H., {et~al.} 2014, \pasp, 126, 100

\bibitem[{{Stumpe} {et~al.}(2012){Stumpe}, {Smith}, {Van Cleve}, {Twicken}, {Barclay}, {Fanelli}, {Girouard}, {Jenkins}, {Kolodziejczak}, {McCauliff}, \& {Morris}}]{2012PASP..124..985S}
{Stumpe}, M.~C., {Smith}, J.~C., {Van Cleve}, J.~E., {et~al.} 2012, \pasp, 124, 985

\bibitem[{{Szab{\'o}} {et~al.}(2023){Szab{\'o}}, {K{\'a}lm{\'a}n}, {Borsato}, {Heged{\H{u}}s}, {M{\'e}sz{\'a}ros}, \& {Szab{\'o}}}]{2023A&A...671A.132S}
{Szab{\'o}}, G.~M., {K{\'a}lm{\'a}n}, S., {Borsato}, L., {et~al.} 2023, \aap, 671, A132

\bibitem[{{Szentgyorgyi} \& {Fur{\'e}sz}(2007)}]{2007RMxAC..28..129S}
{Szentgyorgyi}, A.~H. \& {Fur{\'e}sz}, G. 2007, in Revista Mexicana de Astronomia y Astrofisica Conference Series, Vol.~28, Revista Mexicana de Astronomia y Astrofisica Conference Series, ed. S.~{Kurtz}, 129--133

\bibitem[{Tanaka {et~al.}(2020)Tanaka, Murase, \& Tanigawa}]{tanaka2020final}
Tanaka, H., Murase, K., \& Tanigawa, T. 2020, The Astrophysical Journal, 891, 143

\bibitem[{{Testi} {et~al.}(2022){Testi}, {Natta}, {Manara}, {de Gregorio Monsalvo}, {Lodato}, {Lopez}, {Muzic}, {Pascucci}, {Sanchis}, {Miranda}, {Scholz}, {De Simone}, \& {Williams}}]{Testi2022}
{Testi}, L., {Natta}, A., {Manara}, C.~F., {et~al.} 2022, \aap, 663, A98

\bibitem[{{Thompson} {et~al.}(2018){Thompson}, {Coughlin}, {Hoffman}, {Mullally}, {Christiansen}, {Burke}, {Bryson}, {Batalha}, {Haas}, {Catanzarite}, {Rowe}, {Barentsen}, {Caldwell}, {Clarke}, {Jenkins}, {Li}, {Latham}, {Lissauer}, {Mathur}, {Morris}, {Seader}, {Smith}, {Klaus}, {Twicken}, {Van Cleve}, {Wohler}, {Akeson}, {Ciardi}, {Cochran}, {Henze}, {Howell}, {Huber}, {Pr{\v{s}}a}, {Ram{\'\i}rez}, {Morton}, {Barclay}, {Campbell}, {Chaplin}, {Charbonneau}, {Christensen-Dalsgaard}, {Dotson}, {Doyle}, {Dunham}, {Dupree}, {Ford}, {Geary}, {Girouard}, {Isaacson}, {Kjeldsen}, {Quintana}, {Ragozzine}, {Shabram}, {Shporer}, {Silva Aguirre}, {Steffen}, {Still}, {Tenenbaum}, {Welsh}, {Wolfgang}, {Zamudio}, {Koch}, \& {Borucki}}]{2018ApJS..235...38T}
{Thompson}, S.~E., {Coughlin}, J.~L., {Hoffman}, K., {et~al.} 2018, \apjs, 235, 38

\bibitem[{{Thorngren} {et~al.}(2023){Thorngren}, {Lee}, \& {Lopez}}]{2023ApJ...945L..36T}
{Thorngren}, D.~P., {Lee}, E.~J., \& {Lopez}, E.~D. 2023, \apjl, 945, L36

\bibitem[{{Tinetti} {et~al.}(2022){Tinetti}, {Eccleston}, {Lueftinger}, {Salvignol}, {Fahmy}, \& {Alves de Oliveira}}]{2022EPSC...16.1114T}
{Tinetti}, G., {Eccleston}, P., {Lueftinger}, T., {et~al.} 2022, in European Planetary Science Congress, EPSC2022--1114

\bibitem[{{Tokovinin} {et~al.}(2018){Tokovinin}, {Mason}, {Hartkopf}, {Mendez}, \& {Horch}}]{2018AJ....155..235T}
{Tokovinin}, A., {Mason}, B.~D., {Hartkopf}, W.~I., {Mendez}, R.~A., \& {Horch}, E.~P. 2018, \aj, 155, 235

\bibitem[{Turrini {et~al.}(2023)Turrini, Marzari, Polychroni, Claudi, Desidera, Mesa, Pinamonti, Sozzetti, Mascare{\~n}o, Damasso, {et~al.}}]{turrini2023gaps}
Turrini, D., Marzari, F., Polychroni, D., {et~al.} 2023, Astronomy \& Astrophysics, 679, A55

\bibitem[{{Turrini} {et~al.}(2020){Turrini}, {Zinzi}, \& {Belinchon}}]{turrini2020}
{Turrini}, D., {Zinzi}, A., \& {Belinchon}, J.~A. 2020, \aap, 636, A53

\bibitem[{Twicken {et~al.}(2010)Twicken, Clarke, Bryson, Tenenbaum, Wu, Jenkins, Girouard, \& Klaus}]{Twicken2010}
Twicken, J.~D., Clarke, B.~D., Bryson, S.~T., {et~al.} 2010, in Software and Cyberinfrastructure for Astronomy, ed. N.~M. Radziwill \& A.~Bridger, Vol. 7740, International Society for Optics and Photonics (SPIE), 749 -- 760

\bibitem[{{Vissapragada} \& {Behmard}(2025)}]{2025AJ....169..117V}
{Vissapragada}, S. \& {Behmard}, A. 2025, \aj, 169, 117

\bibitem[{{Wilson} {et~al.}(2022){Wilson}, {Goffo}, {Alibert}, {Gandolfi}, {Bonfanti}, {Persson}, {Collier Cameron}, {Fridlund}, {Fossati}, {Korth}, {Benz}, {Deline}, {Flor{\'e}n}, {Guterman}, {Adibekyan}, {Hooton}, {Hoyer}, {Leleu}, {Mustill}, {Salmon}, {Sousa}, {Suarez}, {Abe}, {Agabi}, {Alonso}, {Anglada}, {Asquier}, {B{\'a}rczy}, {Barrado Navascues}, {Barros}, {Baumjohann}, {Beck}, {Beck}, {Billot}, {Bonfils}, {Brandeker}, {Broeg}, {Bryant}, {Burleigh}, {Buttu}, {Cabrera}, {Charnoz}, {Ciardi}, {Cloutier}, {Cochran}, {Collins}, {Col{\'o}n}, {Crouzet}, {Csizmadia}, {Davies}, {Deleuil}, {Delrez}, {Demangeon}, {Demory}, {Dragomir}, {Dransfield}, {Ehrenreich}, {Erikson}, {Fortier}, {Gan}, {Gill}, {Gillon}, {Gnilka}, {Grieves}, {Grziwa}, {G{\"u}del}, {Guillot}, {Haldemann}, {Heng}, {Horne}, {Howell}, {Isaak}, {Jenkins}, {Jensen}, {Kiss}, {Lacedelli}, {Lam}, {Laskar}, {Latham}, {Lecavelier des Etangs}, {Lendl}, {Lester}, {Levine}, {Livingston}, {Lovis}, {Luque}, {Magrin}, {Marie-Sainte}, {Maxted}, {Mayo},
  {McLean}, {Mecina}, {M{\'e}karnia}, {Nascimbeni}, {Nielsen}, {Olofsson}, {Osborn}, {Osborne}, {Ottensamer}, {Pagano}, {Pall{\'e}}, {Peter}, {Piotto}, {Pollacco}, {Queloz}, {Ragazzoni}, {Rando}, {Rauer}, {Redfield}, {Ribas}, {Ricker}, {Rieder}, {Santos}, {Scandariato}, {Schmider}, {Schwarz}, {Scott}, {Seager}, {S{\'e}gransan}, {Serrano}, {Simon}, {Smith}, {Steller}, {Stockdale}, {Szab{\'o}}, {Thomas}, {Ting}, {Triaud}, {Udry}, {Van Eylen}, {Van Grootel}, {Vanderspek}, {Viotto}, {Walton}, \& {Winn}}]{2022MNRAS.511.1043W}
{Wilson}, T.~G., {Goffo}, E., {Alibert}, Y., {et~al.} 2022, \mnras, 511, 1043

\bibitem[{{Wright} {et~al.}(2018){Wright}, {Newton}, {Williams}, {Drake}, \& {Yadav}}]{2018MNRAS.479.2351W}
{Wright}, N.~J., {Newton}, E.~R., {Williams}, P. K.~G., {Drake}, J.~J., \& {Yadav}, R.~K. 2018, \mnras, 479, 2351

\bibitem[{{Zechmeister} \& {K{\"u}rster}(2009)}]{2009A&A...496..577Z}
{Zechmeister}, M. \& {K{\"u}rster}, M. 2009, \aap, 496, 577

\bibitem[{{Zechmeister} {et~al.}(2018){Zechmeister}, {Reiners}, {Amado}, {Azzaro}, {Bauer}, {B{\'e}jar}, {Caballero}, {Guenther}, {Hagen}, {Jeffers}, {Kaminski}, {K{\"u}rster}, {Launhardt}, {Montes}, {Morales}, {Quirrenbach}, {Reffert}, {Ribas}, {Seifert}, {Tal-Or}, \& {Wolthoff}}]{2018A&A...609A..12Z}
{Zechmeister}, M., {Reiners}, A., {Amado}, P.~J., {et~al.} 2018, \aap, 609, A12

\bibitem[{{Ziegler} {et~al.}(2020){Ziegler}, {Tokovinin}, {Brice{\~n}o}, {Mang}, {Law}, \& {Mann}}]{2020AJ....159...19Z}
{Ziegler}, C., {Tokovinin}, A., {Brice{\~n}o}, C., {et~al.} 2020, \aj, 159, 19

\bibitem[{{Zingales} {et~al.}(2025){Zingales}, {Malavolta}, {Borsato}, {Turrini}, {Bonfanti}, {Polychroni}, {Mantovan}, {Nardiello}, {Nascimbeni}, {Lanza}, {Bekkelien}, {Sozzetti}, {Broeg}, {Naponiello}, {Lendl}, {Bonomo}, {Simon}, {Desidera}, {Piotto}, {Mancini}, {Hooton}, {Bignamini}, {Egger}, {Maggio}, {Alibert}, {Locci}, {Delrez}, {Biassoni}, {Fossati}, {Cabona}, {Lacedelli}, {Carleo}, {Leonardi}, {Andreuzzi}, {Brandeker}, {Cosentino}, {Correia}, {Claudi}, {Alonso}, {Damasso}, {Wilson}, {B{\'a}rczy}, {Pinamonti}, {Baker}, {Barkaoui}, {Barrado Navascues}, {Barros}, {Baumjohann}, {Beck}, {Beichman}, {Benz}, {Bieryla}, {Billot}, {Bosch-Cabot}, {Bouma}, {Ciardi}, {Collier Cameron}, {Collins}, {Crossfield}, {Csizmadia}, {Cubillos}, {Davies}, {Deleuil}, {Deline}, {Demangeon}, {Demory}, {Derekas}, {Dragomir}, {Edwards}, {Ehrenreich}, {Erikson}, {Falk}, {Fortier}, {Fridlund}, {Fukui}, {Gandolfi}, {Gazeas}, {Gillon}, {Gonzales}, {G{\"u}del}, {Guerra}, {G{\"u}nther}, {Heitzmann}, {Helling}, {Howell}, {Isaak},
  {Jenkins}, {Kiss}, {Korth}, {Lam}, {Laskar}, {Lecavelier des Etangs}, {Magrin}, {Matson}, {Matthews}, {Maxted}, {McDermott}, {Munari}, {Mordasini}, {Narita}, {Olofsson}, {Ottensamer}, {Pagano}, {Pall{\'e}}, {Peter}, {Pollacco}, {Queloz}, {Ragazzoni}, {Rando}, {Ratti}, {Rauer}, {Ribas}, {Salmon}, {Santos}, {Scandariato}, {Seager}, {S{\'e}gransan}, {Smith}, {Schlieder}, {Schwarz}, {Shporer}, {Sousa}, {Stalport}, {Steinberger}, {Sulis}, {Szab{\'o}}, {Twicken}, {Udry}, {Van Grootel}, {Venturini}, {Villaver}, {Walton}, \& {Winn}}]{zingales2025}
{Zingales}, T., {Malavolta}, L., {Borsato}, L., {et~al.} 2025, \aap, 695, A273

\end{thebibliography}

\begin{appendix}


\section{Additional figures and tables}\label{sec:appendix}

\begin{figure}[h!]
\centering
\includegraphics[width=0.5\textwidth]{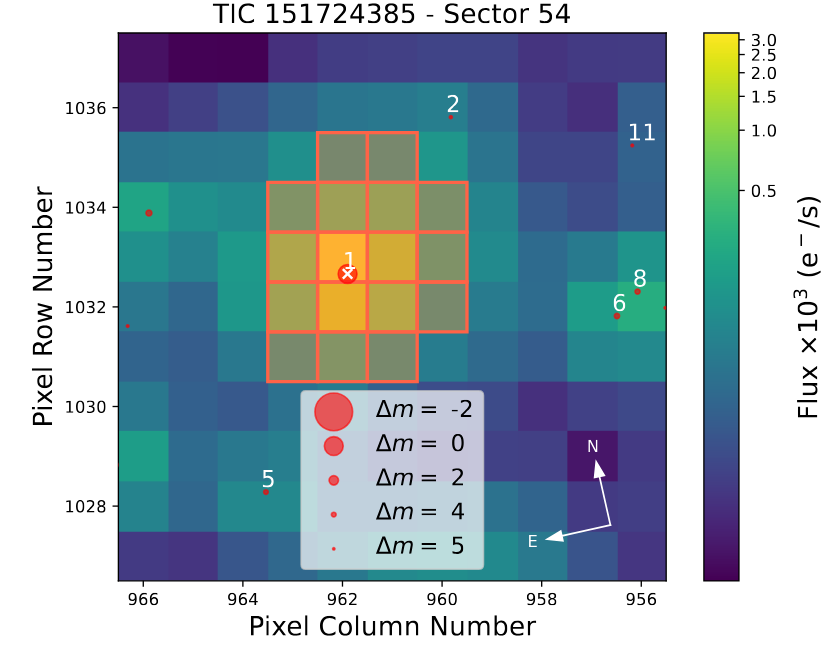}
\includegraphics[width=0.5\textwidth]{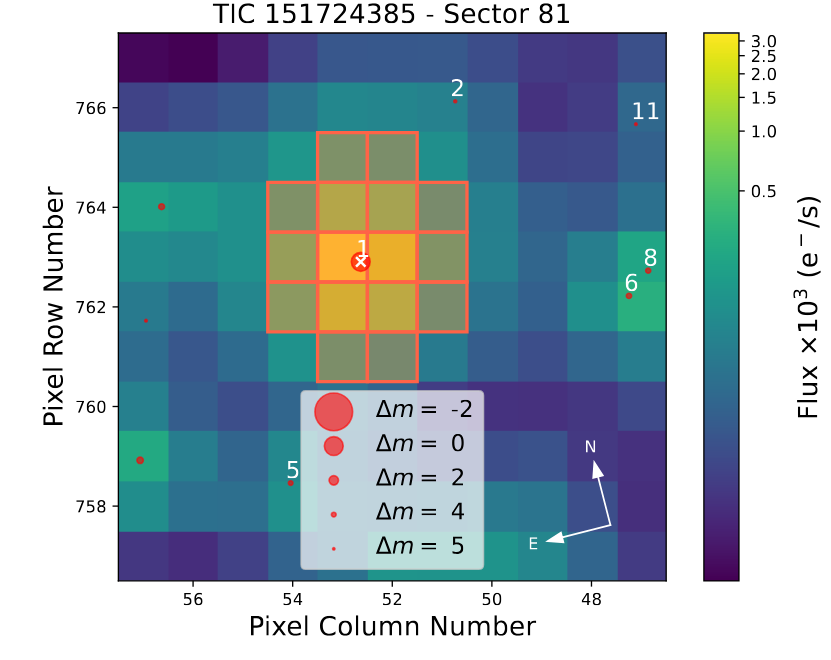}
\includegraphics[width=0.5\textwidth]{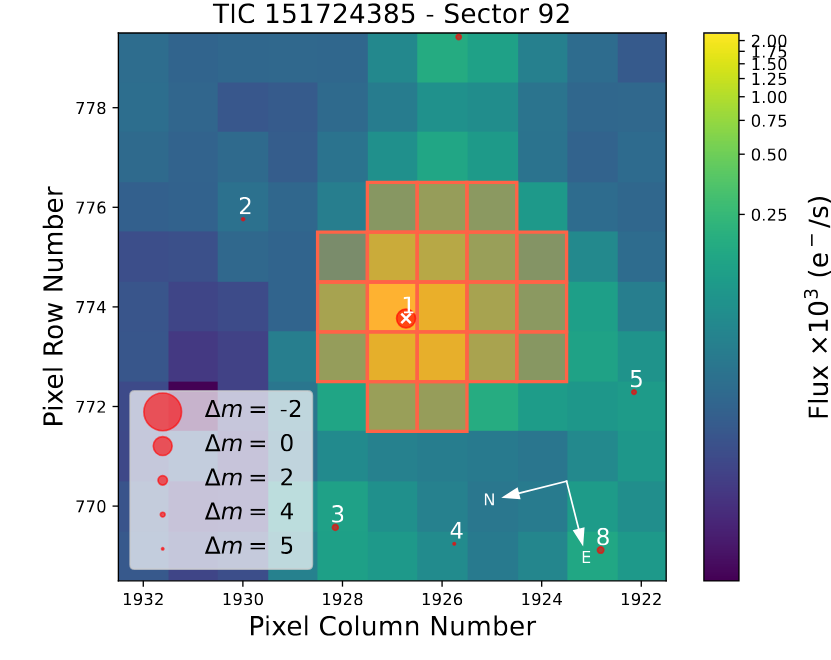}
\caption{Target pixel files from the TESS observation of Sector~54 (top panel), Sector~81 (middle panel), and Sector~92 (bottom panel). TOI-5795 is indicated by a white cross. The SPOC pipeline aperture is shown by shaded red squares. Gaia~DR3 catalogue \citep{2023A&A...674A...1G} is overlaid with symbol sizes proportional to the magnitude difference with TOI-5795.}
\label{fig:tpfs}
\end{figure}

\begin{table}[h!]
\centering %
\caption[]{TRES and HARPS radial-velocities and relative errors of TOI-5795.} %
\label{tab:RV_data} %
\begin{tabular}{l r c c}
\hline\hline \\[-8pt] %
~~~~~BJD$_{\rm TDB}$ & RV~~~ & $\sigma_{\rm RV}$ & \textbf{$\sigma_{\rm RV \, with \, jtter}$}\\
~~~~~~$[{\rm days}]$ & [m\,s$^{-1}$] & [m\,s$^{-1}$] & [m\,s$^{-1}$] \\
\hline  \\[-6pt] %
\multicolumn{1}{l}{{\bf HARPS}} \\ [2pt] %
2460490.8329 & $-9.20$~~~~  & 1.98 & 4.24\\
2460492.6482 & $-6.26$~~~~  & 2.27 & 4.39\\
2460493.7177 & $9.37$~~~~   & 2.23 & 4.36\\
2460495.7240 & $-13.45$~~~~ & 4.29 & 5.69\\
2460497.7079 & $-8.41$~~~~  & 2.29 & 4.39\\
2460498.6601 & $-1.96$~~~~  & 1.96 & 4.23\\
2460499.7998 & $3.12$~~~~   & 2.09 & 4.29\\
2460500.7406 & $5.94$~~~~   & 2.30 & 4.39\\
2460503.6988 & $-0.26$~~~~  & 3.66 & 5.24\\
2460504.6547 & $-11.44$~~~~ & 2.73 & 4.64\\
2460516.6357 & $-10.17$~~~~ & 3.87 & 5.39\\
2460517.7241 & $2.39$~~~~   & 2.48 & 4.49\\
2460519.7068 & $-3.49$~~~~  & 2.91 & 4.75\\
2460530.6715 & $10.72$~~~~  & 3.00 & 4.80\\
2460531.6555 & $-2.31$~~~~  & 3.84 & 5.37\\
2460532.6626 & $-10.49$~~~~ & 2.63 & 4.58\\
2460533.5801 & $-5.16$~~~~  & 2.89 & 4.73\\
2460534.6866 & $-10.83$~~~~ & 2.19 & 4.34\\
2460535.6932 & $1.77$~~~~   & 1.70 & 4.12\\
\multicolumn{1}{l}{{\bf TRES}} \\ [2pt] %
2459864.5963 & $-59$~~~~  & 30 \\
2459879.6927 & $-31$~~~~ & 21 \\
2460269.6029 & $-23$~~~~ & 31 \\
2460271.5714 & $21$~~~~  & 30 \\
2460272.5841 & $45$~~~~  & 25 \\
2460281.5673 & $81$~~~~  & 28 \\
2460282.5692 & $-23$~~~~ & 15 \\
2460296.5677 & $61$~~~~  & 27 \\
2460554.8071 & $69$~~~~  & 33 \\
2460562.7231 & $63$~~~~  & 33 \\
2460577.6931 & $-7$~~~~ & 24 \\
2460608.5804 & $-56$~~~~ & 19 \\ [2pt] 
\hline %
\end{tabular}
\end{table}

\clearpage
\section{Detection sensitivity}
\label{sec:sensitivity}
\begin{figure}[!hb]
    \centering
    \includegraphics[width=0.49\textwidth]{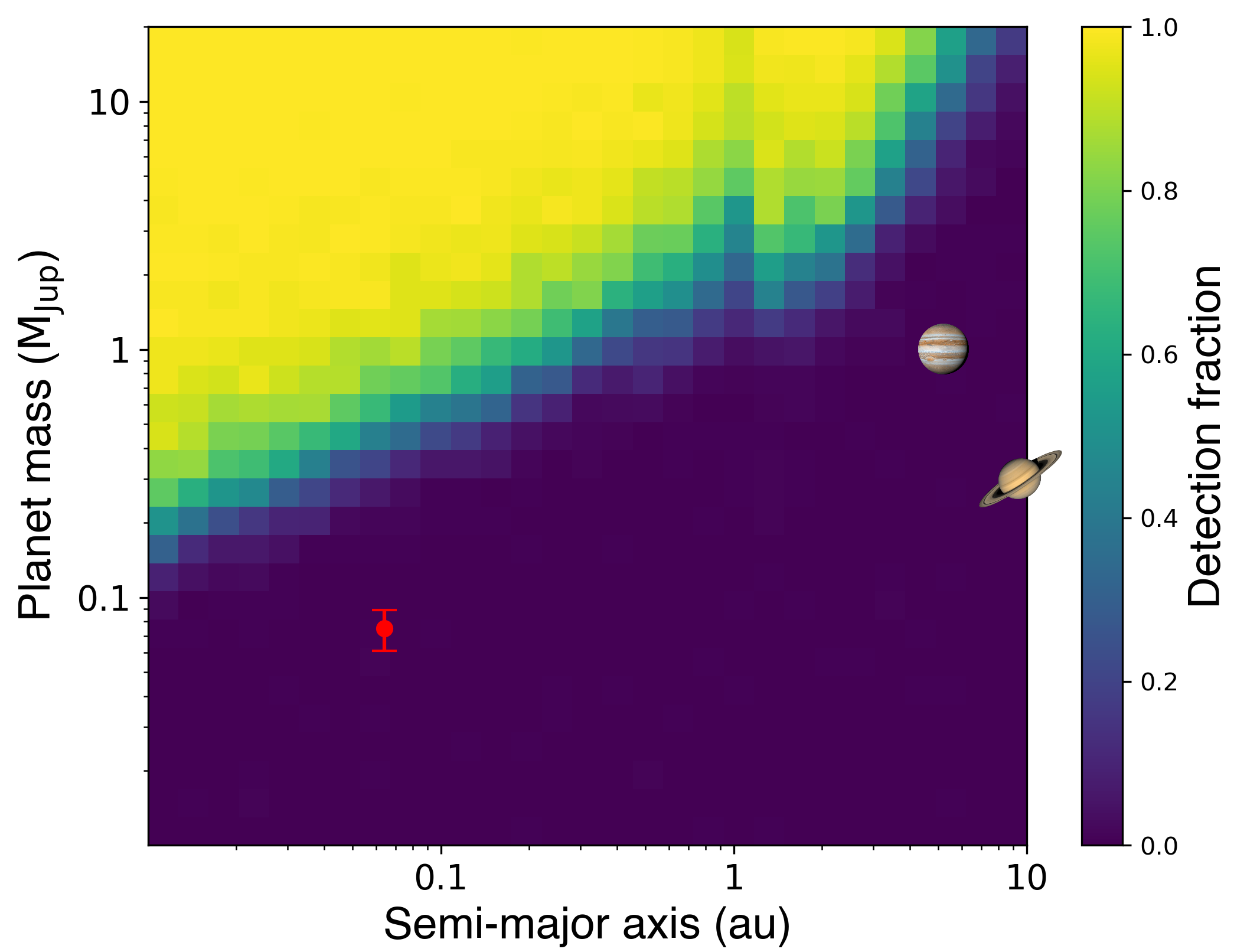}
    \includegraphics[width=0.49\textwidth]{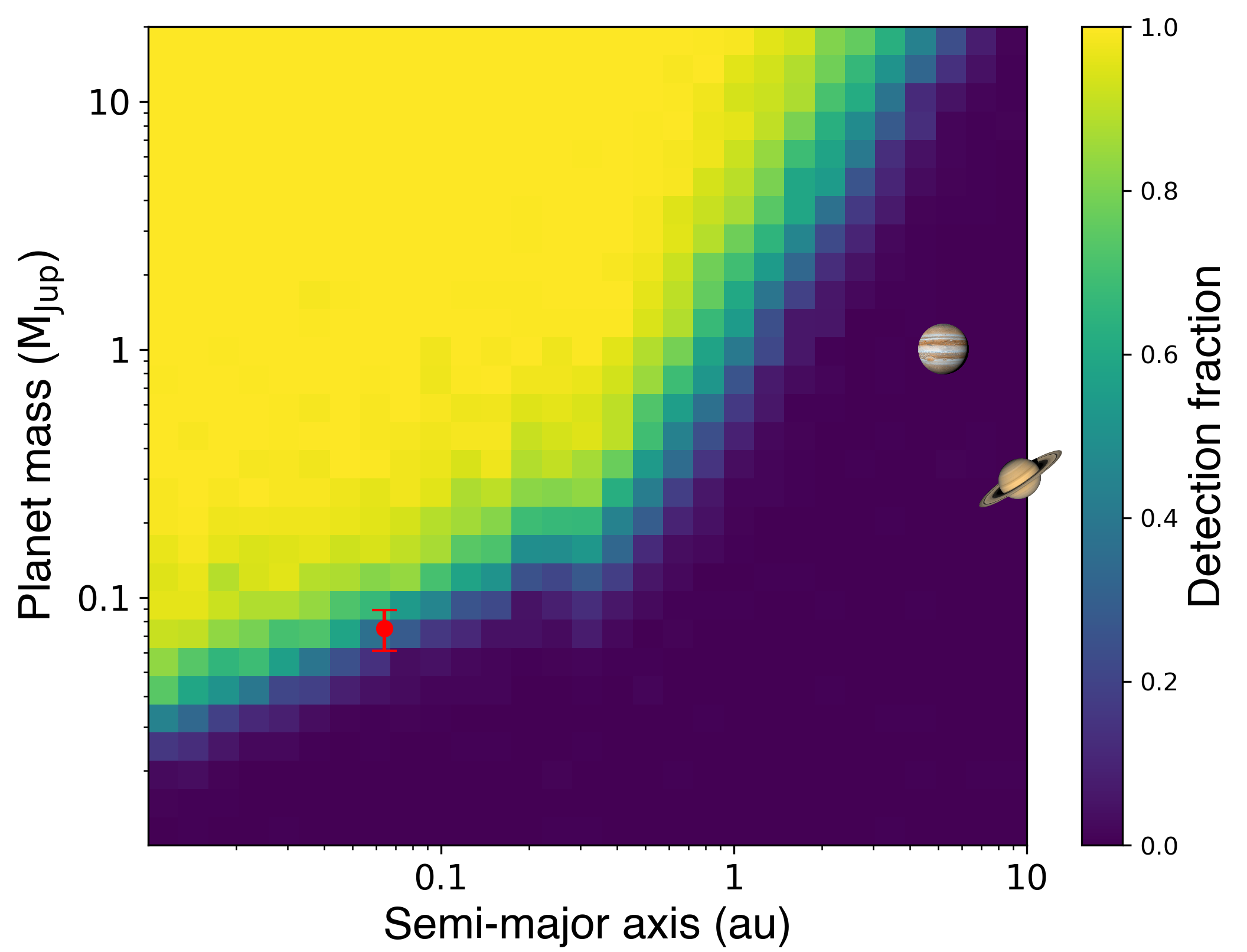}
    \caption{TRES (left) and HARPS (right) RV detection maps for TOI-5795. The color scale expresses the detection fraction, while the red circle marks the position of TOI-5795\,b. Jupiter and Saturn are shown for comparison.}
    \label{fig:completeness}
\end{figure}

We estimated the completeness of both the TRES and HARPS RV time series by performing injection-recovery simulations, in which RVs with synthetic planetary signals were injected at the times of our observations. 
The details of the simulation can be found in \citealt{2025A&A...693A...7N} and the results are shown in Fig.\,\ref{fig:completeness}. Thanks to the HARPS dataset, we are sensitive to planets up to roughly the mass of Jupiter placed at 1\,au, though their signals would only appear as linear and quadratic trends, while TRES is slightly more sensitive to distant, very massive ($\sim10\,M_{\mathrm{jup}}$) planets, despite a lower resolution, owing to its longer timeline.

\end{appendix}

\end{document}